\documentclass[a4paper, 12pt, twoside]{report}

\usepackage{graphicx}
\usepackage{subfigure}
\usepackage[english,ngerman]{babel}
\usepackage[utf8]{inputenc}
\usepackage{amsfonts}
\usepackage{amssymb}
\usepackage{multirow}
\usepackage[errorstop]{feynmf}
\usepackage{float}
\usepackage{empheq}
\usepackage{enumerate}
\usepackage{longtable}
\usepackage{fancyhdr}
\usepackage[toc,page]{appendix}
\usepackage[hang]{caption}
\usepackage{url}
\usepackage[breaklinks=true]{hyperref}
\usepackage{listings}
\usepackage{lmodern}
\usepackage[T1]{fontenc}
\usepackage{tikz}
\usetikzlibrary{shapes,arrows}
\usepackage{amsmath,bm,times}

\numberwithin{equation}{chapter}

\allowdisplaybreaks

\newcommand{\bsf}{\sffamily\bfseries}

\topfigrule
\botfigrule

\begin{document}

\selectlanguage{english}
\unitlength = 1mm

\pagestyle{empty}

{\centering

\LARGE
\textsc{Faculty of Physics and Astronomy}

\vskip5mm

\Large
\textsc{University of Heidelberg}

\vfill

\normalsize

Diploma thesis\\[5pt]
in Physics\\
\vspace{.7cm}
submitted by\\[5pt]
\textbf{Florian Beutler}\\[5pt]
born in Waiblingen\\
\vspace{.7cm}
2008

\par}

\cleardoublepage

{\centering

\vspace{1cm}

{\bsf\large
Thermal description of hadron production in $e^+e^-$ collisions
\par}

\vfill

This diploma thesis has been carried out by Florian Beutler at the \\
Institute of Physics\\
under the supervision of\\
Prof.~Dr.~Johanna Stachel 

\par}

\cleardoublepage

\begin{otherlanguage}{ngerman}
\section*{Thermische Beschreibung der Hadronenproduktion in $e^{+}e^{-}$ Kollisionen}
Die vorliegende Arbeit beschäftigt sich mit der Hadronenproduktion in $e^{+}e^{-}$ Kollisionen bei verschiedenen Schwerpunktsenergien im Rahmen des Statistischen Models eines Hadronen Resonanz Gases. Das Model, welches für diese Analyse verwendet wird, basiert auf dem kanonischen Ensemble mit der exakten Erhaltung von fünf Quantenzahlen. Die entsprechende kanonische Zustandsgleichung wird in Quantenstatistik hergeleitet und der kanonische Faktor wird, als eines der grundlegenden Eigenschaften der kanonischen Beschreibung, detailiert untersucht. Die Parameter des Models werden durch einen $\chi^2$-Fit an die neuesten gemessenen Multiplizitäten bei $\sqrt{s}$=10, 29-35, 91 und 130-200 GeV gewonnen. Wie in Schwerionenkollisionen werden die Messwerte mit der reinen thermischen Produktion verglichen. In einem zweiten Szenario werden harte Kollisionen berücksichtigt, die eine Verstärkung der Produktion von schweren Quarks zur Folge haben. Die Ergebnisse zeigen, dass innerhalb der hohen Genauigkeit der Experimente von wenigen Prozent, keines der Datensätze zufriedenstellend mit dem gewählten Ansatz erklärt werden kann. Deshalb ist es fraglich, ob die Teilchen Produktion in $e^+e^-$ Kollisionen thermischen Ursprung hat. Allerdings wurde auch beobachtet, dass die Messwerte für charm und bottom Teilchen, die in harten Kollisionen erzeugt werden durch grosse kanonische Verstärkung solcher Teilchen in schweren Jets nahezu unabhängig von den Modell Parametern sehr gut erklärt werden können.
\end{otherlanguage}
{%
\vspace{-0.3cm}
\selectlanguage{english}
\section*{Thermal description of hadron production in $e^{+}e^{-}$ collisions}
This thesis gives a comprehensive analysis of hadron production in $e^{+}e^{-}$ collisions at different center of mass energies in the framework of the Statistical Model of the hadron resonance gas. The model used for the analysis is formulated in the canonical ensemble with exact conservation of five quantum numbers. The corresponding canonical partition function in quantum statistics is derived and the canonical factor as the striking feature of the canonical framework will be investigated in detail. The parameters of the underlying model were determined using a fit to the average multiplicities of the latest measurements at $\sqrt{s}$=10, 29-35, 91 and 130-200 GeV. The measurements are compared to the pure thermal production like in heavy ion collisions as well as to the production in a second scenario accounting for hard collisions and the resulting enhancement of heavy quark. The results demonstrate that, within the accuracy of the experiments, none of the data sets is satisfactorily described with this approach, calling into question the notion that particle production in $e^{+}e^{-}$ collisions is thermal in origin. Nevertheless it is observed that the charm and bottom particle yields produced in hard collisions can be explained by large canonical enhancements of such particles in heavy jets, almost independent of the model parameters.
}

\cleardoublepage

\pagestyle{plain}
\pagenumbering{roman} \tableofcontents

\cleardoublepage
\setcounter{page}{1}

\chapter{Introduction}
\pagestyle{fancy}
\pagenumbering{arabic}

Quantum Chromodynamics (QCD) is a non-Abelian gauge theory describing the strong interaction which takes place between color charged particles called quarks and gluons. The strong coupling constant is by far not a real constant, but depends on energy. Unlike the electromagnetic coupling constant in QED the strong coupling constant is increasing with decreasing energy. Therefore the strong interaction has some special features like confinement and asymptotic freedom. Asymptotic freedom means that as distance scales decrease and energy scales increase the coupling constant for the strong force gets smaller and smaller. Soon after this discovery \cite{Politzer:1973fx,Gross:1973id}, it was realized that this implies that at very high temperatures and/or very high densities the interactions between quarks grow smaller and smaller. These interactions could become small enough so that the quarks no longer bind together to form nucleons but instead roam freely through the system \cite{Collins:1974ky}.\\
These free quarks have free color charges, just as the particles in a electromagnetic plasma have free electric charges. Hence, just as the charged particles in the traditional plasma screen the Coulomb interaction, free color charges screen the strong interaction \cite{Kaczmarek:2005zn}. This screening washes out the long length scale interactions and the quark confinement is reversed, what means that the quarks behave like free particles. At normal length and energy scales, free quarks are never observed and hence this prediction of free quarks is essentially a prediction of an entirely new form of matter being known as the Quark Gluon Plasma (QGP) \cite{Satz:1985vb}.\\
Understanding the Quark Gluon Plasma could lead not only to new understandings of QCD, but also to new cognitions in cosmology and astrophysics. The conditions right after the big bang would have been ideal for the formation of a QGP, and there are also suggestions that color-charged matter might be important in neutron stars.\\
In the recent years, a successful effort to solve the QCD equations numerically on a (space-time) lattice has brought deeper insight into the subject of phase transition(s) from hadronic to quark gluon matter \cite{Karsch:2003jg, Laermann:2003cv, Karsch:2001cy}. So far it is not yet clear whether the transition is a true singular behavior of thermodynamic variables or just a rapid crossover.\\
The only possibility of modern particle physics to produce a Quark Gluon Plasma is the collision of heavy ions at ultra-relativistic energies.

\section*{Heavy ion collisions}
\addcontentsline{toc}{section}{Heavy ion collisions}

\begin{figure}[]
\begin{center}
\includegraphics[width=12cm,height=8cm]{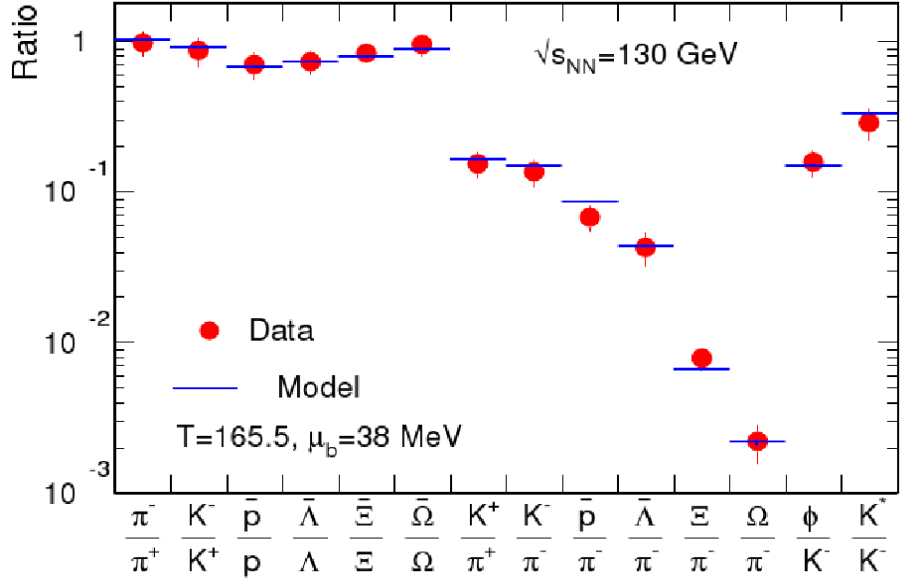}
\caption{Result of a $\chi^2$-fit to different particle ratios measured at $\sqrt{s}$=130 GeV at RHIC \cite{Andronic:2005yp}.}
\label{fig:heavy}
\end{center}
\end{figure}The analysis of hadron yields measured in central heavy ion collisions from AGS up to RHIC energies has shown \cite{heppe} that hadron multiplicities can be described very well with a hadro-chemical equilibrium approach governed by the chemical freeze-out temperature T, the fireball volume V and the baryo-chemical potential $\mu_B$. The RHIC result at $\sqrt{s}$=130 GeV is shown in Figure~\ref{fig:heavy} for a certain set of particle ratios.\\
The temporal evolution of a (central) nucleus-nucleus collision at ultra-relativistic energies is understood to proceed through the following stages \cite{BraunMunzinger:2003zd}: 
\begin{enumerate}
 \item Liberation of quarks and gluons due to the high energy deposited in the overlap region of the two nuclei.
 \item Equilibration of quarks and gluons.
 \item Crossing of the phase boundary and hadronization.
 \item freeze out.
\end{enumerate}
Reaching an equilibrium state is interesting, as it opens the door to a thermal description. A natural question though is how this equilibrium is achieved. Considerations about collision rates and timescale of the hadronic fireball expansion \cite{Stock:1999hm} imply that at SPS, RHIC, and LHC the equilibrium cannot be established in the hadronic medium and that it is the phase transition which drives the particles densities and ensures chemical equilibrium \cite{wetterich}.\\
Applying a statistical model which assumes an equilibrium and testing experimental data against model predictions is one way of testing reality against a thermally and chemically equilibrated fireball at the point of hadro-chemical freeze-out. The equilibrium behavior of thermodynamical observables can be evaluated by averaging over statistical ensembles. The equilibrium distribution is thus obtained by averaging over the complete accessible phase space. Furthermore, the ensemble which corresponds to thermodynamic equilibrium is the one for which the phase space density is uniform over the accessible phase space. In this sense filling the accessible phase space uniformly is both a necessary and a sufficient condition for equilibrium \cite{BraunMunzinger:2003zd}.

\section*{$e^+e^-$ collisions}
\addcontentsline{toc}{section}{$e^+e^-$ collisions}

The discussion above showed that there is a good understanding of the system behavior in heavy ion collisions. An interesting question arising here is whether this statistical behavior is a unique feature of high energy nucleus-nucleus collisions or whether it is also applicable in elementary collisions like, e.g., $e^{+}e^{-}$.  Previous publications indicated that indeed hadron production in $e^{+}e^{-}$ collisions at 14-43 GeV \cite{Becattini:1995if,Becattini:1996gy} and 91 GeV \cite{Becattini:1995if,Becattini:1997uf} can be described well within a Thermal Model provided that local quantum number conservation is properly implemented. The main results of these investigations are that the temperature values deduced are almost constant near T=160 MeV and that the volume increases with energy, while strangeness is under-saturated. These results were
taken, together with the results for nucleus-nucleus collisions where a similar temperature is reached at high energies, as evidence for the interpretation that the thermodynamical state is not reached via a dynamical equilibration among constituents. It is rather a generic fingerprint of hadronization \cite{Stock:1999hm,heinz} or a feature of the excited QCD vacuum \cite{castorina}.\\ \\
In this thesis the latest multiplicity measurements of $e^+e^-$ collisions at $\sqrt{s}$=10 GeV, 29-35 GeV, 91 GeV and 130-200 GeV summarized and published by the Particle Data Group (PDG) \cite{Yao:2006px} are used. Since the aim is a precision calculation for a small system, a fully canonical form of the Thermal Model \cite{BraunMunzinger:2003zd,Hagedorn:1984uy} is applied, conserving baryon number, electric charge, strangeness, charmness, and bottomness. In order to reach a precision comparable to that of the data from the LEP collider (a few percent) within the model, quantum statistics is included in the formulation.\\
In chapter \ref{ch:stat} the basics of statistical mechanics and thermodynamics are briefly introduced with focus on the equations which are relevant for this study. The following chapter \ref{ch:theo} discusses the understanding of an $e^+e^-$ collision, with a short overview over the existing fragmentation models. The last part of this chapter introduces the Thermal Model, in the grand-canonical as well as in the canonical ensemble relevant for this work. The corresponding canonical partition function is derived in chapter \ref{sec:themodel}. In contrary to earlier studies \cite{Becattini:1995if,Cleymans:1997ib}, the calculation of the partition function is carried out in quantum statistics instead of Boltzmann statistics. In chapter \ref{sec:program} the C++ program, developed within this thesis to simulate the particle production in a collision, is explained and the main steps and approximations are investigated. Section \ref{sec:about} demonstrates the influence of the canonical factor and shows the necessity for the canonical approach used in this work. In chapter \ref{sec:fit} the fit procedure and the treatment of strong decays is explained. Furthermore the incidence of weak decays within a collision evolution is discussed and a significant influence to the particle yields is found. Finally all results are presented and discussed in chapter \ref{sec:results}. The appendix contains a fully quantum statistical derivation of the canonical partition function (for bosons and fermions) and a discussion of more technical aspects of the program development.\\
The work presented in this thesis has been published in \cite{Andronic:2008ev,Andronic:2009sv,Redlich:2009xx}\footnote{The results of this work and the results published in the paper are not always identical. In such a case the values in the paper are the more recent ones.}.

\newpage

\section*{Units and conventions}
\addcontentsline{toc}{section}{Units and conventions}

The common unit of length in hadron physics is
\begin{equation}
1\; \text{fm} = 1\; \text{Fermi} = 10^{-15}\text{m}
\end{equation}
and for the time scale at relativistic velocities $v \approx c$
\begin{equation}
\frac{1\text{fm}}{c} = \frac{1}{2.99792458}10^{-23}\text{s}
\end{equation}
Energies are treated in units of
\begin{equation}
1\;\text{MeV} = 10^6\cdot 1.602\cdot 10^{-19}\text{AsV} = 1.602\cdot 10^{-13}\text{J}
\end{equation}
In all thermodynamic equations the Boltzmann constant is set to one
\begin{equation}
k_B \equiv 1
\end{equation}
Therefore the temperature gets the dimension of energy
\begin{equation}
1\;\text{MeV} = \frac{1\;\text{MeV}}{k_B} = \frac{1.60218\cdot 10^{-13}\text{Nm}}{1.38066\cdot 10^{-23}\text{Nm/K}} = 1.16044\cdot 10^{10}\text{K}
\end{equation}
Also the speed of light is set to one
\begin{equation}
c\equiv1
\end{equation}
and the Planck constant becomes
\begin{equation}
\hbar = 1.05457\cdot 10^{-34}\text{Nms}\frac{1\;\text{MeV}}{1.60218\cdot 10^{-13}\text{Nm}}\frac{1\;\text{fm/c}}{10^{-15}m/c} = 197.327\;\text{MeV}\;\text{fm}
\end{equation}

\chapter{Statistical mechanics and thermodynamics}
\label{ch:stat}

In this chapter the basic properties of an ideal quantum gas are discussed. The equations presented in section \ref{sec:Quantumstat} are used later to derive the partition function of a hadron gas. In section \ref{sec:stat} basic thermodynamical relations are introduced.
%For detailed derivations or further informations please use \cite{Reif, Wikipedia}.

\section{Quantum statistics}
\label{sec:Quantumstat}

Particles of the same species are indistinguishable, for that reason it is not possible to specify the microstate of each particle. Instead a microstate of an ideal gas is specified by the occupation numbers $n_k$, the number of particles in each single particle energy $\epsilon_k$. If the value of the occupation number $n_k$ is known, the total energy of a microstate $\vert s\rangle$ can be written as
\begin{equation}
E_s = \sum_k n_k\epsilon_k
\end{equation}
The set $\{n_{k}\}$ completely specifies one microstate of the system. The \textbf{canonical} partition function $\mathcal{Z}^C$ of an ideal gas can be expressed in terms of the occupation numbers as
\begin{eqnarray}
\mathcal{Z}^C(V, T) &=& \sum_{s}e^{-\beta E_s} = \sum_s e^{-\beta \sum_k n_k\epsilon_k}\cr
&=& \sum_{s}\prod_ke^{-\beta n_k\epsilon_k}
\end{eqnarray}
where $\beta=1/T$\footnote{In the rest frame of the system and $k_B \equiv 1$}, $V$ is the volume of the system and the occupation numbers satisfy the condition
\begin{equation}
N = \sum_k n_k
\end{equation}
with the total particle number $N$.\\
According to the spin-statistics theorem all particles are classified into two groups. Particles with zero or integer spin such as deuterons ($^2$H$^+$) or pions are bosons and follow Bose-Einstein statistics. Particles with half-integer spin such as electrons, protons, and neutrons are fermions and obey Fermi-Dirac statistics. The difference between fermions and bosons results in the allowed values for the occupation number $n_k$. Fermions are restricted by
\begin{equation}
n_{k} = 0 \quad \text{or}\quad n_k = 1
\end{equation}
which is the Pauli exclusion principle. For noninteracting particles two identical fermions cannot be in the same single particle state. In contrast, the occupation numbers n$_k$ for identical bosons can take zero or any positive integer value
\begin{equation}
n_{k} = 0, 1, 2, \dots
\end{equation}
For fermions, the canonical partition function for the $k^{th}$ energy level is
\begin{equation}
\mathcal{Z}^C_{k}(V, T) = \sum_{n_k = 0}^1 e^{-\beta n_k \epsilon_k} = 1 + e^{-\beta \epsilon_k}
\label{eq:QS1}
\end{equation}
whereas for bosons
\begin{equation}
\mathcal{Z}^C_k(V, T) = \sum_{n_k=0}^{\infty}e^{-\beta n_k\epsilon_k} = 1 + e^{-\beta\epsilon_k} + e^{-2\beta\epsilon_k} + \dots = \sum_{n_k=0}^{\infty} (e^{-\beta\epsilon_k})^{n_k}
\end{equation}
This geometrical series is convergent for $e^{-\beta\epsilon_k} < 1$, thus
\begin{equation}
\mathcal{Z}^C_k(V, T) = \dfrac{1}{1 - e^{-\beta\epsilon_k}}
\label{eq:QS2}
\end{equation}
To get the average particle number at energy level $k$, one has to differentiate eq. (\ref{eq:QS1}) and (\ref{eq:QS2})
\begin{equation}
\langle n_k^C\rangle = -\frac{1}{\beta}\frac{\partial\ln \mathcal{Z}^C_k}{\partial\epsilon_k} = -\frac{1}{\beta}\frac{\partial\ln(1\mp e^{-\beta\epsilon_k})^{\mp1}}{\partial\epsilon_k} = \frac{1}{e^{\beta\epsilon_k}\pm1}
\label{eq:qs1}
\end{equation}
These functions are called the Fermi-Dirac (+1) and Bose-Einstein (-1) distribution functions.\\
For a system with more then one subsystems $k$, the partition function of the whole system is the product of all subsystem partition functions. This means that for fermions it can be written
\begin{equation}
\mathcal{Z}^C_F(V, T) = \prod_k\mathcal{Z}^C_k(V, T) = \prod_k(1 + e^{-\beta\epsilon_k})
\end{equation}
and for bosons
\begin{equation}
\mathcal{Z}^C_B(V, T) = \prod_k\mathcal{Z}^C_k(V, T) = \prod_k\dfrac{1}{1 - e^{-\beta\epsilon_k}}
\end{equation}
In a similar way to the definition of the canonical partition function for the canonical ensemble, one can define a grand-canonical partition function for a grand-canonical ensemble. The only difference is a fugacity term
\begin{equation}
z = e^{\beta\mu}
\end{equation}
taking into account that the particle number in a grand-canonical ensemble is not fixed.\\
The partition function of the grand-canonical ensemble is then given by a weighted sum of canonical partition functions with different number of particles $N$
\begin{equation}
\mathcal{Z}^{GC}(z,V,T) = \sum^{\infty}_{N=0}z^NZ^C(V,T) = \sum^{\infty}_{N=0}\sum_{s} z^Ne^{-\beta E_s}
\end{equation}
For a system of bosons or fermions, it is often mathematically easier to treat the number of particles of the system as an intrinsic property of each state, $\vert s\rangle$ (specified by the set $\{n_k\}$). Therefore the partition function can be written as
\begin{equation}
\mathcal{Z}^{GC}(z, V, T) = \sum_{s} z^{N_{s}}e^{-\beta E_s} = \sum_{s}\prod_k z^{n_k}e^{-\beta n_k\epsilon_k}
\end{equation}
where $N_s$ is not fixed anymore.\\
Now the discussion of the canonical ensemble (above) can be applied to derive the grand-canonical distribution functions 
\begin{equation}
\langle n_k^{GC} \rangle = \frac{1}{\beta} \frac{\partial \ln \mathcal{Z}^{GC}_k}{\partial \mu} = \frac{1}{z^{-1}e^{\beta\epsilon_k}\pm1} = \frac{1}{e^{\beta(\epsilon_k-\mu)}\pm1}
\end{equation}
with $-$ for bosons and $+$ for fermions.

\subsection{Boltzmann statistics: The classical limit of Quantum statistics}
\label{sec:Boltzmann}

For fermions and bosons, the quantum statistical canonical partition functions can be written as
\begin{subequations}
\begin{eqnarray}
\ln \mathcal{Z}^C_B(V,T) &=& \ln\left(\prod_k\dfrac{1}{1 - e^{-\beta\epsilon_k}}\right) = -\sum_k\ln (1-e^{-\beta \epsilon_k})\\
\ln \mathcal{Z}^C_F(V,T) &=& \ln\left(\prod_k(1 + e^{-\beta\epsilon_k})\right) = \sum_k \ln (1+e^{-\beta \epsilon_k})
\label{eq:bolzmann1}
\end{eqnarray}
\end{subequations}
where the results of the last section are used. The classical Boltzmann limit arises from the first term of the Taylor expansion of the logarithms
\begin{equation}
\ln(1\pm x)^{\pm1} \approx x + \mathcal{O}(x^2)\quad\quad (x \ll 1)
\label{eq:bolzmann1.2}
\end{equation}
i.e., when it is possible to consider the exponential term as small compared to unity ($e^{-\beta\epsilon_k} \ll 1$). In this case the classical Maxwell-Boltzmann limit is reproduced for fermions and bosons:
\begin{equation}
\ln \mathcal{Z}^C_{cl}(V,T) = \sum_k e^{-\beta \epsilon_k}
\label{eq:bolzmann2}
\end{equation}\\
Furthermore it is also possible to apply the Boltzmann approximation directly to the distribution functions, introduced in (\ref{eq:qs1})
\begin{equation}
\langle n_{F/B} \rangle = \frac{1}{e^{\beta\epsilon}\pm1}
\label{eq:boltzmann3}
\end{equation}
where the plus sign refers to fermions, and the minus sign to bosons. The classical Boltzmann approximation comes by neglecting the term $\pm1$ in the denominator. Therefore the Boltzmann condition is again
\begin{equation}
e^{\beta\epsilon} \gg 1
\end{equation}
This is given for sufficiently low temperature or sufficiently high energy (mass) where for both, fermions and bosons, one obtains the same value
\begin{equation}
\langle n_{cl}\rangle = e^{-\beta\epsilon}
\end{equation}
It follows again that the classical limit of the quantum distribution functions, for Fermi-Dirac as well as for Bose-Einstein, reduces to the Maxwell-Boltzmann distribution.

\section{Statistical and thermodynamical relations}
\label{sec:stat}

If the partition function of a system is known, it is possible to calculate all macroscopic properties of the system. For instance the average energy $E$ is
\begin{equation}
E = -\frac{1}{\mathcal{Z}}\frac{\partial \mathcal{Z}}{\partial \beta} = -\frac{\partial\ln \mathcal{Z}}{\partial\beta}
\label{eq:stat5}
\end{equation}
The free energy is defined as
\begin{equation}
F = -T\ln \mathcal{Z}
\end{equation}
and links statistical mechanics to thermodynamics. A pure thermodynamical definition of $F$ is
\begin{equation}
F = E - TS
\end{equation}
With the fundamental thermodynamic relation\footnote{The particle number conservation is not treated here} $dE = TdS - pdV$, one finds
\begin{equation}
dF = -SdT-pdV
\end{equation}
Thus the entropy is
\begin{equation}
S = -\left.\frac{\partial F(V,T)}{\partial T}\right|_{V}
\label{eq:stat4}
\end{equation}
and the pressure of the system is
\begin{equation}
p = -\left.\frac{\partial F(V,T)}{\partial V}\right|_{T}
\label{eq:stat3}
\end{equation}
In conclusion, the partition function $\mathcal{Z}$ is the key to all thermodynamical quantities.

\chapter{Theoretical description of $e^+e^-$ collisions}
\label{ch:theo}

The fragmentation process in an $e^+e^-$ collision is still an open issue, because a pure analytical QCD approach is not able to access this area. This gave rise to many phenomenological models. After the discussion of the $e^+e^-$ annihilation process some of these models are described. In this context the Thermal Model will be introduced.

\section{Characteristics of an $e^+e^-$ collision}
\label{sec:char}

\begin{figure}[]
\begin{center}
\includegraphics[width=8cm,height=8cm]{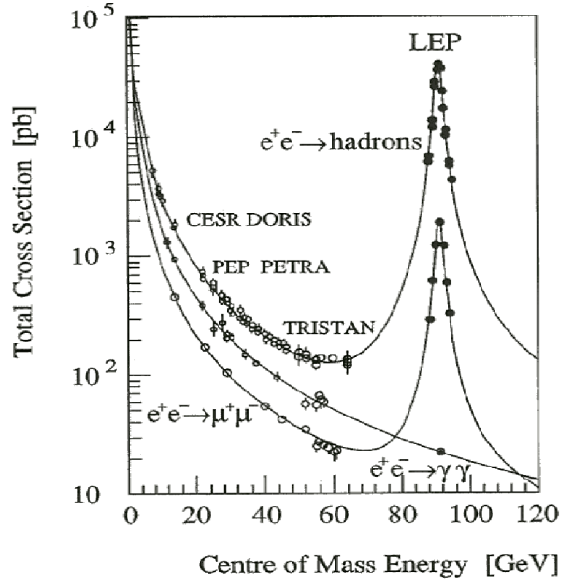}
\vspace{-0.2cm}
\caption{Measurements of the different $e^+e^-$ cross sections at LEP. The measurements at lower energies are also shown for CESR, DORIS, PEP, PETRA and TRISTAN~\cite{:2005ema}.}
\label{fig:crosssection}
\end{center}
\end{figure}

% Unlike other types of collisions, the $e^+e^-$ interaction has the advantage of offering a clean framework for study. Electrons and positrons being point-like particles, and interacting only via the electroweak interaction.\\
For a low center of mass (c.m.) energy $\sqrt{s}$, the process $e^+e^- \rightarrow f\overline{f}$ is dominated by a single virtual photon exchange. When the c.m. energy of the $Z^0$ resonance is reached ($\sqrt{s} \approx$ 91 GeV), $Z^0$ exchange becomes the dominating process (see Figure~\ref{fig:crosssection}). The most probable result of an $e^+e^-$ collision near the $Z^0$ resonance is multihadron production due to the large branching ratio of $Z^0 \rightarrow q\overline{q}$ ($ \approx 70\%$). Other channels, such as $e^+e^- \rightarrow e^+e^-$ (Bhabha scattering), $e^+e^- \rightarrow \mu^+\mu^-$ and $e^+e^- \rightarrow \tau^+\tau^-$, occur less frequently~\cite{Yao:2006px}. The further discussion will focus on the $Z^0$ decay to quarks, disregarding the leptonic decays, because this is the essential process needed in this work.\\
The process $e^+e^- \rightarrow$ hadrons can be divided into four stages. In Figure~\ref{fig:hadronization1} these stages are shown:
\begin{figure}[]
\begin{center}
\includegraphics[width=11cm,height=8cm]{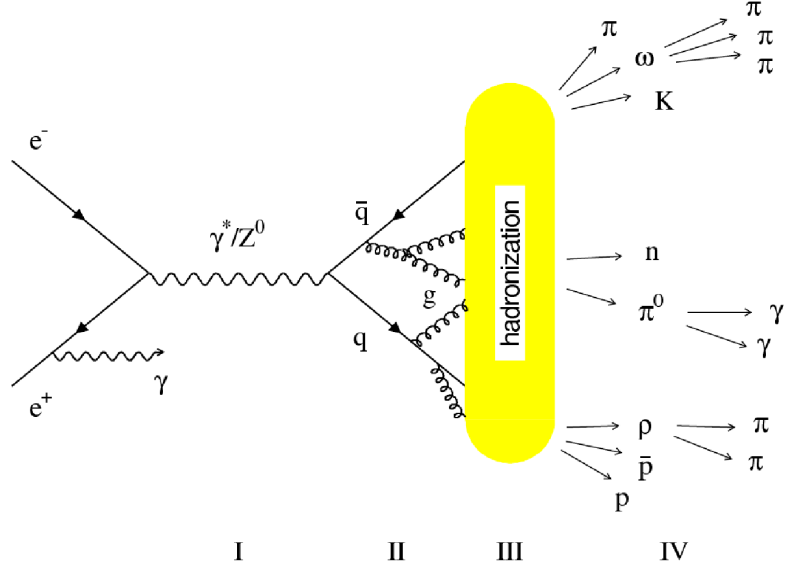}
\caption{The process $e^+e^- \rightarrow$ hadrons \cite{Knowles:1997dk}.}
\label{fig:hadronization1}
\end{center}
\end{figure}
\begin{enumerate}[I.]
 \item The formation of the $Z^0$ in an $e^+e^-$ collision, and its subsequent decay into $q\bar{q}$ are determined by the electroweak part of the Standard Model.
 \item As the quark and antiquark separate, gluons may be emitted, which may radiate further gluons or generate $q\bar{q}$ pairs. This process can be calculated in QCD perturbative theory, as long as the strong coupling, $\alpha_s$, remains significantly smaller than one.
 \item The particle shower increases, with decreasing momentum exchange Q and increasing $\alpha_s$. At some point $\alpha_s$ will reach the non-perturbative region ($\alpha_s \approx 1$). Finally the quarks and gluons will form colorless hadrons, but the transition from quarks to hadrons is not describable in perturbative theory, and one has to resort to various more or less ad hoc models to describe the process, usually called hadronization or fragmentation.
 \item In order to compare theory with data one also has to count for decays of unstable hadrons in photons, leptons and stable hadrons.
\end{enumerate}

\subsection*{The electroweak stage}

In the first stage, the $e^+e^-$ pair annihilates into the $Z^0$ resonance according to electroweak theory. This phenomenon may be accompanied by the emission of photons (initial-state radiation) prior to the annihilation. This electroweak correction reduces the c.m. energy of the $e^+e^-$ collision and therefore the total effective mass of the hadronic final state. Following its creation, the vector boson $Z^0$ decays into a quark-antiquark pair ($\approx 70\%$). All these phenomena are described in the framework of the very successful electroweak model.\\
The partial width for $Z^0 \rightarrow q\bar{q}$ is given by \cite{Knowles:1997dk}
\begin{equation}
\Gamma_{q\bar{q}} = \frac{G_Fm_Z^3\beta_q}{2\pi\sqrt{2}}\left[\left(1+2\eta^2\right)((g^q_V)^2+(g^q_A)^2)-6\eta^2(g^q_A)^2\right]
\end{equation}
with
\begin{equation}
\beta_q = \sqrt{1-4\eta^2};\quad\quad \eta = \frac{m_q}{m_Z}
\end{equation}
and with the vector and axial-vector currents
\begin{align}
g_V &= I_3 - 2Q\sin^2\theta_W\\
g_A &= I_3
\label{eq:vecaxi}
\end{align}
For up-type quarks with $I_3$=1/2 and $Q$=2/3 one gets
\begin{align}
g^u_V &= \frac{1}{2} - \frac{4}{3}\sin^2\theta_W = 0.18\\
g^u_A &= \frac{1}{2}
\end{align}
This gives $\Gamma_u$=280 MeV.\\
For down-type quarks with $I_3$=-1/2 and $Q$=-1/3 one gets
\begin{align}
g^d_V &= -\frac{1}{2} + \frac{2}{3}\sin^2\theta_W = -0.34\\
g^d_A &= -\frac{1}{2}
\end{align}
This yields $\Gamma_d$=370 MeV.\\
The total hadronic width from the three down-type quarks ($d, s, b$) and the two up-type quarks ($u, c$) is therefore $\Gamma_{\text{hadron}}$=1.67 GeV\footnote{The top quark is not produced at $\sqrt{s}$=91 GeV.}. The relative hadronic branching ratios for the up-type quarks and down-type quarks are
\begin{equation}\label{eq:eweak12}
\begin{split}
\Gamma_{u\overline{u}}/\Gamma_{\text{hadron}} &\approx 17\%\\
\Gamma_{d\overline{d}}/\Gamma_{\text{hadron}} &\approx 22\%
\end{split}
\end{equation}
what will be an important point in the discussion of heavy particle production in $e^+e^-$ collisions, later in this work.

\subsection*{The perturbative QCD stage}

\begin{figure}[]
\begin{center}
\includegraphics[width=6cm,height=6.5cm]{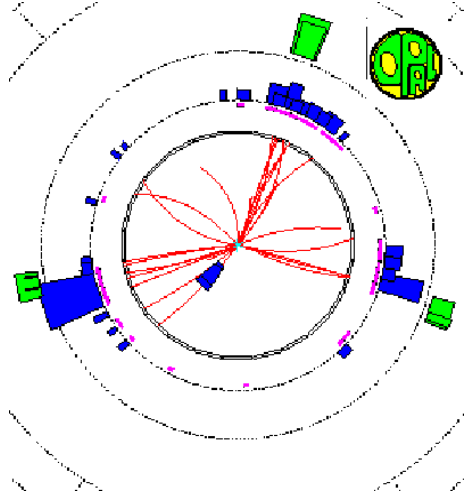}
\caption{Three jet event at the OPAL detector (LEP).}
\label{fig:3jet}
\end{center}
\end{figure}

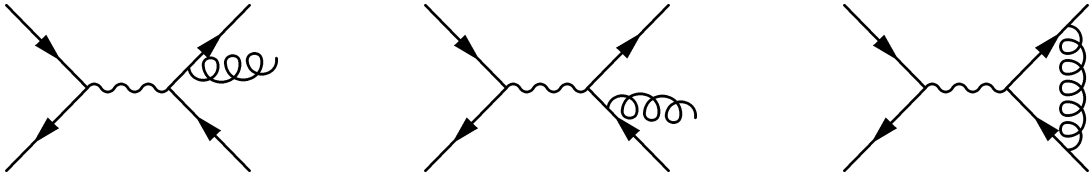
\begin{figure}[]
\begin{center}
\begin{fmffile}{feynman-v1}
\begin{fmfchar*}(40,22)
\fmfleft{i1,i2,i3,i4}\fmflabel{$e^+$}{i1} \fmflabel{$e^-$}{i4}
\fmfright{o1,o2,o3,o4}\fmflabel{$\bar{q}$}{o1} \fmflabel{$q$}{o4} \fmflabel{$g$}{o3}
\fmf{plain,tension=2}{i4,v1}
\fmf{fermion}{v1,v2}
\fmf{plain,tension=2}{v2,v3,v4}
\fmf{fermion}{v4,v5}
\fmf{plain,tension=2}{v5,i1}
\fmf{boson,label=$Z^0$,label.side=left}{v3,v6}
\fmf{plain,tension=2}{o1,v9}
\fmf{fermion}{v9,v7}
\fmf{plain,tension=2}{v7,v6,v8}
\fmf{fermion}{v8,v10}
\fmf{plain,tension=2}{v10,o4}
%\fmffixed{2}{v3,v6}
\fmffreeze
%\fmfshift{.1h}{v3}
\fmf{gluon}{v8,o3}
\end{fmfchar*}
\hspace{1.3cm}
\begin{fmfchar*}(40,22)
\fmfleft{i1,i2,i3,i4}\fmflabel{$e^+$}{i1} \fmflabel{$e^-$}{i4}
\fmfright{o1,o2,o3,o4}\fmflabel{$\bar{q}$}{o1} \fmflabel{$q$}{o4} \fmflabel{$g$}{o2}
\fmf{plain,tension=2}{i4,v1}
\fmf{fermion}{v1,v2}
\fmf{plain,tension=2}{v2,v3,v4}
\fmf{fermion}{v4,v5}
\fmf{plain,tension=2}{v5,i1}
\fmf{boson,label=$Z^0$,label.side=left}{v3,v6}
\fmf{plain,tension=2}{o1,v9}
\fmf{fermion}{v9,v7}
\fmf{plain,tension=2}{v7,v6,v8}
\fmf{fermion}{v8,v10}
\fmf{plain,tension=2}{v10,o4}
%\fmffixed{2}{v3,v6}
\fmffreeze
%\fmfshift{.1h}{v3}
\fmf{gluon}{o2,v7}
\end{fmfchar*}
\hspace{1.3cm}
\begin{fmfchar*}(40,22)
\fmfleft{i1,i2,i3,i4}\fmflabel{$e^+$}{i1} \fmflabel{$e^-$}{i4}
\fmfright{o1,o2,o3,o4}\fmflabel{$\bar{q}$}{o1} \fmflabel{$q$}{o4}
\fmf{plain,tension=2}{i4,v1}
\fmf{fermion}{v1,v2}
\fmf{plain,tension=2}{v2,v3,v4}
\fmf{fermion}{v4,v5}
\fmf{plain,tension=2}{v5,i1}
\fmf{boson,label=$Z^0$,label.side=left}{v3,v6}
\fmf{plain,tension=2}{o1,v9}
\fmf{fermion}{v9,v7}
\fmf{plain,tension=2}{v7,v6,v8}
\fmf{fermion}{v8,v10}
\fmf{plain,tension=2}{v10,o4}
%\fmffixed{2}{v3,v6}
\fmffreeze
%\fmfshift{.1h}{v3}
\fmf{gluon,label=$g$,label.side=right,label.dist=3mm}{v9,v10}
\end{fmfchar*}\notag
\end{fmffile}
\vspace{0.3cm}
\caption{The leading order Feynman diagrams of an $e^+e^- \rightarrow Z \rightarrow q\bar{q}$ process.}
\label{feynman1}
\end{center}
\end{figure}

In the second stage, the initial $q\overline{q}$ may radiate gluons according to the theory of Quantum Chromodynamics. The gluons may radiate other gluons or $q\overline{q}$ pairs, giving rise to a cascade process. This stage is responsible for the formation of hadronic jets as seen in Figure \ref{fig:3jet} for a 3 jet event as it appears in the OPAL detector. The probability ratio for observing events with a certain number of jets can be written in terms of increasing powers of the strong coupling constant $\alpha_s$ as \cite{Altarelli:1989hv}
\begin{equation}
\text{2 jets : 3 jets : 4 jets }= \mathcal{O}(\alpha_s^0) : \mathcal{O}(\alpha_s^1) : \mathcal{O}(\alpha_s^2)
\end{equation}
This means that at $\sqrt{s}$=91 GeV the majority of the $e^+e^-$ events have a two-jet structure\footnote{$\alpha_s$ at $\sqrt{s}$=91 GeV is $0.12\pm0.0031$ \cite{Bethke:2004bp}.}. Due to the small value of $\alpha_s$, this stage can be described by perturbative QCD. Furthermore the prediction of a cross-section $\sigma$ (either the total hadronic cross-section, or some differential cross-section) can be written in the general form
\begin{equation}
\sigma = \sigma_0 + A\alpha_s + B\alpha_s^2 + C\alpha_s^3 + \cdots
\end{equation}
To compute the leading coefficient, A, it is necessary to consider the Feynman diagrams shown in Figure \ref{feynman1}. The first two diagrams represent the radiation of a gluon from either quark or antiquark, leading to three-parton final states, and clearly yield cross-sections proportional to $\alpha_s$. Although the square of the matrix element for the third loop diagram is $O(\alpha_s^2)$, it leads to the same $q\bar{q}$ final state as the leading order diagram with no gluon emission, and the interference term between the two diagrams is $O(\alpha_s)$ \cite{Green}.

\subsection*{The non perturbative QCD stage}

In the third stage, the partons fragment into colorless hadrons. Usually, this stage is called hadronization. This process cannot be described as a power expansion in the strong coupling constant, since $\alpha_s \gtrsim 1$ (soft processes).\\
Since the description of the final-state particles cannot be accessed analytically, a second approach is to use phenomenological models. Three of them are described later in this chapter.

\subsection*{Particle decays}

The last stage in Fig.~\ref{fig:hadronization1} represents the decay of unstable hadrons into experimentally observable particles (mostly pions). Many of these are well understood from low energy experiments, though in the case of heavy-flavored hadrons there are uncertainties since not all decay modes have been measured.

\section{Fragmentation models}
\label{sec:frag}

Quarks and gluons cannot be observed as free particles but are confined in color neutral hadrons. The process of hadron formation is not well understood, because it falls into the domain of non-perturbative QCD and it has not been possible to derive
it from first principles. Therefore, several phenomenological models aimed at describing quantitatively the fragmentation process have been developed.\\
In quantitative terms, one has the picture of the color field lines between a quark and antiquark. As the quark and antiquark move apart their kinetic energy is transformed into increasing field energy, which can generate $q\bar{q}$ pairs, and turn into hadrons finally. There are basically three models: independent fragmentation, string fragmentation, and cluster fragmentation. The basic ideas of these models are summarized in the following subsections.

\subsection{Independent fragmentation}
\label{subsec:independent}

In the framework of independent fragmentation it is assumed that each parton hadronizes on its own in the following iterative scheme. A quark $q$ fragments into a hadron $q\bar{q}'$ and a quark $q'$. The $q'$ then proceeds to fragment in the same way, and the process is repeated until insufficient energy remains. The sharing of energy-momentum is governed by a fragmentation function. Flavor and transverse momentum are conserved in each break-up.\\
There are a number of difficulties with this scheme; for example, the result of independent fragmentation normally depends on the coordinate frame and is thus not Lorentz invariant. A fully Lorentz-invariant scheme has been proposed in \cite{Montvay:1979py} but it is hard to implement. Furthermore it does not conserve energy, momentum and flavor exactly (they are conserved locally but not global), so that these have to be patched up at the end.

\subsection{String fragmentation}
\label{subsec:string}

The quark and antiquark from the $Z^0$ move apart in opposite directions. The color field between them forms a narrow flux tube called string. In the Lund model \cite{Andersson:1978vj} the string is idealized to an one-dimensional massless relativistic string with energy density $\kappa \simeq$ 1 GeV/fm. The potential rises linearly with the distance between the charges and at some point it becomes possible to form a new quark-antiquark pair from the field energy.\\
In the classical picture, a $q\bar{q}$ pair with a mass or transverse momentum relative to the string is produced a certain distance apart, with the field energy between the quarks transforming into the transverse momentum. In quantum mechanics, they are produced at the same point (to conserve flavor locally) and then tunnel out. The tunneling probability for a common transverse mass $m^2_{\perp} = m_q^2+p_{\perp}^2$, of the $q\bar{q}$ pair is \cite{Casher:1978wy}
\begin{equation}
\exp\left(-\frac{\pi m^2_{\perp}}{\kappa}\right) = \exp\left(-\frac{\pi m_q^2}{\kappa}\right)\exp\left(-\frac{\pi p^2_{\perp}}{\kappa}\right)
\end{equation}
The tunneling also ensures that heavy quark and strange quark production is suppressed within the string. Heavy quarks in $e^+e^-$ annihilation thus either come from perturbative gluon splitting or are the primary quarks from the $Z^0$ decay. The ratios of the different quark flavors given by the tunneling probability are $u : d : s : c \approx 1 : 1 : 0.3 : 10^{-11}$.\\
Mesons are produced when a quark and an antiquark from adjacent string breaks are combined. The baryon production is usually modeled by the diquark \cite{Andersson:1981ce} or the popcorn \cite{Andersson:1984af} mechanism (see the references for details).\\
The different break-ups of the string are assumed to be independent of each other. It is therefore possible to start at one end of the string, make a break-up and be left with a meson and a shorter string. The procedure can be iterated until the available energy is used up (the termination requires some extra treatment that will not be discussed here).\\
The fractions $z$ of the available energy, taken by the hadrons, are distributed according to a probability distribution function $f(z)$. The constraint that the result should be independent of the choice from which end to start leads to the Lund symmetric fragmentation function
\begin{equation}
f(z) \propto z^{-1}(1-z)^a\exp(-bm^2_{\perp}/z)
\end{equation}
Other functions are available\footnote{Such as the Peterson fragmentation function which is designed for the fragmentation of heavy quarks~\cite{Peterson:1982ak}.}. The parameters $a$ and $b$ are chosen in order to fit the data. In the case of gluon bremsstrahlung, the situation is more complicated.\\
Overall, the string model presents a more consistent and covariant picture than the independent fragmentation model and is generally in better agreement with experimental results. The Lund string model is implemented in the PYTHIA \cite{Sjostrand:2006za} event generator (merged with the former JETSET \cite{Sjostrand:1993yb} generator). The most recent version is PYTHIA 8.1 \cite{Sjostrand:2007gs}. One drawback of PYTHIA is the large number of variable parameters, but it is considered to be the most faithful simulation currently available\footnote{\url{http://home.thep.lu.se/~torbjorn/Pythia.html}}.

\subsection{Cluster fragmentation}
\label{subsec:cluster}

The model of cluster fragmentation \cite{Webber:1983if} takes a very different approach compared to the string model. It is based on the observation that, at the end of the perturbative evolution of a parton shower, a given color and the corresponding anticolor generally occur on partons which lie quite close together in phase space. One can replace each gluon by a $q\bar{q}$ pair, and can then identify the color singlet combinations amongst all the quark and antiquark. Each of these is considered to form a cluster, with flavor quantum numbers determined by the $q\bar{q}$. The mass of a cluster is typically a few GeV. Each cluster is then allowed to decay into two mesons or a baryon-antibaryon pair, with the appropriated net flavor.\\ 
This model typically has fewer arbitrary tunable parameters than the string model, but it gives quite a good account of the data. The most popular event generator which uses the cluster fragmentation approach is the HERWIG \cite{Corcella:2000bw} event generator\footnote{\url{http://projects.hepforge.org/herwig/}}.

\section{The Thermal Model of ultrarelativistic particle collisions}

The Thermal (Statistical) Model \cite{Becattini:2004td} is a model of hadronization, aiming to reproduce the quantitative features of the hadronization process. This model has been recognized as a powerful approach to describe particle production yields in heavy-ion collisions \cite{BraunMunzinger:2003zd,Andronic:2005yp}.\\
% Experiments with ultra-relativistic nucleus-nucleus collisions provide a unique opportunity to study the properties of strongly interacting matter under the extreme condition of high energy densities. Hadronic multiplicities, and their spectra in particular, carry information about the nature of the medium from which they originated.\\
To be able to apply the framework of thermodynamics to a system of two colliding objects, the creation of a fireball in complete thermodynamical equilibrium at the collision point is assumed. The basic idea of the model is very simple and lies in two assumptions~\cite{Becattini:2004td}:
\begin{enumerate}
 \item In the late stage of a high energy collision, the cluster (fireball) decays into hadrons in a purely statistical fashion. The decay happens at a critical temperature $T_c$ of the fireball.
 \item All multihadronic states within the cluster compatible with its quantum numbers are equally likely.
\end{enumerate}
The analysis of particle yields measured in central heavy ion collisions from AGS up to RHIC energies has shown
\cite{agssps,satz,heppe,cley,beca1,rhic,nu,beca2,rapp,becgaz,Andronic:2005yp} that hadron multiplicities can be described very well with a hadro-chemical equilibrium approach which is governed by the chemical freeze-out temperature T, the baryo-chemical potential $\mu_B$, and the fireball volume V.

In the limit of high temperature and/or large system size, the grand-canonical treatment is adequate. However, if the number of particles in a collision fireball is small, either due to low temperature (e.g. in nuclei-nuclei collisions at small energies of a few AGeV) or due to small system size (elementary particle collisions), then the canonical ensemble has to be used \cite{Becattini:1995if,Hagedorn:1984uy}.

\subsection{Grand-canonical ensemble}

In the grand-canonical ensemble the fireball is characterized by the temperature $T$ ($\beta$=1/T), its volume $V$ and the chemical potentials $\vec{\mu}$ which are assumed to be uniform over the whole volume. These parameters determine the partition function
\begin{equation}
\mathcal{Z}^{GC}(\beta,V,\vec{\mu}) = \sum_{\text{states}}e^{-\beta(\epsilon_s - \vec{\mu}_s)}
\label{eq:can1}
\end{equation}
In the "normal" grand-canonical approach the particle number $N$ is not fixed, but conserved in the mean, by the chemical potential $\mu$=$\mu_NN$. At relativistic energies the particle number is not fixed at all (because particles can be produced from the available energy). Instead of conserving the particle number, the chemical potential is used to conserve the quantum numbers of the system. Therefore $\vec{\mu}$ is a combination of the different quantum number chemical potentials, e.g. $\vec{\mu}$=$(\mu_N,\mu_S,\mu_{I_3},\mu_C,\mu_B)$ for the baryon number $N$, strangeness $S$, isospin $I_3$, charmness $C$ and bottomness $B$.\\
\textbf{Note: It is crucial to understand, that the words grand-canonical and canonical in our context refer exclusively to the quantum number conservation laws, because the particle number is not conserved, neither in heavy ion, nor in $e^+e^-$ collisions.}\\
In the hadronic fireball of non-interacting particles and resonances, the partition function $\mathcal{Z}^{GC}$ is the product of the particle partition functions of all particles $j$ (see chapter \ref{ch:stat})
\begin{equation}\label{eq:GC1}
\mathcal{Z}^{GC}(\beta,V,\vec{\mu}) = \prod_jz_j^1(\beta,V,\vec{\mu})
\end{equation}
with,
\begin{equation}
z_j^1(\beta,V,\vec{\mu}) = \frac{g_jV}{2\pi^2}\int^{\infty}_0 p^2 dp\;\left[1\pm\exp\left(- \beta(\epsilon_j - \vec{x}_j\cdot \vec{\mu})\right)\right]^{\pm 1}
\label{eq:GC2}
\end{equation}
where $\epsilon_j$ is the relativistic energy of particle species $j$ and $g_j$=$(2J_j+1)$ is the spin degeneracy factor. The vector $\vec{x}_j$=$(N_j,S_j,I_{3j},C_j,B_j)$ contains the quantum numbers of the particle and $\vec{\mu}$=$(\mu_N,\mu_S,\mu_{I_3},\mu_C,\mu_B)$ is the vector of chemical potentials related to the different quantum numbers. The upper (lower) signs refer to fermions (bosons).\\
The partition function (\ref{eq:GC1}) contains all information to obtain the particle multiplicity $\langle n_j^{GC}\rangle$ of particle species $j$ (see section \ref{sec:Quantumstat})
\begin{equation}
\langle n_j^{GC}\rangle = \frac{g_jV}{2\pi^2}\int^{\infty}_0\frac{p^2dp}{\exp[\beta(\epsilon_j-\vec{x}_j\cdot\vec{\mu})]\pm 1}
\label{eq:density}
\end{equation}
For hadron j with baryon number $N_j$, isospin $I_{3j}$, strangeness $S_j$, charmness $C_j$ and bottomness $B_j$, the particle specific chemical potential $\mu_j$ is  
\begin{equation}
\mu_j = \vec{x}_j\cdot\vec{\mu} = \mu_NN_j + \mu_{I_3}I_{3j} + \mu_SS_j + \mu_CC_j + \mu_BB_j
\end{equation}
The chemical potentials related to baryon number ($\mu_N$), isospin ($\mu_{I_{3j}}$), strangeness ($\mu_S$), charmness ($\mu_C$) and bottomness ($\mu_B$) ensure the conservation (on average) of the respective quantum number:
\begin{align}
\text{baryon number: }\quad &V\sum_j n_jN_j = Z + N,\\
\text{strangeness: }\quad& V\sum_j n_jS_j =0,\\
\text{isospin: }\quad& V\sum_j n_jI_{3j} = I_3^{tot}\\
\text{charmness: }\quad& V\sum_j n_jC_j = 0\\
\text{and bottemness: }\quad& V\sum_j n_jB_j = 0
\end{align}
Where the (net) baryon number $Z + N$ ($Z$ = number of protons, $N$ = number of neutrons) and the total isospin $I_3^{tot}$ of the system are input values which need to be specified according to the colliding particles (initial system) studied\footnote{The degree of stopping of the colliding particles, which is energy dependent and cannot be precisely determined experimentally, brings some uncertainty in the choice of initial baryon number and $I_3^{tot}$.}.\\
The temperature $T$, the volume V and the baryo-chemical potential $\mu_B$ are the three independent parameters of the model, while the chemical potentials $\mu_S, \mu_{I_3}, \mu_C, \mu_B$, are fixed by the conservation laws above\footnote{These conservation laws apply strictly only for quantities which are evaluated over the complete phase space.}. $T$, $V$ and $\mu_B$ are obtained from fits to experimental data. When fitting ratios of hadron yields instead of pure yields, the parameter $V$ will cancel out.

\subsection{Canonical ensemble}

The grand-canonical ensemble is the simplest realization of a statistical approach and is suited for systems with a large number of particles. However, for small systems and for low energies, a canonical treatment is mandatory. It leads to a phase space reduction for particle production so-called "canonical suppression".\\
The exact conservation of quantum numbers replaces the chemical potential of the grand-canonical ansatz by delta functions
\begin{equation}
\mathcal{Z}^C(V,\beta) = \sum_{\text{states}}e^{-\beta\epsilon_s}\delta_{\vec{X}_s,\vec{X}}
\label{eq:canensemble1}
\end{equation}
where $\vec{X}=(N,S,Q,C,B)$\footnote{In the canonical case it is often easier to use the electric charge Q instead of the isospin quantum number.} is the initial quantum number vector including all quantum numbers which are conserved and $\vec{X}_s$ contains the quantum numbers of the state $\vert s\rangle$. The delta function forces the exact conservation of quantum numbers.\\
The particle multiplicity in the canonical approach is suppressed, due to the fact, that all quantum numbers have to be conserved exactly (will be discussed in detail in chapter \ref{sec:about}). This canonical suppression is the main difference between the canonical and the grand-canonical ensemble at relativistic energies. A detailed derivation of the canonical partition function is given in chapter \ref{sec:themodel}.\\
The canonical ensemble will find the grand-canonical results as the asymptotic value for large volume and/or large temperature (see chapter \ref{sec:about}).

\subsection{Strangeness suppression mechanism}
\label{subsec:strange}

In the application of the Thermal Model to particle production in heavy-ion and particularly in elementary particle collisions it was found that canonical suppression alone is not sufficient to quantify the observed strange particle yields. Consequently, additional suppression mechanisms were proposed to account for deviations from experimental data.\\
The suppression of strangeness has been parameterized by a factor, $\gamma_S$, that is introduced to suppress hadrons composed of strange and/or anti-strange quarks \cite{Becattini:2003wp,Becattini:2001fg}. $\gamma_s$ is a value between 0 and 1, where 1 means that all strange particles fill there whole phase space. In this description the particle partition function (eq. (\ref{eq:GC2}) in the case of the grand-canonical ensemble) of a particle composed of $n_s$ strange quarks/antiquarks is modified in the grand-canonical as well as in the canonical ensemble by,
\begin{equation}
z^1_j \rightarrow \gamma_s^{n_{s_j}} z^1_j
\end{equation}
where $\gamma_s$ is an additional parameter of the model and $n_{s_j}$ is the absolute strangeness of particle $j$.

\subsection{Thermal Model and $e^+e^-$ collisions}

The reach of an equilibrium is the reason, why the Thermal Model can be used to describe the hadron production of a high energy heavy ion collision. The question is whether this is also applicable to an $e^+e^-$ collision, since it is a quite small system compared to heavy ion collisions.\\
The result of previous investigations \cite{Becattini:1995if,Becattini:1996gy,Becattini:1997uf} was that the temperature values deduced are almost constant near T=160 MeV and that the volume increases with energy, while strangeness is under-saturated. These results were taken, together with the results for nucleus-nucleus collisions where a similar temperature is reached at high energies, as evidence for the interpretation that the thermodynamical state is not reached by dynamical equilibration among constituents but rather is a generic fingerprint of hadronization \cite{Stock:1999hm,heinz} or a feature of the excited QCD vacuum \cite{castorina}.\\
Alternatively, it was argued in \cite{wetterich} that the quark-hadron phase transition drives the equilibration dynamically for nucleus-nucleus collisions. An apparent equilibration in $e^{+}e^{-}$ collisions is not easy to explain in this latter approach.\\
Another difficulty in context of $e^+e^-$ collisions is the jet structure. The fragmentation models of section \ref{sec:frag}, treat the particle production in $e^+e^-$ collisions not separately in two jets. In these models the jets are combined by a string or cluster system, means they are not independent of each other. This is possible, because the particle production is treated on parton level\footnote{If there is no treatment of separate jets in string or cluster models, how do they explain the jet structure in the final state? The jet structure is a result of the first quark pair arising from the $Z^0$ decay. For the radiation of hard gluons (gluons with much energy) small angles are preferred. The propagator factor is
\begin{equation}
\frac{1}{p^2_q+p^2_g} \approx \frac{1}{2E_qE_q(1-\cos\theta_{qg})}
\end{equation}
This means there is a \textbf{preferred energy direction} (= jet), the direction of the original quarks (quarks from the $Z^0$ decay). This leads to a jet structure.}.\\
If the Thermal Model is applied to an $e^+e^-$ collision the assumption of the creation of a fireball in thermal equilibrium is needed. If only one fireball is assumed, it is hard to argue how a two jet structure can be produced. Therefore, the model usually works with the creation of two fireballs. N jet events caused by gluon radiation are neglected, which is reasonable (see section \ref{sec:char}).\\
A further assumption is, that a jet keeps the original quantum numbers throughout its evolution. At this point a distinction is drawn between the so called uncorrelated and the correlated jet scheme:
\begin{itemize}
\item In the uncorrelated jet scheme, both jets have to conserve the initial quantum numbers. In an $e^+e^-$ collision all relevant initial quantum numbers are zero.
\item A more general approach is to allow the sharing of quark pairs, means only both jets together conserve the quantum numbers of the initial system. This is called the correlated jet scheme.
\end{itemize}
In this work both schemes, the correlated and the uncorrelated jet scheme, are treated. In the case of the correlated jet scheme it would make sense to transfer the quantum numbers of one quark originated by the $Z^0$ decay, to one fireball. But then the quantum numbers of the quarks have to be conserved by the different fireballs separately. This will fail, because of the fractional quantum numbers of quarks like e.g. baryon number 1/3. The Thermal Model does not know any quark structure\footnote{One exception is the parameter $\gamma_s$ which is discussed above.}, it knows only particle properties like mass and quantum numbers of the different hadrons and uses the Boltzmann factor to calculate their production possibilities. A fractional baryon number cannot be conserved through the production of hadrons (same for the electric charge). In addition to that the color is not conserved. This means the jets must interact at some stage of the hadronization process. The problem of fractional quantum numbers will be discussed in chapter \ref{sec:results}.

\chapter{The canonical partition function at relativistic energies}
\label{sec:themodel}

The exact treatment of quantum numbers in statistical mechanics has been well established \cite{Hagedorn:1970gh,Shuryak:1973pv,Rafelski:1980gk,Hagedorn:1984uy} for some time now. It is generally obtained by projecting the partition function onto the desired values of the conserved charge by using group theoretical methods \cite{Turko:2000if,Redlich:2003dw}. In this section these methods are introduced and it is shown, how one gets the partition function that accounts for exact conservation of quantum numbers.\\
Later the derivation of a numerically manageable quantum statistical partition function is shown in detail. The partition function includes the conservation of baryon number, strangeness, charmness and bottomness. In section \ref{sec:extent} the equations are extended to conserve also the electric charge.\\
The results of this chapter are the equations used in the C++ program, developed within this thesis, to simulate the Thermal Model. All the results presented in chapter \ref{sec:results} are obtained by using this program.

\section{Canonical description of an internal symmetry for Abelian charges}

The description of a system in a canonical ensemble is based on a fixed number of particles. However, in particle collisions at relativistic energies the number of particles is no longer conserved, as particles can be produced from the system energy. So the concept of the canonical distribution should be modified.\\
The canonical distribution used here is related to the given irreducible representation of the symmetry group. Consider, for example, a gas consisting of particles all of which transform under the same representation $U(g)^{\alpha}$. The transformation properties of the gas will be given by some reducible unitary representation $U(g)$ which is a multiple tensor product of the representations $U(g)^{\alpha}$. One can obtain states transforming under different irreducible unitary representations of the symmetry group by decomposing the representation $U(g)$ into irreducible ones.\\
The usual way of treating the problem of quantum number conservation in statistical physics is by introducing the grand-canonical partition function
\begin{equation}
\mathcal{Z}(\mu_k,V,\beta) = Tr[e^{-\beta(\widehat{H} - \sum_k\mu_k \widehat{X}_k)}]
\label{eq:partition0.3}
\end{equation}
with the Hamiltonian $\widehat{H}$ of the system, $\widehat{X}_k$ the charge operator, $\beta$ the inverse temperature and $\mu_k$ the chemical potential associated with the conserved charge $k$, e.g. baryon number or strangeness. By denoting the states under the trace as $|s\rangle$ such that $\widehat{H}|s\rangle = E_s|s\rangle$ and $\widehat{X}_k|s\rangle = X_{ks}|s\rangle$ one can rewrite eq. (\ref{eq:partition0.3}) to 
\begin{equation}
\mathcal{Z}(\mu_k,V,\beta) = \sum_{\text{states}}e^{-\beta E_s}\; e^{-\beta\sum_k\mu_k X_{ks}}
\label{eq:partition0.2}
\end{equation}
The chemical potentials $\mu_k$ play the role of Lagrange multipliers which are fixed by the condition that the average value of the corresponding quantum number of a thermodynamical system is conserved and has the required value
\begin{equation}
\langle k \rangle = T\frac{\partial \ln \mathcal{Z}(\mu_k,V,\beta)}{\partial \mu_k}
\end{equation}
This method, as is shown in chapter \ref{sec:about}, is only adequate if the number of particles carrying the quantum number is asymptotically large and their fluctuation can be neglected.\\
For simplicity only one quantum number, the baryon number $N$, will be treated in the following equations, in order to explain the transition to exact quantum number conservation. The generalization to more quantum numbers is straightforward.\\
Then equation (\ref{eq:partition0.2}) becomes
\begin{equation}
\mathcal{Z}(\mu_N,V,\beta) = \sum_{\text{states}}e^{-\beta E_s}\; e^{-\beta\mu_N N_s} = \sum_{N=-\infty}^{\infty}\mathcal{Z}_N(V,\beta)\lambda_N^{N}\;
\label{eq:partition0.4}
\end{equation}
where the fugacity $\lambda_N = \exp(\beta\mu_N)$ is introduced and
\begin{equation}
\mathcal{Z}_N(V,\beta) = \sum_{s_N}e^{-\beta E_{s_N}}
\label{eq:partition0.5}
\end{equation}
where $s_N$ is restricted to those states that carry an exact value $N$ of the conserved charge. $\mathcal{Z}_N(V,\beta)$ is the canonical partition function with respect to baryon number conservation. The goal is to calculate $\mathcal{Z}_N(V,\beta)$.\\
Equation (\ref{eq:partition0.5}) can be rewritten with a projection operator
\begin{equation}
\mathcal{Z}_N(V,\beta) = \sum_{s_N}e^{-\beta E_{s_N}} = \sum_{\text{states}} e^{-\beta E_s}P_N
\label{eq:partition0.6}
\end{equation}
where $P_N = P_N^2$ is the projection operator on the states with exact value $N$. For an Abelian symmetry, $P_N$ is the $\delta$-function, $P_N = \delta_{N,N_s}$
\begin{equation}
\mathcal{Z}_N(V,\beta) = \sum_{\text{states}} e^{-\beta E_s}\delta_{N,N_s}
\label{eq:partition0.7}
\end{equation}
The conservation of additive quantum numbers like baryon number, strangeness, electric charge, charmness or bottomness is related to the invariance of the Hamiltonian under the Abelian $U(1)$ internal symmetry group. In many applications it is important to generalize the projection method to symmetries that are related to a non-Abelian Lie group. An example is the special unitary group $SU(N)$ that plays an essential role in the theory of strong interactions. A generalization of the projection method would require to specify the projection operator.\\
However, as all quantum numbers which are important for this work, namely baryon number, strangeness, electric charge, charmness and bottomness are Abelian, the generalization to non-Abelian symmetries will not be discussed further, but the reader is referred to the original papers of Redlich and Turko \cite{Redlich:1981xs,Turko:1981nr,Redlich:1979bf} or a nice summery, given in \cite{BraunMunzinger:2003zd}.

\section[Derivation of the partition function]{Derivation of the partition function: \\Conserving baryon number, strangeness, charmness and bottomness}

In the case of the conservation of the baryon number, strangeness, charmness and bottomness the symmetry group is $U(1)^4$, where each $U(1)$ corresponds to one of the conserved quantum numbers. Generalizing equation (\ref{eq:partition0.7}) to four exactly conserved quantum numbers yields\footnote{From now on the dependence of $\mathcal{Z}$ on $V$ and $\beta$ is not indicated anymore; instead the dependence on the quantum number vector $\vec{X}$ is pointed out.}
\begin{equation}
\mathcal{Z}_{N,S,C,B}(\vec{X}) = \sum_{\text{states}} e^{-\beta E_s}\delta_{\vec{X},\vec{X}_s}
\label{eq:partition0.8}
\end{equation}
with the four dimensional vector $\vec{X}=(N,S,C,B)$. $\delta_{\vec{X}_s,\vec{X}}$ is the usual Kronecker tensor, which forces the sum to be performed only over the states $|s\rangle$ whose quantum numbers $\vec{X}_s$ are equal to the initial set $\vec{X}$.\\
The integral representation of the delta-function above is
\begin{equation}
\delta_{\vec{X}_s,\vec{X}} = \frac{1}{(2\pi)^4}\int^{2\pi}_0\int^{2\pi}_0\int^{2\pi}_0\int^{2\pi}_0d^4\vec{\phi}\; e^{i(\vec{X} - \vec{X}_s)\vec{\phi}}
\label{eq:partition0.9}
\end{equation}
Inserting (\ref{eq:partition0.9}) in equation (\ref{eq:partition0.8}) yields \cite{Becattini:1995if}
\begin{equation}
\mathcal{Z}_{N,S,C,B}(\vec{X}) = \frac{1}{(2\pi)^4}\int^{2\pi}_0\int^{2\pi}_0\int^{2\pi}_0\int^{2\pi}_0d^4\vec{\phi}\; e^{i\vec{X}\vec{\phi}}\sum_{\text{states}}e^{-\beta E_s-i\vec{X}_s\vec{\phi}}
\label{eq:partition2}
\end{equation}
with $\vec{\phi} = (\phi_N,\phi_S,\phi_C,\phi_B)$. Each integral over one of the $\phi$'s represents the conservation of one quantum number.\\
A state is specified by a set of occupation numbers $\lbrace \vec{n}_{j,k}\rbrace$ for each phase space cell $k$ and for each particle species $j$. This means $\vec{X}_s=\sum_{j,k}\vec{x}_j\vec{n}_{j,k}$, where $\vec{x}_j=(N_j,S_j,C_j,B_j)$ is the quantum number vector associated to the $j^{th}$ particle. After summing over these states by using the results of section \ref{sec:Quantumstat}, eq. (\ref{eq:partition2}) becomes
\begin{equation}
\begin{split}
\mathcal{Z}_{N,S,C,B}(\vec{X}) &= \frac{1}{(2\pi)^4}\int d^4\vec{\phi}\; e^{i\vec{X}\vec{\phi}}\prod^{N_B}_{j=1}\prod_k\left(\frac{1}{1-e^{-\beta \epsilon_{kj}-i\vec{x}_j\vec{\phi}}}\right)\\ 
&\quad\prod^{N_F}_{j=1}\prod_k(1+e^{-\beta \epsilon_{kj}-i\vec{x}_j\vec{\phi}})
\end{split}
\end{equation}
with $N_B$ = number of bosons and $N_F$ = number of fermions. It is possible to rewrite this equation to
\begin{equation}\label{eq:partition3}
\begin{split}
\mathcal{Z}_{N,S,C,B}(\vec{X}) &= \frac{1}{(2\pi)^4}\int d^4\vec{\phi}\; e^{i\vec{X}\vec{\phi}}\exp\left\{\sum^{N_B}_{j=1}\sum_k \ln(1-e^{-\beta \epsilon_{kj}-i\vec{x}_j\vec{\phi}})^{-1}\right.\\
&\left.\quad+ \sum^{N_F}_{j=1}\sum_k \ln(1 + e^{-\beta \epsilon_{kj}-i\vec{x}_j\vec{\phi}})\right\}
\end{split}
\end{equation}
To get the mean number (multiplicity) of any hadronic particle one has to assign a fictitious fugacity $\lambda_j$ which multiplies the Boltzmann factor $\exp(-\beta \epsilon_{kj}-i\vec{x}_j\vec{\phi})$. This fugacity is set to one after differentiation
\begin{equation}
\left.\langle n_j\rangle = \frac{\partial\ln \mathcal{Z}}{\partial \lambda_j}\right|_{\lambda_j=1}
\label{eq:partition3.1}
\end{equation}
hence
\begin{equation}\label{eq:partition3.2}
\begin{split}
\langle n_j\rangle &= \frac{1}{\mathcal{Z}_{N,S,C,B}(\vec{X})}\frac{1}{(2\pi)^4}\int d^4\vec{\phi}\; e^{i\vec{X}\vec{\phi}}\exp\Bigg\{\sum^{N_B}_{j=1}\sum_k \ln(1-e^{-\beta \epsilon_{kj}-i\vec{x}_j\vec{\phi}})^{-1}\cr
&\left.\quad+ \sum^{N_F}_{j=1}\sum_k \ln(1+e^{-\beta \epsilon_{kj}-i\vec{x}_j\vec{\phi}})\right\}\sum_k \frac{1}{e^{\beta \epsilon_{kj}-i\vec{x}_j\vec{\phi}}\pm 1}
\end{split}
\end{equation}
where the $+$ is for fermions and $-$ for bosons.\\
In the rest frame of the system, the four-vector $\beta$ reduces to
\begin{equation}
\beta = (1/T,0,0,0)
\label{eq:partition4}
\end{equation}
with the temperature $T$ of the system. Furthermore, the sum over phase space cells in eq. (\ref{eq:partition3}) can be turned into an integration by changing to the continuum limit, where each quantum state refers to a classical phase space volume of $(2\pi\hbar)^3$
\begin{equation}
\sum_k \rightarrow \frac{g_j}{(2\pi\hbar)^3}\int_\Gamma d\Gamma
\label{eq:partition4.1}
\end{equation}
with $d\Gamma = d^3q\;d^3p$ and the spin degeneracy $g_j$. In a homogeneous system the volume integral $\int_Vd^3q$ gives a factor $V$ and (\ref{eq:partition4.1}) is then
\begin{equation}
\sum_k \rightarrow \frac{g_jV}{(2\pi\hbar)^3}\int d^3p
\label{eq:partition5}
\end{equation}
The energy of a relativistic particle under the condition of a non-interacting system is
\begin{equation}
\epsilon_{kj} = \sqrt{p^2_k+m_j^2} 
\label{eq:partition5.1}
\end{equation}
By using (\ref{eq:partition4}), (\ref{eq:partition5}) and (\ref{eq:partition5.1}), eq. (\ref{eq:partition3}) becomes
\begin{empheq}[box=\fbox]{equation}
\label{eq:partition6}
\begin{split}
\mathcal{Z}_{N,S,C,B}(\vec{X}) &= \frac{1}{(2\pi)^4}\int d^4\vec{\phi}\; e^{i\vec{X}\vec{\phi}}\\
&\quad \exp\left\{\sum_j\frac{g_jV}{(2\pi\hbar)^3}\int d^3p \ln(1\pm e^{-\frac{\sqrt{\vec{p}^2+m_j^2}}{T}-i\vec{x}_j\vec{\phi}})^{\pm1}\right\}
\end{split}
\end{empheq}
For $\exp(-\sqrt{\vec{p}^2+m_j^2}/T-i\vec{x}_j\vec{\phi}) \ll 1$ the Boltzmann approximation discussed in section \ref{sec:Boltzmann} can be used\footnote{For kaons the Boltzmann approximation causes an error of about $1\%$, but for pions the error can rise over $10\%$. Therefore all particles with $\vec{x} = (0,0,0,0)$ are calculated in correct quantum statistics (see eq.~(\ref{eq:partition8.12})).}
%From the QCD soft scale we know T $\approx$ $\mathcal{O}$(100)MeV. Therefore we can use the first term of the series of logarithm (Boltzmann approximation) for fermions and bosons (this is valid for all particles but pions, because $m_{\pi} \approx T$)
\begin{equation}
\ln(1\pm e^{-\frac{\sqrt{\vec{p}^2+m_j^2}}{T}-i\vec{x}_j\vec{\phi}})^{\pm1} \simeq e^{-\frac{\sqrt{\vec{p}^2+m_j^2}}{T}-i\vec{x}_j\vec{\phi}}
\label{eq:partition7}
\end{equation}
Inserting (\ref{eq:partition7}) into (\ref{eq:partition6}) yields
\begin{equation}
\mathcal{Z}_{N,S,C,B}(\vec{X}) = \dfrac{Z_{0}}{(2\pi)^4}\int d^4\vec{\phi}\; e^{i\vec{X}\vec{\phi}}\exp\left\{\sum_jz^1_je^{-i\vec{x}_j\vec{\phi}}\right\}
\label{eq:partition8}
\end{equation}
with the particle partition function
\begin{equation}
z^1_j = \frac{g_jV}{(2\pi\hbar)^3}\int d^3p\; e^{-\frac{\sqrt{\vec{p}^2+m_j^2}}{T}}
\label{eq:partition8.01}
\end{equation}
and
\begin{equation}
Z_{0} = \exp\left[\sum_{j}\frac{V}{(2\pi\hbar)^3}\int d^3p \ln\left(1-e^{-\frac{\sqrt{\vec{p}^2+m_j^2}}{T}}\right)^{-1}\right]
\label{eq:partition8.12}
\end{equation}
is the exponential of the sum over all pion-like states $j$, whose relevant quantum numbers are zero (states with $\vec{x}_j = \vec{0}$). With (\ref{eq:partition3.1}) and (\ref{eq:partition8}) one gets
\begin{equation}
\langle n_j\rangle = z^1_j\frac{\mathcal{Z}_{N,S,C,B}(\vec{X}-\vec{x}_j)}{\mathcal{Z}_{N,S,C,B}(\vec{X})}
\label{eq:partition8.22}
\end{equation}
whereas for all particles with $\vec{x}_j=\vec{0}$
\begin{equation}
\langle n_j(\vec{x}_j=\vec{0})\rangle = \frac{g_jV}{(2\pi\hbar)^3}\int d^3p \frac{1}{e^{\frac{\sqrt{\vec{p}^2+m_j^2}}{T}}-1}
\label{eq:partition8.2}
\end{equation}
Assuming that $\sum_{j_{c/b}}z^1_{j_{c/b}} \ll 1$ \textbf{for charm and bottom hadrons}\footnote{The particle partition function $z^1_{j_{c/b}}$ is $\mathcal{O}(10^{-4})$ for charm particles and $\mathcal{O}(10^{-13})$ for bottom particles.}, $\exp\{\sum_{j_{c/b}}z^1_{j_{c/b}}e^{-i\vec{x}_{j_{c/b}}\vec{\phi}}\}$ can be expanded in power of $\sum_{j_{c/b}}z^1_{j_{c/b}}$ up to first order \cite{Becattini:1995if}
\begin{equation}
\exp\left\{\sum_{j_{c/b}}z^1_{j_{c/b}}e^{-i\vec{x}_j\vec{\phi}}\right\} \simeq 1 + \sum_{j_{c/b}}z^1_{j_{c/b}}e^{-i\vec{x}_{j_{c/b}}\vec{\phi}}
\label{eq:partition9}
\end{equation}
Inserting (\ref{eq:partition9}) in (\ref{eq:partition8}) (for charm and bottom hadrons) one obtains
\begin{equation}\label{eq:partition10}
\begin{split}
\mathcal{Z}_{N,S,C,B}(\vec{X}) & \approx \frac{Z_{0}}{(2\pi)^{4}}\int d^{4}\vec{\phi}\; e^{i\vec{X}\vec{\phi}} e^{f(\vec{\phi})}\\
&\quad + \sum_{j_{c}}z^1_{j_{c}}\frac{Z_{0}}{(2\pi)^{4}}\int d^{4}\vec{\phi}\; e^{i(\vec{X}-\vec{x}_{j_{c}})\vec{\phi}} e^{f(\vec{\phi})}\\
&\quad + \sum_{j_{b}\;\;\&\atop C_{j_b} = 0}z^1_{j_{b}}\frac{Z_{0}}{(2\pi)^{4}}\int d^{4}\vec{\phi}\; e^{i(\vec{X}-\vec{x}_{j_{b}})\vec{\phi}} e^{f(\vec{\phi})}\\
&\quad + \sum_{j_{c}}\sum_{j_{b}\;\;\&\atop C_{j_b} = 0}z^1_{j_{c}}z^1_{j_{b}}\frac{Z_{0}}{(2\pi)^{4}}\int d^{4}\vec{\phi}\; e^{i(\vec{X}-\vec{x}_{j_{c}}-\vec{x}_{j_{b}})\vec{\phi}} e^{f(\vec{\phi})}
\end{split}
\end{equation}
with
\begin{equation}
f(\vec{\phi}) = \sum_jz^1_je^{-i\vec{x}_j\vec{\phi}}
\end{equation}
where the index $j$ runs over all charged hadrons except heavy flavored ones. The index $j_c$ runs over all charm hadrons and $j_b$ runs over all bottom hadrons. In this equation there are $2\pi$-periodic functions (the exponential functions) within $2\pi$-integrals. The contributions of the charm and bottom particles are zero, except if the argument of the exponential function is zero. Therefore the integrals over $\phi_C$ and $\phi_B$ can be replaced by Kronecker deltas
\begin{equation}\label{eq:partition11}
\begin{split}
\mathcal{Z}_{N,S,C,B}(\vec{X}) & \approx Z_{0}\bigg(\frac{1}{(2\pi)^{2}}\int d^{2}\vec{\phi}\; e^{i\vec{X}\vec{\phi}} e^{f(\vec{\phi})} \delta_{C,0}\delta_{B,0}\\
&\quad + \sum_{j_{c}}z^1_{j_{c}}\frac{1}{(2\pi)^{2}}\int d^{2}\vec{\phi}\; e^{i(\vec{X}-\vec{x}_{j_{c}})\vec{\phi}} e^{f(\vec{\phi})} \delta_{C,C_{j_{c}}}\delta_{B,B_{j_c}}\\
&\quad + \sum_{j_{b} \;\;\&\atop C_{j_b} = 0}z^1_{j_{b}}\frac{1}{(2\pi)^{2}}\int d^{2}\vec{\phi}\; e^{i(\vec{X}-\vec{x}_{j_{b}})\vec{\phi}} e^{f(\vec{\phi})} \delta_{C,0}\delta_{B,B_{j_{b}}}\\
&\quad + \sum_{j_{c}}\sum_{j_{b}\;\;\&\atop C_{j_b} = 0}z^1_{j_{c}}z^1_{j_{b}}\frac{1}{(2\pi)^{2}}\int d^{2}\vec{\phi}\; e^{i(\vec{X}-\vec{x}_{j_{c}}-\vec{x}_{j_{b}})\vec{\phi}} e^{f(\vec{\phi})} \delta_{C,C_{j_{c}}}\delta_{B,B_{j_{c}}+B_{j_b}}\bigg)
\end{split}
\end{equation}
Now, $\vec{X}$ and $\vec{x}_j$ are two-dimensional vectors having as components the baryon number and the strangeness, whilst the charmness $C$ and bottomness $B$ appear only in the Kronecker deltas.\\
The integral representation of the partition function above is not convenient for numerical analysis as the integrand is a strongly oscillating function. However, it is possible to rewrite the integrals into sums over Bessel-functions \cite{Cleymans:1990ia,Cleymans:1991yu,Cleymans:1997ib}.\\
The integrals which have to be solved, are
\begin{equation}
\mathcal{I}_{N,S} = \frac{1}{(2\pi)^2}\int^{2\pi}_{0}d^2\vec{\phi}\; e^{i\vec{X}\vec{\phi}}\exp\left\{\sum_j z^1_j e^{-i\vec{x}_j\vec{\phi}}\right\}
\end{equation}
and after rewriting
\begin{equation}\label{eq:partition12}
\begin{split}
\mathcal{I}_{N,S} &= \frac{1}{2\pi}\int^{2\pi}_0 d\phi_N\; e^{iN\phi_N}\exp[Z_p(e^{i\phi_N} + e^{-i\phi_N})]\\
&\quad\frac{1}{2\pi}\int^{2\pi}_0 d\phi_S\; e^{iS\phi_S} \exp[Z_K(e^{i\phi_S} + e^{-i\phi_S})]\\
&\quad\exp[Z_{\Lambda}(e^{i(\phi_N-\phi_S)} + e^{-i(\phi_N-\phi_S)})]\\
&\quad\exp[Z_{\Xi}(e^{i(\phi_N-2\phi_S)} + e^{-i(\phi_N-2\phi_S)})]\\
&\quad\exp[Z_{\Omega}(e^{i(\phi_N-3\phi_S)} + e^{-i(\phi_N-3\phi_S)})]
\end{split}
\end{equation}
where $Z_{K,p,\Lambda,\Xi,\Omega}$ are sums over single particle partition functions having the same quantum numbers (see Table \ref{tabl:model1}).
\begin{table}[]
\begin{center}
\begin{tabular}{ccc}
  \hline
  \hline
  \multicolumn{2}{c}{Quantum numbers} & Notation\\
  \hline
  $N$=0 & $S$=0 & ($Z_0$)\\
  $N$=0 & $S$=1 & $Z_K$\\
  $N$=1 & $S$=0 & $Z_p$\\
  $N$=1 & $S$=-1 & $Z_{\Lambda}$\\
  $N$=1 & $S$=-2 & $Z_{\Xi}$\\
  $N$=1 & $S$=-3 & $Z_{\Omega}$\\
  \hline
  \hline
\end{tabular}
\caption{The possible quantum number combinations for particles ($N$ = baryon number, $S$ = strangeness).}
\label{tabl:model1}
\end{center}
\end{table}
Note that all pion-like particles are already separated in $Z_0$ (see eq.~(\ref{eq:partition8.12})). For the last three terms in eq.~(\ref{eq:partition12}) the following relation of the modified Bessel-function can be used~\cite{Cleymans:1990mn}
\begin{equation}
\exp\left[\frac{x}{2}\left(t+\frac{1}{t}\right)\right] = \sum^{\infty}_{m=-\infty}t^mI_m(x)
\label{eq:partition13}
\end{equation}
and (\ref{eq:partition12}) becomes:
\begin{equation}\label{eq:partition14}
\begin{split}
\mathcal{I}_{N,S} &= \sum^{\infty}_{n_1=-\infty}I_{n_1}(2Z_{\Xi})\sum^{\infty}_{n_2=-\infty}I_{n_2}(2Z_{\Omega})\sum^{\infty}_{n_3=-\infty}I_{n_3}(2Z_{\Lambda})\\ 
&\quad\frac{1}{2\pi}\int^{2\pi}_0 d\phi_N\; e^{i(N+n_1+n_2+n_3)\phi_N}\exp[Z_p(e^{i\phi_N} + e^{-i\phi_N})]\\
&\quad\frac{1}{2\pi}\int^{2\pi}_0  d\phi_S\; e^{i(S-n_1-2n_2-3n_3)\phi_S} \exp[Z_K(e^{i\phi_S} + e^{-i\phi_S})]
\end{split}
\end{equation}
By using
\begin{equation}
\frac{1}{2}(e^{i\phi}+e^{-i\phi}) = \cos\phi
\label{eq:partition15}
\end{equation}
one gets
\begin{equation}\label{eq:partition16}
\begin{split}
\mathcal{I}_{N,S} &= \sum^{\infty}_{n_1=-\infty}I_{n_1}(2Z_{\Xi})\sum^{\infty}_{n_2=-\infty}I_{n_2}(2Z_{\Omega})\sum^{\infty}_{n_3=-\infty}I_{n_3}(2Z_{\Lambda})\\ 
&\quad\frac{1}{2\pi}\int^{2\pi}_0 d\phi_N\; e^{i(N+n_1+n_2+n_3)\phi_N} \exp[2Z_p\cos\phi_N]\cr
&\quad\frac{1}{2\pi}\int^{2\pi}_0  d\phi_S\; e^{i(S-n_1-2n_2-3n_3)\phi_S}\exp[2Z_K\cos\phi_S]
\end{split}
\end{equation}
Applying the following integral representation of the Bessel-function of order $h$
\begin{equation}
I_h(x) = \frac{1}{2\pi}\int^{2\pi}_0 d\phi \exp(x\cos\phi)\exp(-ih\phi)
\label{eq:partition16.2}
\end{equation}
yields
\begin{equation}\label{eq:partition17}
\begin{split}
\mathcal{I}_{N,S} &=
\sum^{\infty}_{n_1=-\infty}I_{n_1}(2Z_{\Xi})\sum^{\infty}_{n_2=-\infty}I_{n_2}(2Z_{\Omega})\sum^{\infty}_{n_3=-\infty}I_{n_3}(2Z_{\Lambda})\\
&\quad I_{-N-n_1-n_2-n_3}(2Z_p)I_{-S+n_1+2n_2+3n_3}(2Z_K)
\end{split}
\end{equation}
Now, the integral can be solved by summing over Bessel-functions. The last step is to replace the integrals in (\ref{eq:partition11}) by (\ref{eq:partition17}) to obtain
\begin{equation}\label{eq:partition18}
\begin{split}
\mathcal{Z}_{N,S,C,B}(\vec{X}) & \approx Z_{0}\bigg(\mathcal{I}_{N,S}\;\delta_{C,0}\delta_{B,0}\\
&\quad +  \sum_{j_{c}}z^1_{j_{c}}\mathcal{I}_{N-N_{j_c},S-S_{j_c}} \delta_{C,C_{j_{c}}}\delta_{B,B_{j_c}}\\
&\quad +  \sum_{j_{b} \;\;\&\atop C_{j_b} = 0}z^1_{j_{b}}\mathcal{I}_{N-N_{j_b},S-S_{j_b}} \delta_{C,0}\delta_{B,B_{j_{b}}}\\
&\quad +  \sum_{j_{c}}\sum_{j_{b}\;\;\&\atop C_{j_b} = 0}z^1_{j_{c}}z^1_{j_{b}}\mathcal{I}_{N-N_{j_c}-N_{j_b},S-S_{j_c}-S_{j_b}} \delta_{C,C_{j_{c}}}\delta_{B,B_{j_{c}}+B_{j_b}}\bigg)
\end{split}
\end{equation}
In the next section the charge conservation is included. The way used to derive a computable equation for $\mathcal{Z}(\vec{X}$) is very similar, therefore only the differences are shown.

\subsection[Model extension]{Model extension: Including conservation of electric charge}
\label{sec:extent}

The general canonical partition function is (eq. (\ref{eq:partition6}) for five quantum numbers)
\begin{equation}\label{eq:partition19}
\begin{split}
\mathcal{Z}_{N,S,Q,C,B}(\vec{X}) &= \frac{1}{(2\pi)^5}\int d^5\vec{\phi}\; e^{i\vec{X}\vec{\phi}}\\
&\quad \exp\left\{\sum_j\frac{g_jV}{(2\pi\hbar)^3}\int d^3p \ln(1\pm e^{-\frac{\sqrt{\vec{p}^2+m_j^2}}{T}-i\vec{x}_j\vec{\phi}})^{\pm1}\right\}
\end{split}
\end{equation}
% \begin{equation}\label{eq:partition19}
% \begin{split}
% \mathcal{Z}_{N,S,Q,C,B}(\vec{X}) &= \frac{1}{(2\pi)^5}\int d^5\vec{\phi}\; e^{i\vec{X}\vec{\phi}}\exp\{\sum^{N_B}_{j=1}\sum_k \ln(1-e^{-\beta \epsilon_{kj}-i\vec{x}_j\vec{\phi}})^{-1}\\
% &+ \sum^{N_F}_{j=1}\sum_k \ln(1 + e^{-\beta \epsilon_{kj}-i\vec{x}_j\vec{\phi}})\}
% \end{split}
% \end{equation}
where $\vec{X}$ is now a five dimensional vector, $\vec{X}=(N,S,Q,C,B)$. To apply the Bessel relations (see (\ref{eq:partition16.2}) and (\ref{eq:partition13})) to the equation above, again Boltzmann approximation has to be used. But because it is now impossible to exclude the $\pi^{\pm}$ from the integrals (see the definition of $Z_0$ in (\ref{eq:partition8.12})) the Boltzmann approximation will give rise to a significant deviation at least for the pion yield.\\
The Boltzmann approximation is just the first term of the series of $\ln(1\pm x)^{\pm1}$ (see eq. (\ref{eq:partition7})). To calculate it accurately (in quantum statistics) one has to use the whole series of the logarithm\footnote{Because of the series used (eq. (\ref{eq:partition19.2})) the quantum statistical calculation is only applied for particles which follow Bose-Einstein statistics. The extension of the equations to fermions is straightforward (see appendix~\ref{ap:fermions}). However, the deviation for fermions caused by the Boltzmann approximation is negligible because of the high mass.}
\begin{equation}
\ln(1-x)^{-1} = \sum^{\infty}_{k=1}\frac{x^{k}}{k}
\label{eq:partition19.2}
\end{equation} 
yields
\begin{equation}\label{eq:partition19.3}
\begin{split}
\mathcal{Z}_{N,S,Q,C,B}(\vec{X}) &= \frac{Z_{0}}{(2\pi)^5}\int d^5\vec{\phi}\; e^{i\vec{X}\vec{\phi}}\\
&\quad\exp\left\{\sum_j z^1_je^{-i\vec{x}_j\vec{\phi}} + \sum_{b}\sum_{k=2}^{\infty}z^k_be^{-ik\vec{x}_b\vec{\phi}}\right\}
\end{split}
\end{equation}
with
\begin{equation}
z^k_j = \frac{g_jV}{k(2\pi\hbar)^3}\int d^3p\; e^{-\frac{\sqrt{\vec{p}^2+m_j^2}}{T}k}
\label{eq:partition19.4}
\end{equation}
where all particles with $\vec{x} = \vec{0}$ are excluded within $Z_0$, which is defined in (\ref{eq:partition8.12}), but here just for $\pi^0$-like particles. The index $j$ runs over all charged particles and $b$ runs only over the charged light bosons\footnote{Charged means particles with $\vec{x}\neq\vec{0}$, not necessarily connected to the electric charge $Q$.}.\\
% While all bosons are calculated in quantum statistics, the calculation of fermions remains in Boltzmann approximation (for the full quantum statistical calculation see appendix \ref{ap:fermions}).\\
Applying (\ref{eq:partition3.1}) to (\ref{eq:partition19.3}) gives for bosons, e.g. the $\pi^{\pm}$
\begin{align}
\langle n_{\pi^{\pm}}\rangle &= \left.\frac{\partial\ln \mathcal{Z}}{\partial \lambda_{\pi^{\pm}}}\right|_{\lambda_{\pi^{\pm}}=1}\notag\\
&= z^1_{\pi^{\pm}}\frac{\mathcal{Z}_{N,S,Q,C,B}(\vec{X}-\vec{x}_{\pi^{\pm}})}{\mathcal{Z}_{N,S,Q,C,B}(\vec{X})} + \sum^{\infty}_{k=2}kz^k_{\pi^{\pm}}\frac{\mathcal{Z}_{N,S,Q,C,B}(\vec{X}-k\vec{x}_{\pi^{\pm}})}{\mathcal{Z}_{N,S,Q,C,B}(\vec{X})}\notag\\
&= \sum^{\infty}_{k=1}kz^k_{\pi^{\pm}}\frac{\mathcal{Z}_{N,S,Q,C,B}(\vec{X}-k\vec{x}_{\pi^{\pm}})}{\mathcal{Z}_{N,S,Q,C,B}(\vec{X})}
\end{align}
whereas for fermions and heavy bosons eq. (\ref{eq:partition8.22}) is still valid.\\
After the approximation for charm and bottom particles (see eq. (\ref{eq:partition9})) the following integrals have to be solved
\begin{equation}
\begin{split}
\mathcal{I}_{N,S,Q} &= \frac{1}{(2\pi)^3}\int^{2\pi}_0 d^3\vec{\phi}\;e^{i\vec{X}\vec{\phi}}\\
&\quad\exp\left\{\sum_j z^1_je^{-i\vec{x}_j\vec{\phi}}+\sum_{b}\sum_{k=1}^{\infty}z^k_be^{-ik\vec{x}_b\vec{\phi}}\right\}
\end{split}
\end{equation}
After rewriting
\begin{equation}\label{eq:partition20}
\begin{split}
\mathcal{I}_{N,S,Q} &= \frac{1}{2\pi}\int^{2\pi}_0 d\phi_N\; e^{iN\phi_N}\exp[Z_n(e^{i\phi_N} + e^{-i\phi_N})]\\
&\quad\frac{1}{2\pi}\int^{2\pi}_0 d\phi_S\; e^{iS\phi_S} \exp[Z_{K^0}(e^{i\phi_S} + e^{-i\phi_S})]\\
&\quad\frac{1}{2\pi}\int^{2\pi}_0 d\phi_Q\; e^{iQ\phi_Q}\exp[Z_{\pi^{\pm}}(e^{i\phi_Q} + e^{-i\phi_Q})]\\
&\quad\exp[Z_{p}(e^{i(\phi_N+\phi_Q)} + e^{-i(\phi_N+\phi_Q)})]\\
&\quad\exp[Z_{\Delta^{\mp}}(e^{i(\phi_N-\phi_Q)} + e^{-i(\phi_N-\phi_Q)})]\\
&\quad\exp[Z_{\Delta^{++}}(e^{i(\phi_N+2\phi_Q)} + e^{-i(\phi_N+2\phi_Q)})]\\
&\quad\exp[Z_{K^{\pm}}(e^{i(\phi_S+\phi_Q)} + e^{-i(\phi_S+\phi_Q)})]\\
&\quad\exp[Z_{\Lambda}(e^{i(\phi_N-\phi_S)} + e^{-i(\phi_N-\phi_S)})]\\
&\quad\exp[Z_{\Sigma^+}(e^{i(\phi_N-\phi_S+\phi_Q)} + e^{-i(\phi_N-\phi_S+\phi_Q)})]\\
&\quad\exp[Z_{\Sigma^-}(e^{i(\phi_N-\phi_S-\phi_Q)} + e^{-i(\phi_N-\phi_S-\phi_Q)})]\\
&\quad\exp[Z_{\Xi^0}(e^{i(\phi_N-2\phi_S)} + e^{-i(\phi_B-2\phi_S)})]\\
&\quad\exp[Z_{\Xi^{\mp}}(e^{i(\phi_N-2\phi_S-\phi_Q)} + e^{-i(\phi_N-2\phi_S-\phi_Q)})]\\
&\quad\exp[Z_{\Omega^{\mp}}(e^{i(\phi_N-3\phi_S-\phi_Q)} + e^{-i(\phi_N-3\phi_S-\phi_Q)})]\\
&\quad\exp\left[\sum^{\infty}_{k=2}Z^k_{\pi^{\pm}}(e^{ik\phi_Q} + e^{-ik\phi_Q})\right]\\
&\quad\exp\left[\sum^{\infty}_{h=2}Z^h_{K^{0}}(e^{ih\phi_S} + e^{-ih\phi_S})\right]\\
&\quad\exp\left[\sum^{\infty}_{l=2}Z^l_{K^{\pm}}(e^{il(\phi_S+\phi_Q)} + e^{-il(\phi_S+\phi_Q)})\right]
\end{split}
\end{equation}
where the Zs are given in Table \ref{tabl:extension}. 
\begin{table}[]
\begin{center}
\begin{tabular}{cccc}
  \hline
  \hline
  \multicolumn{3}{c}{Quantum numbers} & Notation\\
  \hline
  $N$=0 & $S$=0 & $Q$=0 & ($Z_0$)\\
  $N$=0 & $S$=1 & $Q$=0 & $Z_{K^0},\;\;Z^h_{K^{0}}$\\
  $N$=1 & $S$=0 & $Q$=0 & $Z_n$\\
  $N$=0 & $S$=0 & $Q$=1 & $Z_{\pi^{\pm}},\;\;Z^k_{\pi^{\pm}}$\\
  $N$=1 & $S$=0 & $Q$=1 & $Z_{p}$\\
  $N$=1 & $S$=0 & $Q$=-1 & $Z_{\Delta^{\mp}}$\\
  $N$=1 & $S$=0 & $Q$=2 & $Z_{\Delta^{++}}$\\
  $N$=0 & $S$=1 & $Q$=1 & $Z_{K^{\pm}},\;\;Z^l_{K^{\pm}}$\\
  $N$=1 & $S$=-1 & $Q$=0 & $Z_{\Lambda}$\\
  $N$=1 & $S$=-1 & $Q$=1 & $Z_{\Sigma^+}$\\
  $N$=1 & $S$=-1 & $Q$=-1 & $Z_{\Sigma^-}$\\
  $N$=1 & $S$=-2 & $Q$=0 & $Z_{\Xi^0}$\\
  $N$=1 & $S$=-2 & $Q$=-1 & $Z_{\Xi^{\mp}}$\\
  $N$=1 & $S$=-3 & $Q$=-1 & $Z_{\Omega^{\mp}}$\\
  \hline
  \hline
\end{tabular}
\caption{The possible quantum number combinations ($N$ = baryon number, $S$ = strangeness, $Q$ = electric charge).}
\label{tabl:extension}
\end{center}
\end{table}
Using the Bessel relations (\ref{eq:partition13}) and (\ref{eq:partition16.2}) gives
\begin{flalign*}\label{eq:partition22}
\mathcal{I}_{N,S,Q} & = \left[\prod^{10}_{j=1}\sum^{\infty}_{n_j={-\infty}}\right]I_{n1}(2Z_{p})I_{n2}(2Z_{\Delta^{\mp}})I_{n3}(2Z_{\Delta^{++}})I_{n4}(2Z_{K^{\pm}})\\
&\quad I_{n5}(2Z_{\Lambda})I_{n6}(2Z_{\Sigma^+})I_{n7}(2Z_{\Sigma^-})I_{n8}(2Z_{\Xi^0})I_{n9}(2Z_{\Xi^{\mp}})I_{n10}(2Z_{\Omega^{\mp}})\\
&\quad\left[\prod^{\infty}_{k=2}\sum^{\infty}_{n_k=-\infty}\right]I_{n_k}(2Z^{k}_{\pi^{\pm}})\left[\prod^{\infty}_{h=2}\sum^{\infty}_{n_h=-\infty}\right]I_{n_h}(2Z^{h}_{K^0})\left[\prod^{\infty}_{l=2}\sum^{\infty}_{n_l=-\infty}\right]I_{n_l}(2Z^{l}_{K^{\pm}})\\
&\quad I_{-N-n1-n2-n3-n5-n6-n7-n8-n9-n10}(2Z_n)\\
&\quad I_{-S-n4+n5+n6+n7+2n8+2n9+3n10-\sum hn_h-\sum ln_l}(2Z_{K^0})\\
&\quad I_{-Q-n1+n2-2n3-n4-n6+n7+n9+n10-\sum kn_k-\sum ln_l}(2Z_{\pi^{\pm}})
\refstepcounter{equation}\tag{\theequation}
\end{flalign*}
and the final partition function for the conservation of five quantum numbers is
\begin{equation}\label{eq:partition23}
\begin{split}
\mathcal{Z}_{N,S,Q,C,B}(\vec{X}) & \approx Z_{0}\bigg(\mathcal{I}_{N,S,Q}\;\delta_{C,0}\delta_{B,0}\\
&\quad +  \sum_{j_{c}}z^1_{j_{c}}\mathcal{I}_{N-N_{j_c},S-S_{j_c},Q-Q_{j_c}} \delta_{C,C_{j_{c}}}\delta_{B,B_{j_c}}\\
&\quad +  \sum_{j_{b} \;\;\&\atop C_{j_b} = 0}z^1_{j_{b}}\mathcal{I}_{N-N_{j_b},S-S_{j_b},Q-Q_{j_b}} \delta_{C,0}\delta_{B,B_{j_{b}}}\\
&\quad +  \sum_{j_{c}}\sum_{j_{b}\;\;\&\atop C_{j_b} = 0}z^1_{j_{c}}z^1_{j_{b}}\mathcal{I}_{N-N_{j_c}-N_{j_b},S-S_{j_c}-S_{j_b},Q-Q_{j_c}-Q_{j_b}} \delta_{C,C_{j_{c}}}\delta_{B,B_{j_{c}}+B_{j_b}}\bigg)
\end{split}
\end{equation}

\section{Alternative way to solve the canonical partition function}
\label{sec:second}

There is another possibility to calculate the canonical partition function (\ref{eq:partition19.3}) numerically. This way of calculation was first published in \cite{Keranen:2001pr}, but only for Boltzmann statistics, and with the conservation of three quantum numbers. Here the quantum statistical case with the conservation of five quantum numbers will be discussed.\\
The calculation of \cite{Keranen:2001pr} is implemented in THERMUS \cite{Wheaton:2004qb}, a Thermal Model simulator which is used later for a comparison to the results of this work (see chapter \ref{sec:results}).\\
The canonical partition function with the conservation of five quantum numbers is (eq. (\ref{eq:partition19.3}))
\begin{equation}\label{eq:second1}
\begin{split}
\mathcal{Z}_{N,S,Q,C,B}(\vec{X}) &= \frac{Z_0}{(2\pi)^5}\int d^5\vec{\phi}\; e^{i\vec{X}\vec{\phi}}\\
&\quad\exp\left\{\sum_j z^1_je^{-i\vec{x}_j\vec{\phi}} + \sum_{b}\sum_{k=2}^{\infty}z^k_be^{-ik\vec{x}_b\vec{\phi}}\right\}
\end{split}
\end{equation}
with
\begin{equation}
z^k_j = \frac{g_jV}{k(2\pi\hbar)^3}\int d^3p\; e^{-\frac{\sqrt{\vec{p}^2-m_j^2}}{T} k}
\end{equation}
Remember that $j$ runs over all charged hadrons, whereas $b$ runs over the light charged bosons (also in this section the full quantum statistical calculation is treated for bosons only). $Z_0$ contains the contributions of all particles with $\vec{x}_j=\vec{0}$ (see eq. (\ref{eq:partition8.12})).\\
First the approximation (\ref{eq:partition9}) is applied to get rid of the integrals over $\phi_C$ and $\phi_B$. Next, the exponent of the exponential function can be separated into mesons and baryons
\begin{equation}
\begin{split}
\sum_{j} z^1_j e^{-i\vec{x}\vec{\phi}} + \sum_{b}\sum_{k=2}^{\infty}z^k_be^{-ik\vec{x}_b\vec{\phi}}&= \sum_{\text{mesons}\atop (S=Q=0)} \sum_{k=1}^{\infty}z^k_j\\
&\quad+ e^{i\phi_N}\sum_{\text{baryons}}z^1_j e^{i(S_j\phi_S + Q_j\phi_Q)}\\
&\quad+ e^{-i\phi_N}\sum_{\text{baryons}}z^1_j e^{-i(S_j\phi_S + Q_j\phi_Q)}\\
&\quad+ \sum_{\text{mesons}}\sum_{k=1}^{\infty}2z_j^k \cos(kS_j\phi_S + kQ_j\phi_Q)
\end{split}
\end{equation}
where the sums over mesons and baryons extend only over the particles (i.e. not the antiparticles). Defining
\begin{equation}
\omega \equiv \sum_{\text{baryons}}z^1_je^{i(S_j\phi_S+Q_j\phi_Q)} = |\omega|e^{i\arg\omega}
\end{equation}
the two sums over baryons can be written as $2|\omega|\cos(\phi_N+\arg\omega)$. Substituting these expressions into the three remaining integrals of (\ref{eq:second1}) gives
\begin{equation}
\begin{split}
\mathcal{I}_{N,S,Q} &= \frac{1}{(2\pi)^3}\int^{2\pi}_0d\phi_N\;e^{iN\phi_N}\int^{2\pi}_0d\phi_S\;e^{iS\phi_S}\int^{2\pi}_0d\phi_Q\;e^{iQ\phi_Q}\\[1.2ex]
&\quad\exp(2|\omega|\cos(\phi_N+\arg\omega))\\[1.2ex]
&\quad\exp\left\{\sum_{\text{mesons j}}\sum_{k=1}^{\infty}2z_j^k\cos(kS_j\phi_S+kQ_j\phi_Q)\right\}
\end{split}
\end{equation}
Changing variables to $\phi_N' = \phi_N + \arg\omega$ gives
\begin{equation}
\begin{split}
\mathcal{I}_{N,S,Q} &= \frac{1}{(2\pi)^3}\int^{2\pi}_0d\phi_N\;e^{iN(\phi_N'-\arg\omega)}\int^{2\pi}_0d\phi_S\;e^{iS\phi_S}\int^{2\pi}_0d\phi_Q\;e^{iQ\phi_Q}\\[1.2ex]
&\quad\exp(2|\omega|\cos(\phi_N'))\\[1.2ex]
&\quad\exp\left\{\sum_{\text{mesons j}}\sum_{k=1}^{\infty}2z_j^k\cos(kS_j\phi_S+kQ_j\phi_Q)\right\}
\end{split}
\end{equation}
and so
\begin{equation}
\begin{split}
\mathcal{I}_{N,S,Q} &= \frac{1}{(2\pi)^3}\int^{2\pi}_0d\phi_S\;e^{iS\phi_S}\int^{2\pi}_0d\phi_Q\;e^{iQ\phi_Q}e^{-iN\arg\omega}\\[1.2ex]
&\quad\frac{1}{2\pi}\int^{2\pi}_0d\phi_N'\exp(2|\omega|\cos\phi_N')\exp(iN\phi_N')\\[1.2ex]
&\quad\exp\left\{\sum_{\text{mesons j}}\sum_{k=1}^{\infty}2z_j^k\cos(kS_j\phi_S+kQ_j\phi_Q)\right\}
\end{split}
\end{equation}
Using the integral representation of the Bessel-function (\ref{eq:partition16.2})
\begin{equation}
I_h(x) = \frac{1}{2\pi}\int^{2\pi}_0 d\phi \exp(x\cos\phi)\exp(-ih\phi)\notag
\end{equation}
yields
\begin{equation}
\begin{split}
\mathcal{I}_{N,S,Q} &= \frac{1}{(2\pi)^2}\int^{2\pi}_0d\phi_S\;\int^{2\pi}_0d\phi_Q\;e^{iS\phi_S+iQ\phi_Q-iN\arg\omega}\\[1.2ex]
&\quad\exp\left\{\sum_{\text{mesons j}}\sum_{k=1}^{\infty}2z_j^k\cos(kS_j\phi_S+kQ_j\phi_Q)\right\}I_{-N}(2|\omega|)
\end{split}
\end{equation}
Finally, using the fact that the partition function is real (in fact, by symmetry, the imaginary part integrates to zero)
\begin{equation}
\begin{split}
\mathcal{I}_{N,S,Q} &= \frac{1}{(2\pi)^2}\int^{2\pi}_0d\phi_S\;\int^{2\pi}_0d\phi_Q\;\cos(S\phi_S+Q\phi_Q-N\arg\omega)\\[1.2ex]
&\quad\exp\left\{\sum_{\text{mesons j}}\sum_{k=1}^{\infty}2z_j^k\cos(kS_j\phi_S+kQ_j\phi_Q)\right\}I_{-N}(2|\omega|)
\end{split}
\end{equation}
Now a double integration is left which can be performed numerically. The complete partition function with the conservation of five quantum numbers is again (\ref{eq:partition23})
\begin{equation}
\begin{split}
\mathcal{Z}_{N,S,Q,C,B}(\vec{X}) & \approx Z_{0}\bigg(\mathcal{I}_{N,S,Q}\;\delta_{C,0}\delta_{B,0}\\
&\quad +  \sum_{j_{c}}z^1_{j_{c}}\mathcal{I}_{N-N_{j_c},S-S_{j_c},Q-Q_{j_c}} \delta_{C,C_{j_{c}}}\delta_{B,B_{j_c}}\\
&\quad +  \sum_{j_{b} \;\;\&\atop C_{j_b} = 0}z^1_{j_{b}}\mathcal{I}_{N-N_{j_b},S-S_{j_b},Q-Q_{j_b}} \delta_{C,0}\delta_{B,B_{j_{b}}}\\
&\quad +  \sum_{j_{c}}\sum_{j_{b}\;\;\&\atop C_{j_b} = 0}z^1_{j_{c}}z^1_{j_{b}}\mathcal{I}_{N-N_{j_c}-N_{j_b},S-S_{j_c}-S_{j_b},Q-Q_{j_c}-Q_{j_b}} \delta_{C,C_{j_{c}}}\delta_{B,B_{j_{c}}+B_{j_b}}\bigg)
\end{split}
\end{equation}
A comparison of the calculation in section \ref{sec:extent} to the calculation shown in this section is given in appendix \ref{ap:compare}. Both methods agree very well. Nevertheless in this thesis the calculation explained in section \ref{sec:extent} is employed and all computations later on are done this way.

\section{The final partition function}
\label{sec:final}

Up to now only the partition function and the multiplicities within one jet are calculated. As already mentioned an $e^+e^-$ collision, emerges into two jets. Therefore the partition function of the whole system is
\begin{equation}
\widehat{\mathcal{Z}} = \mathcal{Z}(\vec{X})\mathcal{Z}(-\vec{X})
\end{equation}
Deriving this function gives the multiplicities:
\begin{equation}\label{eq:fpf1}
\begin{split}
\langle n_j\rangle = \left.\frac{\partial\ln \widehat{\mathcal{Z}}}{\partial \lambda_j}\right|_{\lambda_j=1} &= \left.\frac{1}{\widehat{\mathcal{Z}}}\frac{\partial \widehat{\mathcal{Z}}}{\partial \lambda_j}\right|_{\lambda_j=1}\\
&= \frac{1}{\widehat{\mathcal{Z}}}\left(\frac{\partial \mathcal{Z}(\vec{X})}{\partial \lambda_j}\mathcal{Z}(-\vec{X}) + \mathcal{Z}(\vec{X})\frac{\partial \mathcal{Z}(-\vec{X})}{\partial \lambda_j}\right)\\
&= \frac{1}{\widehat{\mathcal{Z}}}(z^1_j \mathcal{Z}(\vec{X}-\vec{x}_j)\mathcal{Z}(-\vec{X}) + \mathcal{Z}(\vec{X})z^1_j \mathcal{Z}(-\vec{X}-\vec{x}_j))\notag\\
&= \left\{
\begin{array}{lr}
z^1_j \left(\frac{\mathcal{Z}(\vec{X}-\vec{x}_j)}{\mathcal{Z}(\vec{X})} + \frac{\mathcal{Z}(-\vec{X}-\vec{x}_j)}{\mathcal{Z}(-\vec{X})}\right) &\hspace{2cm} \text{(for fermions)}\\
\sum_{k=1}^{\infty} kz^k_j\left(\frac{\mathcal{Z}(\vec{X}-k\vec{x}_j)}{\mathcal{Z}(\vec{X})} + \frac{\mathcal{Z}(-\vec{X}-k\vec{x}_j)}{\mathcal{Z}(-\vec{X})}\right) &\hspace{2cm} \text{(for bosons)}
\end{array}
\right.
\end{split}
\end{equation}
where the product rule and $\partial \mathcal{Z}(\vec{X})/\partial \lambda_j = z^1_j \mathcal{Z}(\vec{X}-\vec{x}_j)$ was used.\\
In the case of a neutral system, like $e^+e^-$, all initial quantum numbers are zero ($\vec{X} = \vec{0}$) and therefore the equation above becomes
\begin{equation}
\hspace{4cm}
\langle n_j(\vec{X}=\vec{0})\rangle = \left\{
\begin{array}{lr}
2z^1_j \frac{\mathcal{Z}(-\vec{x}_j)}{\mathcal{Z}(\vec{0})} &\hspace{1.9cm} \text{(for fermions)}\notag\\
2\sum^{\infty}_{k=1}kz^k_j \frac{\mathcal{Z}(-k\vec{x}_j)}{\mathcal{Z}(\vec{0})} &\hspace{1.9cm} \text{(for bosons)}
\end{array}
\right.
\end{equation}
If sharing of quark pairs between the two jets is allowed (correlated jet scheme), the vector $\vec{X}$ is not equal to the zero vector, and one has to use the general form of the equations. Furthermore in the correlated jet scheme the multiplicity of particle $j$ is the sum over the production in the different jets
\begin{equation}\label{eq:fpf2}
\langle n_j^{tot} \rangle = \sum_{q}R_q\langle n_j(\vec{X}_q) \rangle
\end{equation}
$R_q$ is the fraction of the specific jet ($\approx 17\%$ for u-type quarks and $\approx 22\%$ for d-type quarks, see equation~(\ref{eq:eweak12})). The vector of quantum numbers $\vec{X}_q$ for a $u$-jet is $\vec{X}_u=(1/3,0,2/3,0,0)$, whereas for a $\overline{b}$-jet it is $\vec{X}_{\overline{b}}=(-1/3,0,1/3,0,1)$.

\chapter{The C++ program}
\label{sec:program}

The equations discussed in the last chapter are implemented in a class based C++ program developed within this thesis in order to calculate the hadron production in particle collisions. A detailed explanation of the program code can be found in a Bachelor thesis \cite{Beutler} developed parallel to this Diploma thesis. Here, just a short overview is given.

\section{Structure of the program code}

The base of the program is a table storing all relevant particle properties i.e., mass, degeneracy, resonance width, quantum numbers etc. The table includes all available particles up to a mass of 3 GeV and all important decay channels for each of them using the most recent PDG publications \cite{Yao:2006px}. Each particle is a member of the class \textbf{SParticle}. See for instance the $\Delta^+$:
\begin{lstlisting}[frame=tb,caption=$\Delta^+$ particle as a member of the class \textbf{SParticle},label=list:2] 
case 233: {strcpy(name,"Delta +"); g=4; m=1232; 
width=120; ispin=0.5; charge=1; threshold=1073;
baryon=1; strange=0; charm=0; bottom=0;
  strcpy(channel[1],"Neutron");  pchan[1]=0.333;
  strcpy(channel[2],"Proton");   pchan[2]=0.667;
  strcpy(channel[3],"Pi +");     pchan[3]=0.333;   
  strcpy(channel[4],"Pi 0");     pchan[4]=0.667;  break;}
\end{lstlisting}
The $\Delta^+$ particle has a degeneracy of 4, a mass of 1232 MeV, a resonance width of 120 MeV etc. The four most important decay channels are the decays to neutrons, protons, $\pi^+$ and $\pi^0$. The corresponding branching ratios are listed in the array \textbf{pchan[]}. The mass spectra of all particles collected in the particle table is shown in Figure \ref{fig:mass}.\\ \\
\pgfdeclarelayer{background}
\pgfdeclarelayer{foreground}
\pgfsetlayers{background,main,foreground}
% Define block styles
\tikzstyle{decision} = [diamond, draw, fill=blue!40, 
    text width=4.5em, text badly centered, node distance=3cm, inner sep=0pt]
\tikzstyle{block} = [rectangle, draw, fill=blue!40, node distance=5cm,
    text width=8em, text centered, rounded corners, minimum height=4em]
\tikzstyle{line} = [draw, -latex']
\tikzstyle{cloud} = [draw, ellipse,fill=gray!60, node distance=6cm,
    minimum height=2em, text width=9em]
\tikzstyle{sensor}=[draw, fill=gray!60, text width=5em, 
    text centered, minimum height=2.5em]
\tikzstyle{ann} = [above, text width=5em]
\tikzstyle{naveqs} = [sensor, text width=6em, fill=gray!60, 
    minimum height=4em, rounded corners]
\def\blockdist{2.3}
\def\edgedist{2.5}

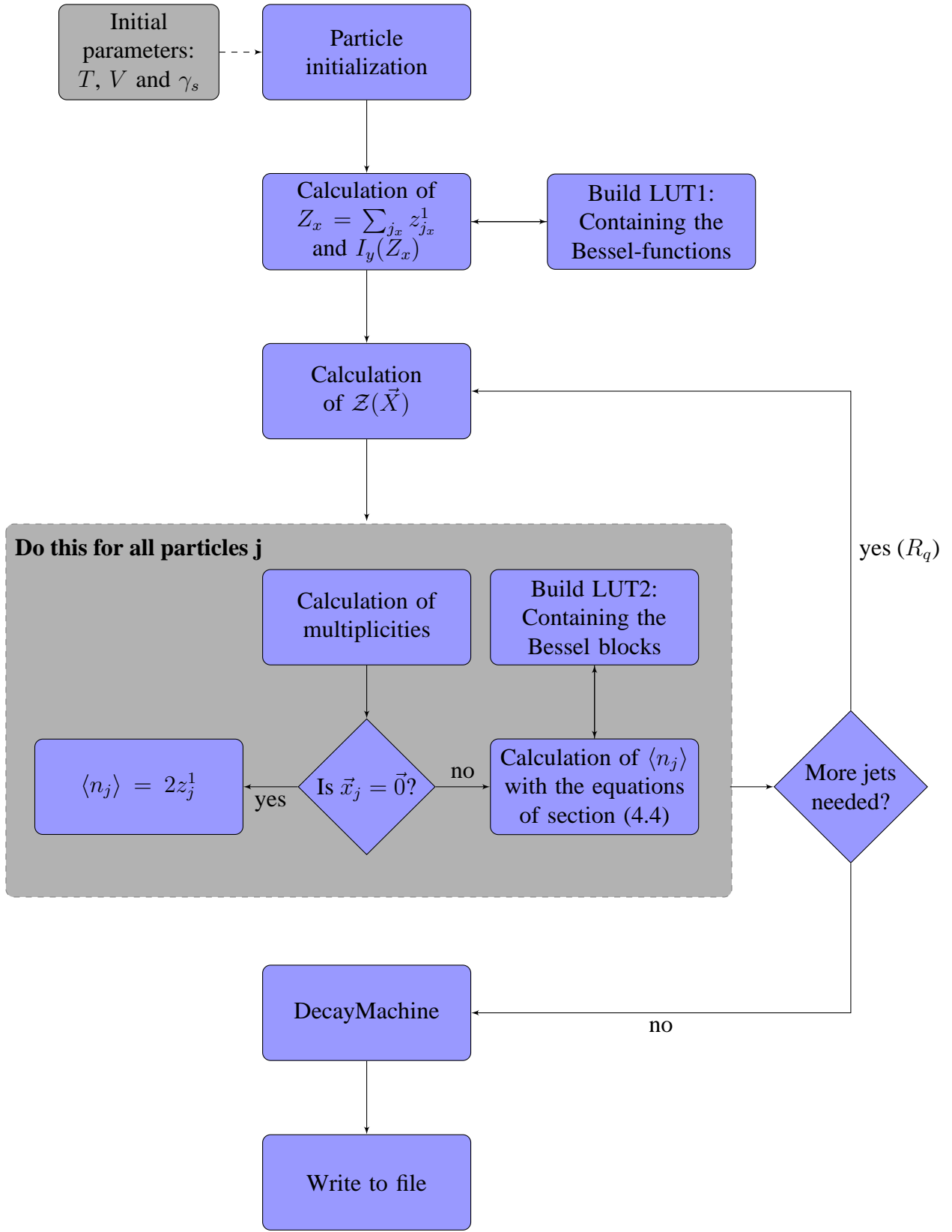
\begin{figure}[]
\begin{center}
\begin{tikzpicture}[node distance = 2cm, auto]
    % Place nodes
    \node [block, node distance = 3cm] (init) {Particle\\initialization};
    \node [naveqs, left of = init, node distance = 4cm] (parameters) {Initial parameters:\\$T$, $V$ and $\gamma_s$};
    \node [block, below of=init, node distance=3cm] (Z) {Calculation of $Z_{x} = \sum_{j_x}z^1_{j_x}$ and $I_y(Z_x)$};
    \node [block, right of=Z] (LUT1) {Build LUT1:\\Containing the Bessel-functions};
    \node [block, below of=Z, node distance=3cm] (calc1) {Calculation of $\mathcal{Z}(\vec{X})$};
    \node [block, below of=calc1, node distance=4cm] (calc2) {Calculation of multiplicities};
    \node [decision, below of=calc2, node distance=3cm] (decide) {Is $\vec{x}_j = \vec{0}$?};
    \node [block, left of=decide, node distance=4cm] (calc3) {$\langle n_j\rangle = 2z^1_j$};
    \node [ann, minimum height=2em, text width=10.3em, above of=calc3, node distance=4.2cm] (label) {\textbf{Do this for all particles j}};
    \node [block, right of=decide, node distance=4cm] (calc4) {Calculation of $\langle n_j\rangle$ with the equations of section (\ref{sec:final})};
    \node [block, above of=calc4, node distance=3cm] (LUT2) {Build LUT2:\\Containing the Bessel blocks};
    \node [decision, right of=calc4, node distance=4.5cm] (cor) {More jets needed?};
    \node [block, below of=decide, node distance=4cm] (decay) {DecayMachine};
    \node [block, below of=decay, node distance=3cm] (write) {Write to file};

    %\path [line] (calc1) -- (decide);
    \path [line] (calc2) -- (decide);
    \path [line] (init) -- (Z);
    \path [line] (Z) -- (calc1);
    %\path [line] (calc3) -| node [near start] {yes} (decide);
    %\path [line] (update) |- (calc1);
    %\path [line] (calc3) -- node {no}(calc4);
    \path [line] (decide) -- node {no}(calc4);
    \path [line] (decide) -- node {yes}(calc3);
    \path [line,dashed] (parameters) -- (init);
    \path [line] (LUT1) -- (Z);
    \path [line] (Z) -- (LUT1);
    \path [line] (LUT2) -- (calc4);
    \path [line] (calc4) -- (LUT2);
    \path [line] (cor) |- node[near start,right] {yes ($R_q$)}(calc1);
    \path [line] (cor) |- node[near end] {no}(decay);
    %\path [line] (calc3) |- (decay);
    \path [line] (decay) -- (write);
    %\path [line,dashed] (calc4) -- node {if correlated jet scheme}(calc1);
    %\path [line] (calc3) -- (decay);

    %\path [line] (calc3) -- (decay);

    \begin{pgfonlayer}{background}
        \path (calc2.north west)+(-4.5,0.8) node (a) {};
        \path (calc2.south -| decide.east)+(+5.2,-4.1) node (b) {};
        \path (calc2.north)+(0,0.7) node (c) {};
        \path (calc2.north)+(0,-5.7) node (d) {};
        \path (calc2.east)+(4.4,-3.0) node (e) {};
        \path[fill=gray!60,rounded corners, draw=black!50, dashed] (a) rectangle (b);
    \end{pgfonlayer}

    \path [line] (calc1) -- (c);
    \path [line] (e) -- (cor);

\end{tikzpicture}
\caption{Flow chart of the program used to calculate the particle multiplicities. For details see text.}
\label{fig:program}
\end{center}
\end{figure}In the following the program behavior will be explained, going through the sequences which have to be passed in a normal run of the program as presented in Figure~\ref{fig:program}. The first step is the initialization of the particles using the particle table described above. Moreover the informations stored in the particle table can be used to calculate the arguments $Z_x$ of the Bessel-functions (see Table~\ref{tabl:model1} and Table~\ref{tabl:extension}). Therefore the program considers all particles, collects their necessary properties and calculates the particle partition function of each of them via
\begin{equation}
z^k_j = \frac{g_jV}{k(2\pi\hbar)^3}\int d^3p\; e^{-\frac{\sqrt{\vec{p}^2+m_j^2}}{T}k}
\label{eq:program2}
\end{equation}
After this procedure these functions are added up for particles with the same quantum number
\begin{equation}
Z_{x} = \sum_{j_x}z^1_{j_x}
\end{equation}
Now the arguments of the Bessel-functions are known. During the calculation there are $\mathcal{O}(10^5)$ calls of the Bessel-function, but it is possible to reduce the number of calls to $\mathcal{O}(10^2)$ by using a lookup table (LUT1). The Bessel-functions are calculated by employing the routines published by "numerical recipes in C++"\cite{numerical:recipes}.\\
In the next step, the partition functions $\mathcal{Z}(\vec{X})$ and $\mathcal{Z}(\vec{X}-k\vec{x}_j)$ are calculated in a loop over all particles by using equation (\ref{eq:partition23}). The function is called \textbf{ParticleCalc()} and needs an array as argument, containing the quantum numbers $\vec{X}$ or $\vec{X}-k\vec{x}$. Now all necessary values are available to calculate the multiplicities
\begin{equation}
\langle n_j\rangle = \sum_{k=1}^{\infty}z^k_j\frac{\mathcal{Z}(\vec{X}-k\vec{x}_j)}{\mathcal{Z}(\vec{X})}
\end{equation}
where $k$ is fixed to 1 for all fermions and heavy bosons (see chapter \ref{sec:themodel}).\\ 
To sum over Bessel-functions (from now on called Bessel blocks) in eq. (\ref{eq:partition23}) like for instance
\begin{equation}\label{eq:program1}
\begin{split} &\left[\prod^{10}_{j=1}\sum^{\infty}_{n_j={-\infty}}\right]I_{n1}(2Z_{p})I_{n2}(2Z_{\Delta^{\mp}})I_{n3}(2Z_{\Delta^{++}})I_{n4}(2Z_{K^{\pm}})\\
&I_{n5}(2Z_{\Lambda})I_{n6}(2Z_{\Sigma^+})I_{n7}(2Z_{\Sigma^-})I_{n8}(2Z_{\Xi^0})I_{n9}(2Z_{\Xi^{\mp}})I_{n10}(2Z_{\Omega^{\mp}})\\
&\left[\prod^{\infty}_{k=2}\sum^{\infty}_{n_k=-\infty}\right]I_{n_k}(2Z^{k}_{\pi^{\pm}})\left[\prod^{\infty}_{h=2}\sum^{\infty}_{n_h=-\infty}\right]I_{n_h}(2Z^{h}_{K^0})\left[\prod^{\infty}_{l=2}\sum^{\infty}_{n_l=-\infty}\right]I_{n_l}(2Z^{l}_{K^{\pm}})\\
& I_{-N-n1-n2-n3-n5-n6-n7-n8-n9-n10}(2Z_n)\\
& I_{-S-n4+n5+n6+n7+2n8+2n9+3n10-\sum hn_h-\sum ln_l}(2Z_{K^0})\\
& I_{-Q-n1+n2-2n3-n4-n6+n7+n9+n10-\sum kn_k-\sum ln_l}(2Z_{\pi^{\pm}})
\end{split}
\end{equation}
is very time consuming. During one run $\mathcal{O}(1000)$ calls of this Bessel blocks are needed, but since the Bessel block depends on the quantum numbers only, it is the same for all particles with identical quantum numbers. Therefore it is very helpful to build a second lookup table (LUT2) containing the results of the previous calculations of the Bessel blocks. This table is build during the runtime. If a Bessel block is needed, it should be checked whether the block is already calculated. The results are stored in a three-dimensional array with the quantum numbers as indices. Therefore each search in the LUT can be done with $\mathcal{O}(1)$. This extension reduces the number of calls from $\mathcal{O}(1000)$ to $\mathcal{O}(10)$.\\
After the calculation of all multiplicities, the decay of the heavy particles should be taken into account. The corresponding routine in the program is called \textbf{DecayMachine()}. It uses the branching ratios provided by the particle table and calculates the contribution of the heavier particles to the lighter ones. Only stable particles do not give decay contributions. The property "stable" is a definition depending on the measured particle yields. With the used measured data \cite{Yao:2006px} all particles with a decay length $c\tau > 10$ cm are stable (see chapter \ref{sec:fit}).\\ \\
In Listings \ref{list:1} the first part of the output of the correlated calculation is shown:
\begin{lstlisting}[frame=tb,caption=program output (only the first part) of the correlated jet scheme,label=list:1] 
root@user:/$ ./therm -cor 157 32 0.8 test.txt

Calculating... T: 157 * V: 32 * gams: 0.8 ...
 *** Total nr. of particles: 474 *** 

We are in a correlated jet scheme
weak decays: 1

kpion: 5
kkaon: 3
kckaon: 3

X[0]: 0.3333
X[1]: 0
X[2]: 0.6666
X[3]: 0
X[4]: 0
Please wait and count the stars:
round: 2
*
Z(X) = 43.59470046778414
******************************
\end{lstlisting}
The routine needs some parameters, namely the scheme (-cor = correlated jet scheme), the model parameters $T$, $V$ and $\gamma_s$ and an output file for the results. The terminal output contains the number of particles (474) and the state of the weak decays (see section \ref{sec:fit}). The three variables $kpion$, $kkaon$ and $kckaon$ give the indices $k$, $h$ and $l$ of the products in eq. (\ref{eq:program1}) and will be explained in the next section.\\
Although the program runs in the correlated jet scheme, the output in Listings \ref{list:1} is restricted to the first jet, an $u$-jet, defined by the quantum numbers $N$=1/3 and $Q$=2/3. The anti-$u$ jet will be calculated in another loop with inverted quantum numbers. The result for the partition function $\mathcal{Z}(\vec{X})$ is also shown\footnote{$Z_0$ is not included in the calculation, since it will cancel out later.}.\\
The numerical implementation of the equations was a big challenge, as the possibility of a fit procedure which is crucial for a data analysis depends strongly on the runtime of the program. With the implementation of the two lookup tables and a precise analysis of the necessary terms in the calculation process (see next section) it was possible to reduce the runtime to an acceptable value. In the correlated jet scheme the program takes 13 minutes (on an Athlon64 2600MHz, L1 2x 128 KB, L2 2x 1024 KB, FSB 1066MHz) for one calculation. The other schemes are much faster.

\section{Numerical and mathematical approximations}
\label{sec:num}

\begin{figure}[]
\begin{center}
\includegraphics[width=11cm,height=8cm]{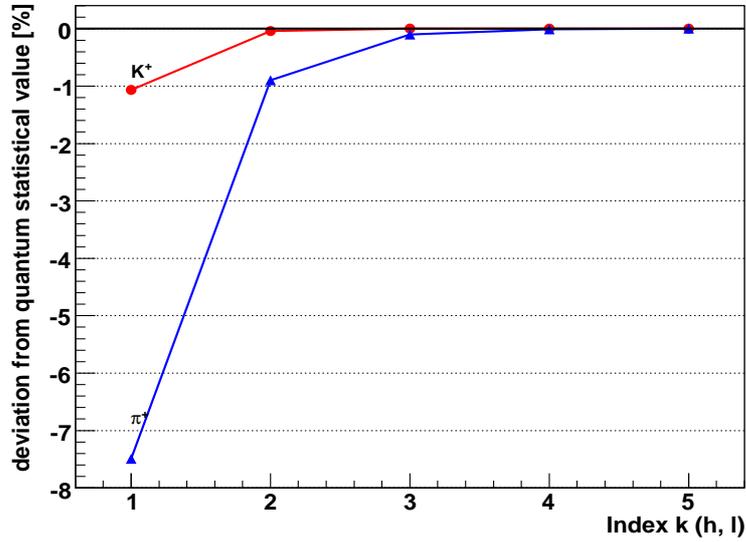}
\caption{Deviation of the initial $\pi^+$ and $K^+$ yields from the quantum statistical value with increasing index $k$ ($h$, $l$).}
\label{fig:dev}
\end{center}
\end{figure}
Equation (\ref{eq:program1}) shows some problematic points for numerical implementations, e.g. terms like
\begin{equation}
\left[\prod^{\infty}_{k=2}\sum^{\infty}_{n_k=-\infty}\right]I_{n_k}(2Z^{k}_{\pi^{\pm}})\left[\prod^{\infty}_{h=2}\sum^{\infty}_{n_h=-\infty}\right]I_{n_h}(2Z^{h}_{K^0})\left[\prod^{\infty}_{l=2}\sum^{\infty}_{n_l=-\infty}\right]I_{n_l}(2Z^{l}_{K^{\pm}})
\end{equation}
This terms ensure the correct quantum statistical calculation for light bosons. But the mathematical calculation needs upper limits for $k$, $h$ and $l$.\\
The contribution of the Bessel-function depends on the argument $Z_x$ and this argument decreases very fast with increasing $k$ ($h, l$), because the index is included in the argument of the exponential function in (\ref{eq:program2}) and is so exponentially suppressed. Therefore just a few terms labeled by the index $k$ ($h, l$) are expected to be needed, but this depends on the model parameters. This is shown in Figure \ref{fig:dev} for a certain parameter set ($T$=157 MeV, $V$=32 fm$^3$ and $\gamma_s$=0.8). It can be seen that for pions not more then five terms of the series of the logarithm (\ref{eq:partition19.2}) are needed to get the correct quantum statistical particle yield and even less terms for kaons. It is also visible that the deviation depends on the mass of the particles and decreases quite fast with increasing mass. For protons the deviation is much below $0.1\%$, what gives the reason to use Boltzmann approximation for fermions.\\
\begin{figure}[H]
\begin{center}
  \subfigure[For Kaon and Omega]{
    \label{fig:kaonomega}
    \includegraphics[width=5cm,height=9.5cm]{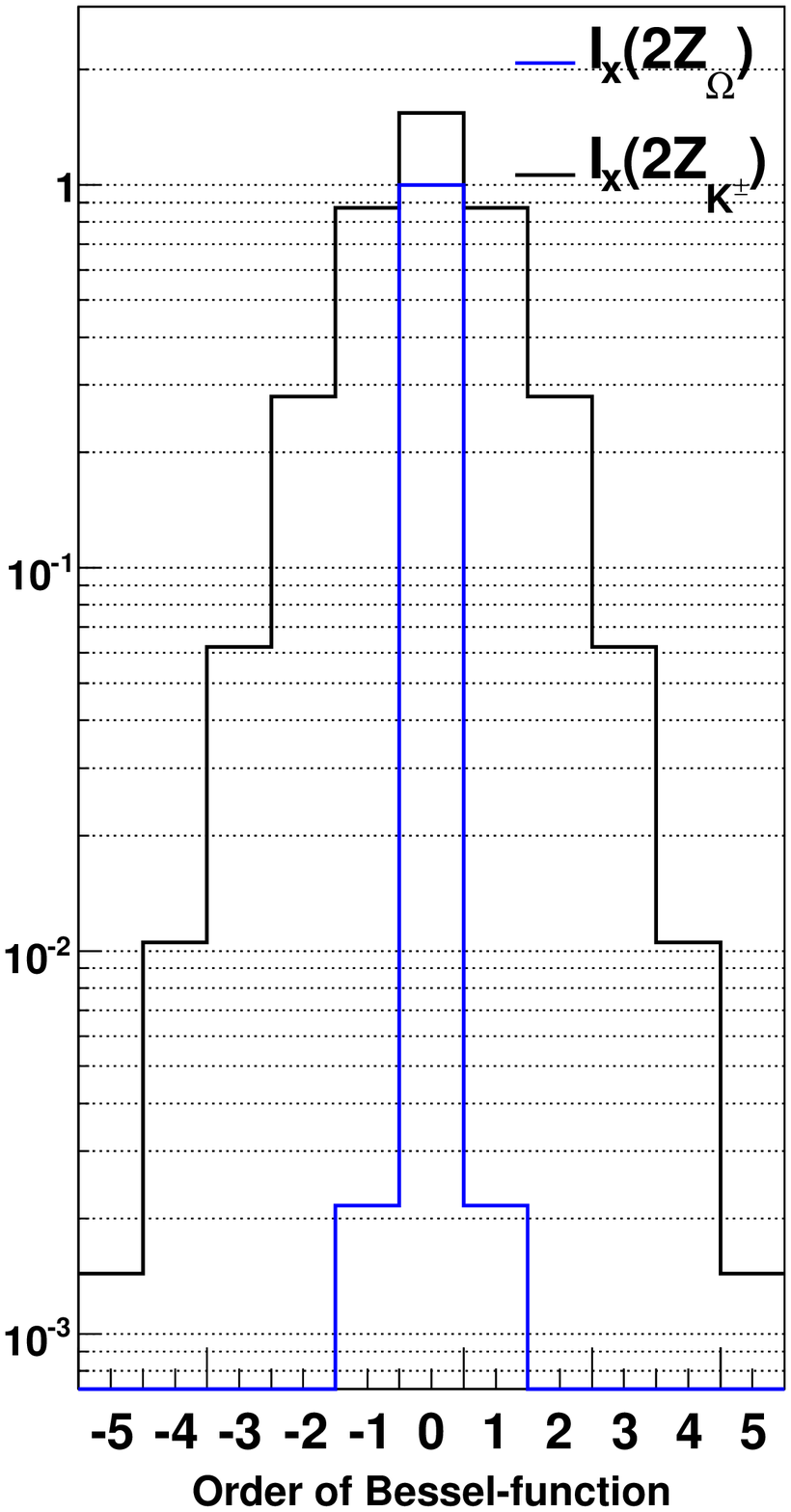} 
  }
  \subfigure[For Pion]{
    \label{fig:pion}
    \includegraphics[width=5cm,height=9.5cm]{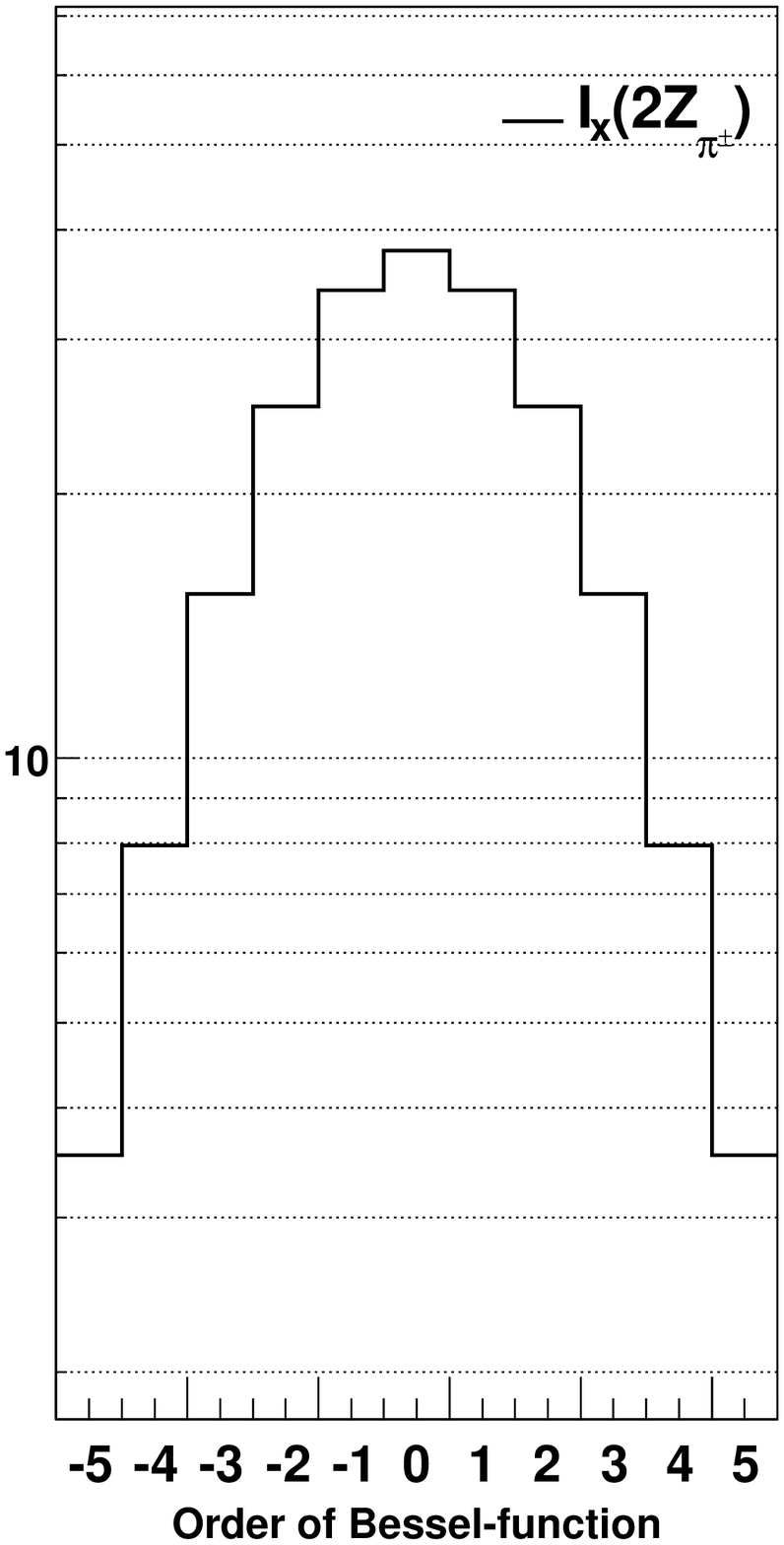} 
  }
  \caption{Bessel-function for different particle species. %(a) $I_x(Z_{K^{\pm}})$ and $I_x(Z_{\Omega})$. (b) $I_x(Z_{\pi^{\pm}})$.
}
  \label{fig:bessel}
\end{center}
\end{figure}The next point deals with the infinite sums of eq. (\ref{eq:program1}) like
\begin{equation}\label{eq:ttt}
\sum^{\infty}_{n={-\infty}}I_{n}(2Z_x)
\end{equation}
because also here a finite index $n$ is needed.\\
The biggest term of the sum is the term with index 0. This is shown in Figure \ref{fig:bessel} for different $Z_x$. With increasing or decreasing index $n$ the value decreases quite fast. All Bessel-functions $I_x(y)$, in the parameter range of this work, are not much larger than 1, which means that the smallest contribution in the product of (\ref{eq:program1}) gives the upper limit to the dimension of the whole Bessel block. This can be used to set limits for the indices of the sums in (\ref{eq:program1}).

\section{Fitting with ROOT MINUIT}

To perform a fit procedure, the program is linked with ROOT MINUIT \cite{minuit}. The MINUIT package acts on a multiparameter object function which will be minimized. This function is called the FCN function for which MINUIT defines the pure abstract base class \textbf{FCNBase} as interface. The FCN function must be implemented in a derived class from \textbf{FCNBase}. The value of the FCN function will in general depend on the parameters of the system. In our case these are the temperature $T$, the volume $V$ and $\gamma_s$. In the case of a fit procedure it is common to use a chi-square or maximum likelihood function to minimize, but in general this is a choice of the user. Here a chi-square function is employed:
\begin{equation} 
\chi^2 = \sum_j\frac{(\langle n_j^{\text{exp}}\rangle-\langle n^{\text{mod}}_j \rangle)^2}{\sigma_j^2} 
\end{equation} 
With this implementation, MINUIT is requested to minimize the FCN with respect to the parameters, that is, find those values of the coefficients which give the lowest value of $\chi^2$ \cite{minuit2}. This function will be discussed further in chapter \ref{sec:fit}.

\chapter{About the necessity for the canonical approach in $e^+e^-$ collisions}
\label{sec:about}

In chapter \ref{sec:themodel} the canonical description of the conservation laws in $e^+e^-$ collisions was used although in heavy ion collision a grand-canonical ansatz seems to be acceptable. The canonical description is valid in the whole parameter range whereas the grand-canonical formalism is the asymptotic realization of the exact canonical approach. However, the increase in complexity from the grand-canonical to the canonical ansatz is large (seen in chapter \ref{sec:themodel}) and rises the question whether it is necessary to use the canonical formalism, or whether it is allowed to perform the calculations within the much simpler grand-canonical framework.\\
%Here just the case of $e^{+}e^{-}$ collision and pp-collisions are investigated. A more general overview to the transit from canonical to grand canonical systems is given in \cite{Keranen:2001pr}.

\section{Grand-canonical conditions}
\label{sec:Basics}

The particle number fluctuation in a grand-canonical ensemble is \cite{Fliessbach}
\begin{equation}
\frac{\sigma_N}{\langle N\rangle} \sim \frac{1}{\sqrt{N}} \rightarrow 0\quad \text{ with }\quad N\rightarrow\infty
\end{equation}
Hence for large $N$ the result of a grand-canonical ensemble is equivalent to a system with constant $N$, i.e. a canonical ensemble. Therefore canonical and grand-canonical ensemble give the same results at the limit of large $N$.\\
The system treated here has the temperature $T$ and the volume $V$ as free parameters. By relating $T$ and $V$ to the number of particles $N$ it can be investigated, at which limits for $T$ and $V$ the grand-canonical treatment is valid.\\
The temperature $T$ and the volume $V$ are included in the particle partition function $z^1_j$ (particle phase space). Moreover $z^1_j$ is proportional to the multiplicity
\begin{equation}
n_j \sim z^1_j \sim\begin{cases}
  V\\
  \exp\{-1/T\}
\end{cases}
\end{equation}
As one can see higher numbers in $T$ or $V$ increase the particle partition functions and therewith the multiplicity of the particles ($N$). Therefore the following limits are equal:
\begin{equation}
\begin{split}
N &\rightarrow \infty\\
V &\rightarrow \infty\\
T &\rightarrow \infty
\end{split}
\end{equation}
This is treated in detail in the next section.

\section{The canonical factor}

\begin{figure}[]
\begin{center}
\includegraphics[width=11cm,height=8cm]{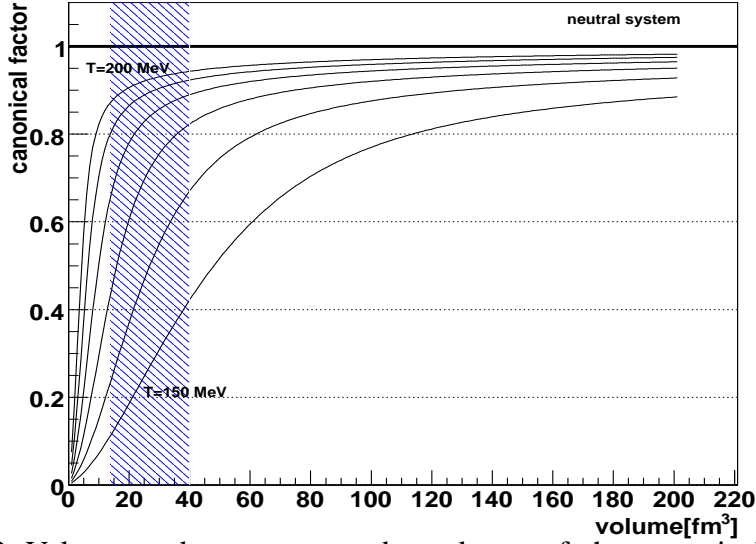}
\vspace{-0.5cm}
\caption{Volume and temperature dependence of the canonical factor for a neutron (antineutron) $\mathcal{Z}^C(1,0,0,0,0)/\mathcal{Z}^C(0,0,0,0,0)$. It starts with $T$=150 MeV and goes - in 10 MeV steps - up to 200 MeV. 
%For higher temperature and higher volume the grand canonical picture is getting valid, because the canonical factor is converging to 1. 
The blue shaded area is the volume region investigated in this thesis. 
%$\gamma_s$ is set to 0.7 in all calculations.
}
\label{fig:canonical}
\end{center}
\end{figure}

The grand-canonical approach gives the following equation for the multiplicity of particle species $j$ \cite{Andronic:2005yp}
\begin{equation}
\langle n^{GC}_j\rangle = \frac{g_jV}{(2\pi\hbar)^3}\int\limits^{\infty}_{0} d^3p\; \frac{1}{e^{\beta(\epsilon_j-\mu_j)}\pm1}
\label{eq:necessity1}
\end{equation}
with spin degeneracy $g_j$, momentum $p$, total energy $\epsilon_j$ and chemical potential $\mu_j$\footnote{$\mu_j$ is a combination of different chemical potentials, depending on the conservation laws treated. E.g. $\mu_j = \mu_BB_j - \mu_SS_j - \mu_{I_3}I^3_j$ where the quantities $B_j$, $S_j$ and $I^3_j$ are the baryon number, strangeness and three-component of the isospin quantum numbers.}. In the $e^+e^-$ system all chemical potentials are zero, and eq. (\ref{eq:necessity1}) becomes
\begin{equation}
\langle n^{GC}_j\rangle = \frac{g_jV}{(2\pi\hbar)^3}\int\limits^{\infty}_{0} d^3p\; \frac{1}{e^{\beta \epsilon_j}\pm1} \approx \frac{g_jV}{(2\pi\hbar)^3}\int\limits^{\infty}_0d^3p\; e^{-\beta \epsilon_j}
\label{eq:necessity2}
\end{equation}
where the Boltzmann approximation was used in the last step. The right hand side of this equation is identical to the canonical particle partition function of chapter \ref{sec:themodel}, equation (\ref{eq:partition8.01}).\\
The canonical approach gives for the multiplicities (see eq. (\ref{eq:partition8.22}))\footnote{For simplicity everything is treated in pure Boltzmann statistics.}
\begin{equation}
\langle n^{C}_j\rangle \approx \frac{\mathcal{Z}^C(\vec{X}-\vec{x}_j)}{\mathcal{Z}^C(\vec{X})} \frac{g_jV}{(2\pi\hbar)^3}\int\limits^{\infty}_{0} d^3p\; e^{-\beta \epsilon_j}
\label{eq:necessity3}
\end{equation}
Therefore the only difference between the grand-canonical and the canonical description is the canonical factor $\mathcal{Z}^C(\vec{X}-\vec{x}_j)/\mathcal{Z}^C(\vec{X})$ (\textbf{in neutral systems}). This "correction" factor depends on the thermal parameters of the system and the quantum numbers of the particles only (i.e. the correction for the $\Delta^+$ and the proton are the same). It is also obvious, that for neutral particles ($\vec{x}_j = 0$) the canonical factor equals unity and eq. (\ref{eq:necessity3}) is exactly the grand-canonical result.
%This factor gives the test, whether canonical or grand canonical calculation is necessary.\\
The canonical factor is a typical feature of the canonical approach due to the requirement of exact conservation of the initial set of quantum numbers. This factor suppresses or enhances the particle production according to the vicinity of their quantum numbers to the initial $\vec{X}$ vector.\\
\begin{figure}[]
\begin{center}
\hspace{-1cm}
\includegraphics[width=15cm]{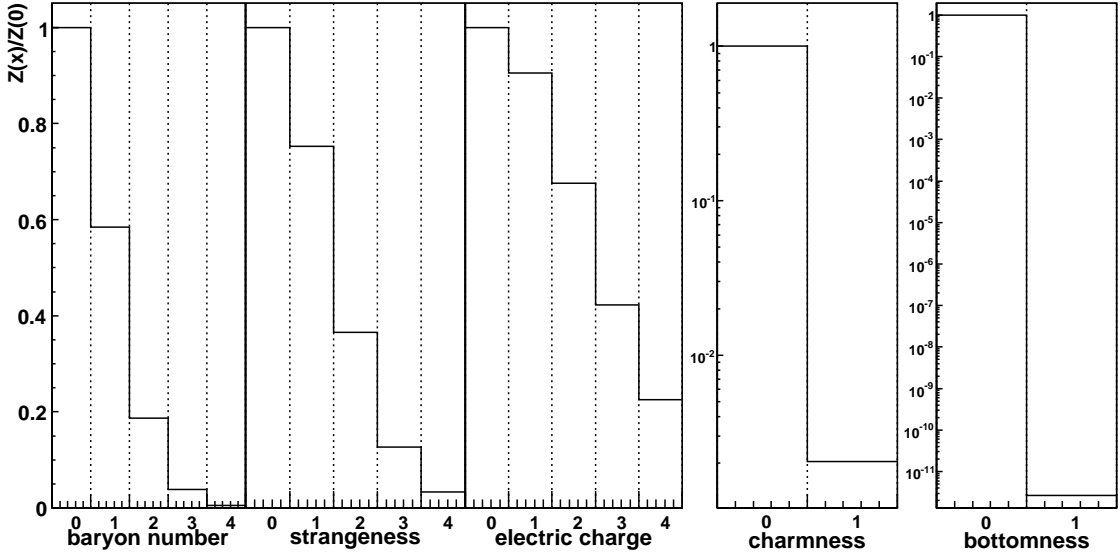}
\caption{Development of the partition function with increasing baryon number, strangeness, electric charge, charmness and bottomness.}
%The calculations are done by fixed T=154MeV, V=40$fm^3$, $\gamma_s$=0.76.}
\label{fig:Zvsall}
\end{center}\end{figure}The behavior of $\mathcal{Z}^C(\vec{X})$ as a function of the baryon number, strangeness, electric charge, charmness and bottomness for suitable T, V and $\gamma_s$ values is shown in Fig. \ref{fig:Zvsall}. For instance, it is evident that the baryo-canonical factor $\mathcal{Z}^C(N,0,0,0,0)/\mathcal{Z}^C(0,0,0,0,0)$ connected with an initially neutral system plays a major role in determining the baryon multiplicities.\\
The ultimate physical reason of charged particle ($\vec{x}_j \neq 0$) suppression with respect to neutral ones ($\vec{x}_j=0$) in a completely neutral system ($\vec{X}=0$), is that for every charged particle created, there has to be an anticharged particle, so that the conservation law is fulfilled. In a finite system this pair creation mechanism is the more unlikely the more massive is the lightest particle necessary to compensate the first particles quantum numbers. For instance, once a baryon is created, at least one anti-baryon must be generated, which is rather unlikely since its mass is much greater than the temperature and the total energy is finite. On the other hand, if a non-strange charged meson is generated, just a pion is needed to balance the total electric charge. Its creation is clearly less unlikely with respect to the creation of a baryon as the energy to be spent is lower. This argument illustrates why the dependence of $\mathcal{Z}^C(\vec{X})$ on the electric charge is much weaker than that on baryon number and strangeness (see Fig. \ref{fig:Zvsall}) \cite{Becattini:2008tx}.\\ 
These canonical suppression effects do not occur in a grand-canonical framework. Equation (\ref{eq:necessity2}) shows that in a completely neutral system, all chemical potentials are zero and consequently charged particles do not undergo any suppression with respect to neutral ones.\\
With increasing temperature the probability of the production of additional particles to balance the net quantum number is increasing, too. Thus the canonical factor should converge to unity. This is shown in Figure~\ref{fig:canonical}. One can also see the conversion against 1 with increasing volume. This is already mentioned in section~\ref{sec:Basics}, with a slightly different argument, but both explanations have the same origin, the increasing number of particles.
Therefore the general behavior of the canonical factor at the limits found in section \ref{sec:Basics} is
\begin{align}
\lim_{\substack{V \to \infty \\ T \to \infty}}\left(\frac{\mathcal{Z}^C(\vec{X}-\vec{x}_j)}{\mathcal{Z}^C(\vec{X})}\right) &= e^{N_j\mu_N/T}e^{S_j\mu_S/T}e^{Q_j\mu_Q/T}e^{C_j\mu_C/T}e^{B_j\mu_B/T}\\
&= 1\tag{in $e^+e^-$collisions}\notag
\end{align}
and the system transits from the canonical to the grand-canonical ensemble.\\
Figure \ref{fig:canonical} shows an example for a baryon without strangeness or electric charge (neutron). It can be seen that the canonical factor contributes a non-negligible effect to the particle yield at the volume and temperature region discussed in this thesis (blue shaded area, see chapter \ref{sec:results}). This justifies the canonical treatment in chapter \ref{sec:themodel}.\\
A detailed look at special particles in a neutral system, shows that there is no difference between the suppression of a particle and the corresponding antiparticle. The production of a baryon needs the additional production of an antibaryon, and the production of an antibaryon needs a baryon, so one finds
\begin{equation}
\mathcal{Z}^C(N,S,Q,C,B) = \mathcal{Z}^C(-N,-S,-Q,-C,-B)
\end{equation}
Therefore we have the same yields for particles and antiparticles in a neutral system.
\begin{figure}[]
\begin{center}
\includegraphics[width=11cm,height=8cm]{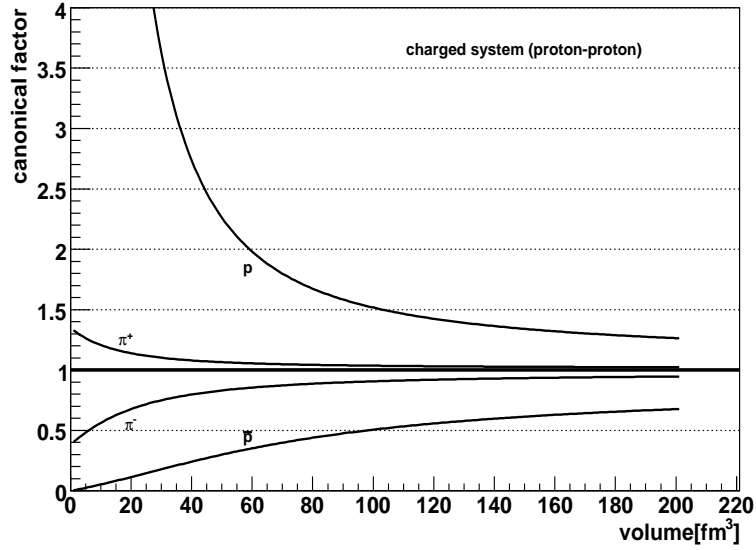}
\vspace{-0.4cm}
\caption{Canonical factor of $p, \overline{p}$, $\pi^+$ and $\pi^-$ in a proton-proton system. 
%The canonical factor of the protons is much larger than for the pions, because for pions only the charge conservation is working whereas for the protons the charge and the baryon number conservation contributes. 
%The temperature T is set to 160MeV and $\gamma_s$ = 0.7 in all calculations.
}
\label{fig:canonical-factor}
\end{center}\vspace{-0.7cm}
\end{figure}

\subsection{Charged systems}

\begin{figure}[]
\begin{center}
\includegraphics[width=11cm,height=8cm]{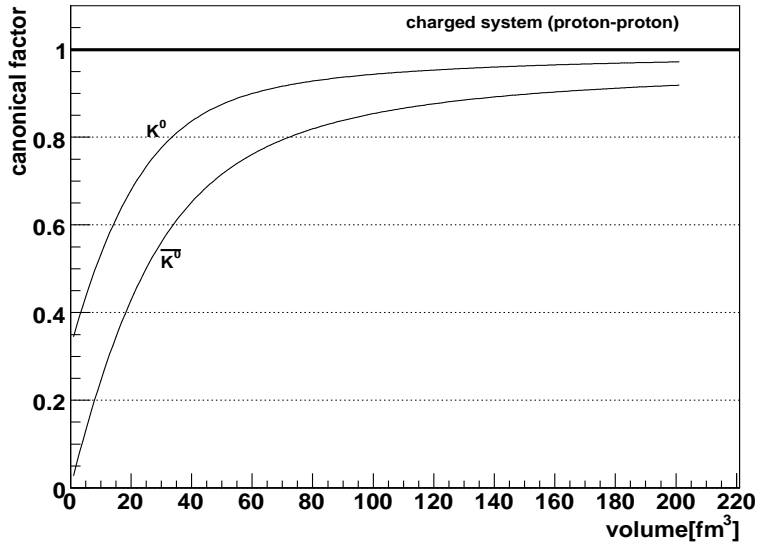}
\vspace{-0.2cm}
\caption{Canonical factors for $\text{K}^0$ and $\overline{\text{K}}{}^0$ in the proton-proton system. The $\overline{\text{K}}{}^0$ is more suppressed than the $\text{K}^0$, as it is possible to use baryons to balance the strangeness for the $\text{K}^0 $ (see text).%, and baryons are enhance in a pp system (T = 160MeV and $\gamma_s$ = 0.7).
}
\label{fig:kaons}
\vspace{-0.2cm}
\end{center}\end{figure}

In a "charged" system like for example the collision of two protons with the initial quantum number vector $\vec{X} = (2,0,2,0,0)$, it is not possible to use the direct analogy between canonical and grand-canonical systems as done in the neutral case in eqs. (\ref{eq:necessity2}) and (\ref{eq:necessity3}), because the chemical potentials are not equal to zero anymore. Nevertheless the suppression effect of the canonical factor can be investigated. But this effect changed now dramatically for some particles. Figure~\ref{fig:canonical-factor} shows an enhancement for protons, instead of a suppression. The explanation is the following: The canonical factors are still intended to conserve the initial quantum numbers. To conserve a baryon number of two, more baryons and less antibaryons are needed. The same effect can be seen at the pion yields. The $\pi^+$ is enhanced, due to the positive electric charge in the initial system. But pions are produced easily, therefore the effect is much smaller than for protons.\\
It is also interesting to look at the canonical factors of $\text{K}^0$ and $\overline{\text{K}}{}^0$ in a proton-proton system. The initial system contains no strangeness, therefore one would expect for both kaon canonical factors the same behavior as in a neutral system. The canonical factors are plotted in Fig. \ref{fig:kaons}. There is a suppression for both - kaon and antikaon - but the suppression for the antikaon is larger. This can be explained by looking at the details of quantum number conservation. To balance a quantum number, it is convenient to look for the lightest particle which can do the job, because this particle has the biggest probability. But in a pp system it has to be taken into account, that baryons are enhanced and anti-baryons are suppressed. To balance the strangeness S = 1 arising by the production of a $\text{K}^0$ it is still the easiest way to produce a $\overline{\text{K}}{}^0$. But its also possible to produce for instance a $\Lambda$. This $\Lambda$ is normally strongly suppressed because of the high mass, but now it delivers the baryon number N = 1 and is therefore enhanced. In order to balance the $\overline{\text{K}}{}^0$, the system has to produce an $\overline{\Lambda}$ and antibaryons are strongly suppressed. This causes the difference in the $\text{K}^0$ and $\overline{\text{K}}{}^0$ canonical factors.

\section{Heavy particles in the canonical ensemble}
\label{sec:heavy2}

All particles containing charm or bottom quarks are heavy and therefore suppressed by the Boltzmann factor. Thus the thermal production of heavy particles is very small\footnote{The particle partition functions of the $D^0$ and $B^0$ are $\mathcal{O}(10^{-4})$ and $\mathcal{O}(10^{-13})$, respectively.}. In the canonical ensemble the canonical factor affects the particle production and depends on the mass of the lightest particle which has to be produced to balance the quantum numbers. In the case of charm particles, the system has to produce another charm particle and the lightest possibility is the $D^0$ with a mass of 1864 MeV. Therefore the canonical suppression is expected to be very high.
\begin{figure}[]
\begin{center}
\includegraphics[width=11cm,height=8cm]{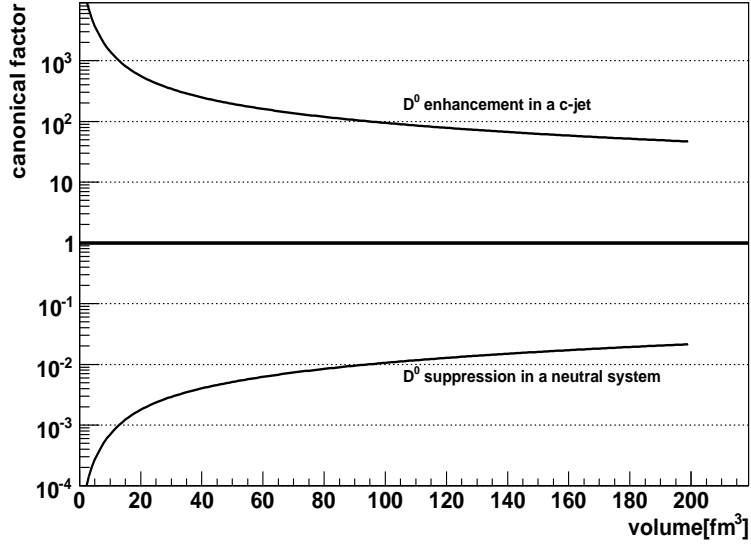}
\vspace{-0.2cm}
\caption{The canonical factor for $\text{D}^0$ in a neutral system and in a c-jet.}
\label{fig:enhancedcharm}
\vspace{-0.2cm}
\end{center}\end{figure}This is shown in Figure \ref{fig:enhancedcharm}.\\
In a neutral system the canonical factor for a $D^0$ is 
\begin{equation}
\frac{\mathcal{Z}^C(\vec{X}-\vec{x}_{D^0})}{\mathcal{Z}^C(\vec{X})} = \frac{\mathcal{Z}^C(0-C_{D^0})}{\mathcal{Z}^C(0)} \approx \mathcal{O}(0.01)-\mathcal{O}(0.001)
\end{equation}
The exact value depends on the parameters (the given range is close to the parameters in $e^+e^-$ collisions), but in any case there is a strong suppression for charm particles in a neutral system by the canonical factor, in addition to the suppression of the Boltzmann factor caused by the high mass.\\
However, as already discussed in the last section, the canonical factor can also enhance a particle yield if this is necessary for quantum number conservation. In a system with initial charmness ($C$=1), like a $c$-jet in an $e^+e^-$ collision, the situation is \textbf{exactly} inverse to what is shown above
\begin{equation}
\frac{\mathcal{Z}^C(\vec{X}-\vec{x}_{D^0})}{\mathcal{Z}^C(\vec{X})} = \frac{\mathcal{Z}^C(C-C_{D^0})}{\mathcal{Z}^C(C)} \approx \mathcal{O}(100)-\mathcal{O}(1000)
\end{equation}
In a charmed system all charm particles are strongly enhanced because they are needed to fulfill the initial charmness. This is also shown in Figure \ref{fig:enhancedcharm} and obviously the canonical factor plays a dominant role in the production of heavy particles\footnote{In the case of bottom particles the canonical factor in a system with bottomness ($B$=1) is $\approx \mathcal{O}(10^{10})$, because of the heavy mass of the bottom particles.}.

\section{Other quantities}
\label{sec:other}

The grand-canonical partition function is\footnote{The general case $\mu\neq0$ is treated here.}
\begin{equation}
\ln \mathcal{Z}^{GC} = \sum_j \frac{g_jV}{(2\pi)^3} \int d^3p\; e^{-\beta(\epsilon_j-\mu_j)}
\end{equation}
where the Boltzmann approximation is applied. By using eqs. (\ref{eq:stat5}), (\ref{eq:stat4}) and (\ref{eq:stat3}) it is possible to calculate the average energy, entropy and pressure of the system
\begin{align}
E^{GC} &= -\frac{\partial \ln \mathcal{Z}^{GC}}{\partial \beta}\Bigg|_{\mu=0} = \sum_j \frac{g_jV}{(2\pi)^3}\int d^3p\;\epsilon_je^{-\beta(\epsilon_j-\mu)} = \sum_j E_j^{GC}\\[2ex]
S^{GC} &= \frac{\partial (T\ln \mathcal{Z}^{GC})}{\partial T} = \ln \mathcal{Z}^{GC} + \frac{1}{T}\sum_j E_j^{GC}\\[2ex]
P^{GC} &= \frac{\partial(T\ln \mathcal{Z}^{GC})}{\partial V} = T\sum_j \frac{g_j}{(2\pi)^3}\int d^3p\;e^{-\beta(\epsilon_j-\mu)} = \sum_j P_j^{GC}
\label{eq:other1}
\end{align}
The canonical partition function is (eq. (\ref{eq:partition19.3}) for simplicity in pure Maxwell-Boltzmann statistics)
\begin{equation}
\mathcal{Z}^C(\vec{X}) = \frac{Z_{0}}{(2\pi)^5}\int d^5\vec{\phi}\; e^{i\vec{X}\vec{\phi}} \exp\left\{\sum_j z^1_je^{-i\vec{x}_j\vec{\phi}}\right\}\notag
\end{equation}
with the particle partition function
\begin{equation}
z^1_j = \frac{g_j V}{(2\pi\hbar)^3} \int d^3p\; e^{-\beta \epsilon_j}\notag
\end{equation}
In the canonical case, the energy, entropy and pressure are
\begin{align}
E^{C} &= -\frac{\partial\ln \mathcal{Z}^C}{\partial\beta} =  \sum_j \frac{\mathcal{Z}^C(\vec{X} - \vec{x}_j)}{\mathcal{Z}^C(\vec{X})}E_j^{GC}\big|_{\mu_j = 0}\\[2ex]
S^{C} &= \frac{\partial (T\ln \mathcal{Z}^C)}{\partial T} = \ln \mathcal{Z}^{C} + \sum_j\frac{\mathcal{Z}^C(\vec{X} - \vec{x}_j)}{\mathcal{Z}^C(\vec{X})} \frac{E_j^{GC}\big|_{\mu_j = 0}}{T}\\[2ex]
P^{C} &= \frac{\partial(T\ln \mathcal{Z}^C)}{\partial V} = \sum_j\frac{\mathcal{Z}^C(\vec{X} - \vec{x}_j)}{\mathcal{Z}^C(\vec{X})}P_j^{GC}\big|_{\mu_j = 0}
\end{align}
At each quantity the canonical factor supplements the grand-canonical quantity, very similar to the situation for the multiplicities where 
\begin{equation}
\langle n^C_j\rangle = \frac{\partial \ln \mathcal{Z}^C}{\partial \lambda_j}\Bigg|_{\lambda_j=1} = \frac{\mathcal{Z}^C(\vec{X} - \vec{x}_j)}{\mathcal{Z}^C(\vec{X})} \langle n^{GC}_j \rangle\big|_{\mu_j=0}
\end{equation}
was found (see chapter~\ref{sec:themodel}).

\chapter{Measurement properties and the fit procedure}
\label{sec:fit}

In this chapter the fit procedure to describe the experimental particle spectra is presented. The contribution of strong and weak decays is discussed as well as the strangeness under-saturation parameter and the influence of the particle width.

\section{Particle width, strangeness under-saturation and the treatment of strong decays}

Within the calculation of the particle partition function, the particle width has to be taken into account. This is important, because the number of light particles coming from the decay of resonances is increased by the finite resonance width\footnote{The effect is decreasing with increasing temperature.}. The finite widths of resonances are taken into account by an additional integration over the particle mass in the multiplicity calculation, with a Breit-Wigner distribution as a weight~\cite{Andronic:2005yp}
\begin{equation}
z^1_j = \frac{g_jV}{(2\pi\hbar)^3}\frac{1}{N_{BW}}\int\limits^{\infty}_{M_0}dm\frac{\Gamma^2_j}{(m-m_j)^2+\Gamma^2/4}\;\int d^3p\;\gamma_s^{n_{s_j}}e^{-\frac{\sqrt{\vec{p}^2+m_j^2}}{T}}
\end{equation}
where $m_j$ is the nominal mass and $\Gamma_j$ is the width of particle $j$. The energy is then calculated for every value of $m$ in the integration step. Here, $N_{BW}$ is the normalization of the Breit-Wigner distribution such that the integral over the Breit-Wigner factor gives 1. $M_0$ is the threshold defined by the masses of the particles in the dominant decay channels. For instance, at the $\Delta^{++}$, there should be 0 strength (Breit-Wigner amplitude) below a threshold defined by the sum of the masses of the dominant decay channels (proton and pion).\\
Following the approach of ref.\cite{Becattini:1995if}, an additional parameter $\gamma_s$ is introduced into the particle partition function to account for a possible deviation of strange particle yields from their chemical equilibrium values (see section~\ref{subsec:strange}). If a hadron contains $n_s$ strange valence quarks, its production is reduced by a factor $\gamma_s^{n_s}$\footnote{Please take into account that for the quantum statistical calculation with terms $k > 1$, $\gamma_s^{kn_s}$ has to be used.}. This parameter is also applied to neutral mesons such as $\eta, \eta', \phi, \omega, f_2(1270)$ and $f_2'(1525)$ according to the fraction of $s\overline{s}$ content in the meson itself. The relevant fraction is determined by using mixing formulas quoted in \cite{Yao:2006px}.\\
The multiplicity calculation basically proceeds in two steps. First, a primary hadron yield, $\langle n_h^{th}\rangle$ is calculated using eqs. (\ref{eq:partition8.22}) and (\ref{eq:partition23}). As a second step all resonances in the hadron gas which are unstable against strong decays are allowed to decay into lighter stable hadrons, using appropriate branching ratios (Br) and multiplicities (M) for the decay $j \rightarrow h$ published by the PDG \cite{Yao:2006px}. The abundances in the final state are thus determined by
\begin{equation}
\langle n^{\text{mod}}_h\rangle = \langle n^{th}_h\rangle + \sum_j \langle n^{\text{mod}}_j\rangle \text{Br}(j \rightarrow h)M(j\rightarrow h)
\end{equation}
where the sum runs over all particles contained in the hadron gas and $h$ refers to a hadron. For many particles the decays give significant contributions to the final yields. For example the measured $\pi^0$ yield consists of about $25\%$ initially produced $\pi^0$, whilst the rest is from (mostly strong) decays (for $e^+e^-$ collisions at 91 GeV).\\
A stable particle is a particle which does not give any decay contributions because it lives long enough to be detected itself, or at least forms a secondary vertex which can be separated from the primary vertex, such that the original particle can be reconstructed. The proton, neutron and the three pion states are treated as stable particles. The weak decaying strange particles are discussed in the next section.

\section{Weak decays}
\label{sec:weak}

Particles which decay by the strong or electromagnetic interaction preserve the strangeness quantum number. The decay process for e.g. the $\Lambda$ particle must violate this rule, since there is no lighter baryon which contains a strange quark, so the strange quark must be transformed into another quark in the decay process. That can only happen via the weak interaction where strangeness is not conserved, and this leads to a much longer lifetime.
\begin{equation}
\left.%
\begin{array}{ccccccc}
&uds &\rightarrow& uud &+& d\bar{u}\notag\\
&\Lambda &\rightarrow& p &+& \pi^-\\ \\
&uds &\rightarrow& udd &+& \frac{u\bar{u}+d\overline{d}}{\sqrt{2}}\notag\\
&\Lambda &\rightarrow& n &+& \pi^0\\
\hline
\text{Strangeness:}&-1 &\neq& 0 &+& 0
\end{array}
\right\} \textrm{dominant $\Lambda$ decays}
\end{equation}
The long lifetime of the $\Lambda$ rises the question whether there are $\Lambda$s or its decay products that are seen in the detector. The PDG data used in this analysis include decay products from all resonances with decay lengths $c\tau < 10$ cm \cite{Yao:2006px}. Therefore approximately 30\% of the $\Lambda$s will survive and give no decay contributions. There is a similar behavior in the case of $K^{+}$, $K_s^0$, $\Xi^0$, $\Xi^-$, $\Sigma^-$, $\Sigma^+$ and $\Omega^-$ (and the corresponding antihyperons). This fact is accounted for in the present calculations. The exact values implemented in the model are shown in Table \ref{tabl:weak}. The importance of the effect can be seen at the $K^{+}$, which is an almost stable particle (in the detector sense of stable), because of the long lifetime, whereas almost 100\% of the $\Sigma^+$ decay.
\begin{table}[]
\begin{center}
\begin{tabular}{cllclclclc}
  \hline
  \hline
  \multicolumn{2}{c}{\textbf{particle}} & \multicolumn{2}{c}{\textbf{10 GeV}} & \multicolumn{2}{c}{\textbf{29-35 GeV}} & \multicolumn{2}{c}{\textbf{91 GeV}} & \multicolumn{2}{c}{\textbf{130-200 GeV}} \\
  \hline
  &$K^{+}$ & $\approx3\%$* & --- & 2.86\% & \cite{Braunschweig:1988hv} & 2.66\% & \cite{Akers:1994ez} & $\approx 0 \%$* & --- \\
  &$K_s^0$ & 97.92\% & \cite{Althoff:1984iz} & 97.87\% & \cite{Althoff:1984iz} & 98.19\% & \cite{Abreu:1994rg} & $\approx 100\%$* & ---\\
  &$\Lambda$ & 73.19\% & \cite{Althoff:1984iz} & 73.74\% & \cite{Braunschweig:1988wh} & 72.44\% & \cite{Alexander:1996qj} & $\approx 70 \%$* & ---\\
  &$\Sigma^+$ & $\approx100\%$* & --- & $\approx100\%$* & --- & 98.53\% & \cite{Alexander:1996qi} & $\approx 100 \%$* & ---\\
  &$\Sigma^-$ & $\approx90\%$* & --- & $\approx90\%$* & --- & 89.63\% & \cite{Alexander:1996qi} & $\approx 90 \%$* & ---\\
  &$\Xi^-$ & $\approx90\%$* & --- & 88.06\% & \cite{Braunschweig:1988wh} & 87.34\% & \cite{Alexander:1996qj} & $\approx 90 \%$* & ---\\
  &$\Xi^0$ & $\approx70\%$* & --- & $\approx70\%$* & --- & $\approx70\%$* & --- & $\approx 70 \%$* & ---\\
  &$\Omega^-$ & $\approx100\%$* & --- & $\approx100\%$* & --- & $\approx100\%$* & --- & $\approx 100 \%$* & ---\\
  \hline
  \hline
\end{tabular}
\caption{Decay rates of the weak decaying particles (within $c\tau$=10 cm). The momentum spectra were extracted from the given references. The values with * are estimates, because no momentum spectrum was available. The lifetimes are taken from the PDG tables \cite{Yao:2006px}.}
\label{tabl:weak}
\end{center}
\end{table}

\section{The effect of uncertainties in the particle table}

\begin{table}[]
\begin{center}
\begin{tabular}{c|ccc}
\hline
\hline
& \multicolumn{3}{|c}{\textbf{mass cut}}\\
\textbf{particle} & \textbf{1.7 GeV} & \textbf{2 GeV} & \textbf{3 GeV}\\
\hline
$\pi^{\pm}$ & 17.508  & 17.864 & 18.077\\
$\pi^0$ & 10.134 & 10.346 & 10.470\\
K$^{\pm}$ & 1.982 & 2.019 & 2.039\\
K$^0$ & 1.913 & 1.943 & 1.962\\
$\eta$ & 1.063 & 1.103 & 1.110\\
p & 0.809 & 0.873 & 0.895\\
$\Lambda$ & 0.326 & 0.348 & 0.354\\
$\Sigma^{+}$ & 0.0828 & 0.0885 & 0.0901\\
$\Xi^-$ & 0.0170 & 0.0197 & 0.0201\\
$\Omega$ & 0.00124 & 0.00124 & 0.00124\\
\hline
\hline
\end{tabular}
\caption{Particle yields at $\sqrt{s}$=91 GeV for different mass cuts. Please take into account that above 2 GeV the mass spectra is incomplete and the higher mass cut includes only the known states of~\cite{Yao:2006px}.}
\label{tabl:masscut}
\end{center}
\end{table}The particle table is surely one of the biggest uncertainties of the model, because
\begin{itemize}
 \item Above 2 GeV the particle spectra is not fully measured. It is assumed that it increases exponentially \cite{Broniowski:2000bj,Broniowski:2000hd,Broniowski:2004yh}.
 \item The decay channels, especially for heavy particles, are not well known.
\end{itemize}
Fortunately the importance of the particles (for our results), decreases with increasing mass because the Boltzmann factor suppresses all heavy particles. Therefore it is acceptable to use an upper cut of the spectra with the assumption that all particles above this cut can be neglected. Nevertheless, in the correlated scheme, large enhancements for heavy particles can occur. Such results are connected to large uncertainties.\\
To investigate the effect of an upper mass cut, the calculation was performed with different mass cut values. The results are shown in Table~\ref{tabl:masscut}. The effect seems to be small, but it is of the order of the experimental errors.\\
\begin{figure}[]
\begin{center}
\includegraphics[width=11cm,height=8cm]{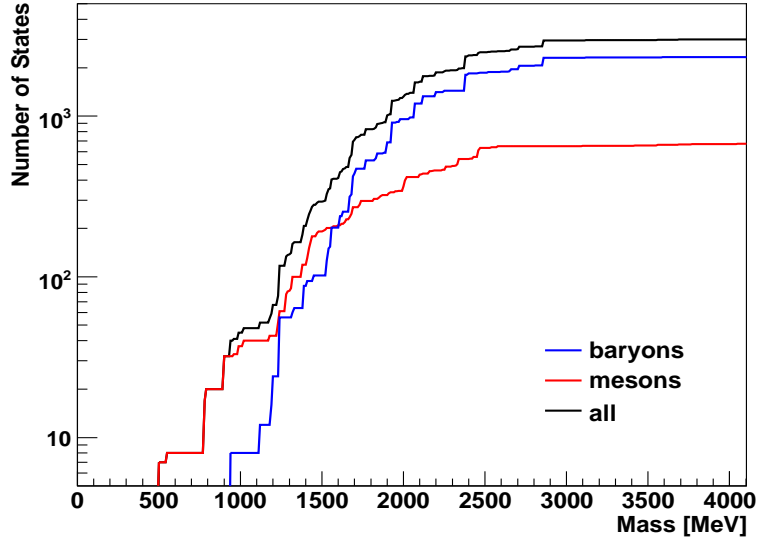}
\caption{Mass spectra of the used mesons (red) and baryons (blue).}
\label{fig:mass}
\end{center}
\end{figure}The set of hadron species used in this work was updated compared to~\cite{Andronic:2005yp} with the most recent information available from the PDG~\cite{Yao:2006px} and includes all resonances listed there. Concerning the decays, it was ensured that the decay tables are symmetric for particles and antiparticles and the decay widths always add up to the total width, even if particular channels are not measured. As the branching ratios are not well known for some of the high-mass states, the known Br of the nearest state with the same quantum numbers are used. The particle spectra is shown in Figure~\ref{fig:mass} and it is clearly visible that the expected exponential increase does not hold for masses above 2 GeV. This is caused by the incompleteness of the measurements.

\section{The fit function}

After calculating the particle yields using the equations of chapter \ref{sec:themodel} and taking into account all the influences discussed in this chapter, the particle yields can be compared to the measurements. All fit procedures are carried out by minimizing (least $\chi^2$-fit)
\begin{equation} 
\chi^2 = \sum_j\frac{(\langle n_j^{\text{exp}}\rangle-\langle n^{\text{mod}}_j \rangle)^2}{\sigma_j^2} 
\end{equation} 
as a function of the three parameters $T$, $V$ and $\gamma_s$, taking account of the experimental uncertainties $\sigma_j$. $\sigma_j$ contains only the errors of \cite{Yao:2006px}. There is no method applied to include the uncertainties in mass, width and branching ratios as additional systematic errors like done in \cite{Becattini:1995if,Becattini:1996gy,Becattini:1997uf}.\\
The value $\langle n_j^{\text{mod}}\rangle$ depends on the used scheme (correlated or uncorrelated) as explained in section \ref{sec:final}. The schemes used in this analysis are discussed in the next chapter.

\chapter{Results and discussions}
\label{sec:results}

In this chapter the results of the data analysis of $e^+e^-$ collisions at $\sqrt{s}$=10 GeV, 29-35 GeV, 91 GeV and 130-200 GeV are discussed. The model is based on the canonical partition function (in quantum statistics) with the conservation of five quantum numbers as derived in section \ref{sec:extent}. The fits include all light particles measured as given in appendix \ref{ap:data}. Heavy particles will be treated separately in section \ref{sec:heavy}.

\section{The model}

The analysis is focusing on thermal production and is neglecting the issue of hard collisions. Therefore an uncorrelated jet scheme is used, i.e. each jet is treated as a fireball with vanishing quantum numbers as fixed by the entrance channel. It is clear at this point that hadrons from jets with heavy quarks ($c$ and $b$) will be greatly underestimated by the model because of the large Boltzmann suppression factors. In this approach the issue of equilibration is effectively addressed only for hadrons with light quarks ($u$, $d$, $s$).\\
It is important to recognize that the measured yields of these hadrons contain the contribution from the e$^+$e$^-$ annihilation events into $c\bar{c}$ and $b\bar{b}$. Heavy-quark production is indeed significant and is very precisely measured, in particular at the $Z^0$ mass ($\sqrt{s}$=91.2 GeV), where the measurements are well described by the Standard Model \cite{Yao:2006px}. Hence, heavy-quark production is manifestly non-thermal in origin. Therefore two scenarios were considered:
\begin{enumerate}[i.]
 \item Fit to the data as measured.
 \item Subtraction of the contribution originating from charm and bottom decays from the yields of hadrons carrying light quarks based on available data for charm and bottom hadron production (and their branching ratios) at 91 GeV.
\end{enumerate}
This approach differs from that used by Becattini et al. (see \cite{Becattini:2008tx} for the most recent results). There, e$^+$e$^-\rightarrow c\bar{c}$ and e$^+$e$^-\rightarrow b\bar{b}$ events are treated in a correlated jet scheme (a 2-jet initial state which carries quantum numbers $C$=$\pm$1 or $B$=$\pm$1). The fractions of hadronic events with $b$ or $c$ quarks, as well as those with lighter quarks, are thus external input values, unrelated with the Thermal Model (see Table II in ref.~\cite{Becattini:2008tx}). For the light quark sector, the difference between calculations with uncorrelated and correlated quark-antiquark schemes are investigated later in this chapter.\\
The second scenario, namely subtracting the contribution originating from charm and bottom decays, is investigated only at $\sqrt{s}$=91 GeV as a case study, as the measurements needed to allow the subtraction are complete only at this energy.

\section{Comparison to other calculations}

\begin{table}[]
\begin{center}
\begin{tabular}{|c|c|c|c|c|c|c|}
\hline
& \multicolumn{4}{|c|}{this work} & \multicolumn{2}{|c|}{THERMUS}\\
\cline{2-7}
particle & \multicolumn{2}{|c|}{final} & \multicolumn{2}{|c|}{initial} & final 
& initial\\
\cline{2-5}
 & QS & BS & QS & BS & & \\ \hline
$\pi^{+}$ & 16.50 & 16.00 & 5.37 & 4.97 & 14.67 & 4.96 \\
$\pi^0$  & 9.73 & 9.39 & 3.10 & 2.80 & 8.50 & 2.80 \\
K$^+$ & 2.039 & 2.02 & 0.988 & 0.978 & 1.94 & 0.980 \\
K$^0$ & 1.963 & 1.949 & 0.971 & 0.961 & 1.889 & 0.963 \\
$\eta$ & 1.11 & 1.10 & 0.486 & 0.482 & 0.909 & 0.487 \\
$\rho^0$(770) & 1.133 & 1.13 & 0.774 & 0.771 & 1.06 & 0.774 \\
K$^{*0}$(892) & 0.594 & 0.591 & 0.456 & 0.454 & 0.583 & 0.454 \\
p & 0.689 & 0.683 & 0.235 & 0.234 & 0.675 & 0.234 \\
$\phi$(1020) & 0.133 & 0.133 & 0.129 & 0.129 & 0.133 & 0.129 \\
$\Lambda$ & 0.321 & 0.318 & 0.0711 & 0.0708 & 0.241 & 0.0688 \\
$\Sigma^{+}$(1385) & 0.0391 & 0.0387 & 0.0313 & 0.0309 & 0.0319 & 0.0303 \\
$\Xi^-$ & 0.0201 & 0.0198 & 0.0116 & 0.0114 & 0.0193 & 0.0110 \\
$\Omega$ & 0.00124 & 0.00122 & 0.00124 & 0.00122 & 0.00118 & 0.00118 \\
\hline
\end{tabular}
\caption{Comparison of particle yields obtained with the code used in this analysis (in quantum statistics (QS) and Boltzmann statistics (BS)) and with the THERMUS code \cite{Wheaton:2004qb} for both initial production and for the final values (after strong and electromagnetic decays). The parameters listed correspond to the best fit at 91 GeV ($T$=157 MeV, $V$=32 fm$^3$ and $\gamma_s$=0.80). No contribution from weak decays is included.}
\label{tabl:comparison}
\end{center}
\end{table}Before engaging in the present data analysis a comparison of results from the model described above with those obtained using the THERMUS  code \cite{Wheaton:2004qb} was performed\footnote{The THERMUS code is available publicly and can be downloaded from the web: \url{http://hep.phy.uct.ac.za/THERMUS/SourceDownload.html}}. The results for a particular set of parameters are presented in Table~\ref{tabl:comparison}. The deviation caused by the Boltzmann approximation is also shown in this Table. Here and in the following the quoted yields include the corresponding antiparticle states (i.e. the yield labeled $\pi^+$ is actually the sum of the yields of $\pi^+$ and $\pi^-$) and are the sum over the two jets, as summarized and published by the PDG \cite{Yao:2006px}. The agreement between the particle yields obtained with both codes is very good, lending strong support also to the numerical implementation of the methods used in this analysis. Note that, in the calculations used here, quantum statistics is employed, whereas in the THERMUS code the Boltzmann approximation is used for the canonical ensemble. This causes an error of about 3\% for the final pion yield\footnote{The differences discussed in section \ref{sec:num} are from the initial yields.}, i.e. larger than the uncertainty in the data. As expected, the effect on all other hadron yields is smaller. If the code used in this analysis is restricted to Boltzmann approximation the yields, prior to strong decays, for mesons and non-strange baryons are in agreement with THERMUS results at the percent level, while for strange baryons the yields are systematically higher by about 2\% compared to those obtained with THERMUS. Inspecting the yields after strong decays, one notices a discrepancy between results from both calculations and for most of the hadrons. The reason is a more complete set of hadron yields used in the calculations.\\
The measured hadron yields summarized in \cite{Yao:2006px} include decay products from all resonances with decay lengths $c\tau<$10 cm. The effect to weak decaying particles is appropriately taken into account in the present calculations as shown in Table \ref{tabl:weak}. Because of weak decays the final yields are increased by 10\% for charged pions (8\% for $\pi^0$) and 23\% for protons. The magnitude of this contribution underlines its importance for the analysis of data.\\
In the resonance gas model the results are determined by the basic thermal parameters, temperature T and volume V (the volume corresponding to one jet).

\subsection{Comparison to former results}

\begin{table}[]
\begin{center}
\begin{tabular}{cccc}
\hline
\hline
\multicolumn{4}{c}{\textbf{$T$=166.7 MeV, $V$=17.5 fm$^3$, $\gamma_s$=0.698}}\\
\hline
\textbf{particle} & \textbf{results of \cite{Becattini:1995if}} &  \textbf{This work} & \textbf{THERMUS}\\
\hline
$\pi^{+}$ & 8.72  & 6.33(6.87) & 5.55\\
$\pi^0$ & 9.83 & 7.55(8.05) & 6.47\\
K$^+$ & 1.06 & 0.653(0.653) & 0.607\\
$\rho^0$ & 1.16 & 0.948(0.948) & 0.878\\
p & 0.484 & 0.301(0.377) & 0.290\\
$\phi$(1020) & 0.167 & 0.0916(0.0916) & 0.0873\\
$\Lambda$ & 0.152 & 0.120(0.130) & 0.0864\\
$\Xi^-$ & 0.0110 & 0.00557(0.00560) & 0.00524\\
$\Omega$ & 0.000782 & 0.000260(0.000260) & 0.000247\\
\hline
\hline
\end{tabular}
\caption{Model comparison, to THERMUS \cite{Wheaton:2004qb} and \cite{Becattini:1995if}. The values in brackets include weak decays, whereas THERMUS yields do not include weak decays.}
\label{table:compare}
\end{center}
\end{table}
The comparison of the present results with those published in \cite{Becattini:1995if,Becattini:1997uf,Becattini:1996gy} is rather unsatisfactory, with discrepancies exceeding 50\% for many particle yields (see Table \ref{table:compare}). For the parameters of the best fit of ref.~\cite{Becattini:1995if} in the uncorrelated jet scheme ($T$=166.7 MeV, $V$=17.5 fm$^3$, $\gamma_s$=0.698) for $\sqrt{s}$=91 GeV, the $\pi^+$ yield from the code used in this work is 6.87, while that published in \cite{Becattini:1995if} is 8.72, whereas for the $\Omega$ baryon the corresponding yields are 2.60$\cdot10^{-4}$ and 7.82$\cdot10^{-4}$, respectively. This is also The THERMUS results are included in Table \ref{table:compare} to show the good agreement with the results of this work. Each deviation between the results obtained in this work and the THERMUS results can be explained by a more complete set of particles.\\
The model differences of this work to \cite{Becattini:1995if} are:
\begin{enumerate}
 \item Conservation of five quantum numbers (\cite{Becattini:1995if} neglects the conservation of electric charge).
 \item Quantum statistics is applied, whereas \cite{Becattini:1995if} uses Boltzmann statistics.
 \item The particle table includes all particles published by the PDG (2008) with a mass cut at almost 3 GeV, whereas in \cite{Becattini:1995if} a mass cut at 1.7 GeV is used based on the data of 1996.
\end{enumerate}
Unfortunately the differences of the values obtained in this work to the results of \cite{Becattini:1995if} cannot be explained by the three points mentioned above and are not understood so far.

\section{Fit to the experimental data}
\label{sec:resultfit}

\begin{figure}[t]
\centering\includegraphics[width=13cm, height=12cm]{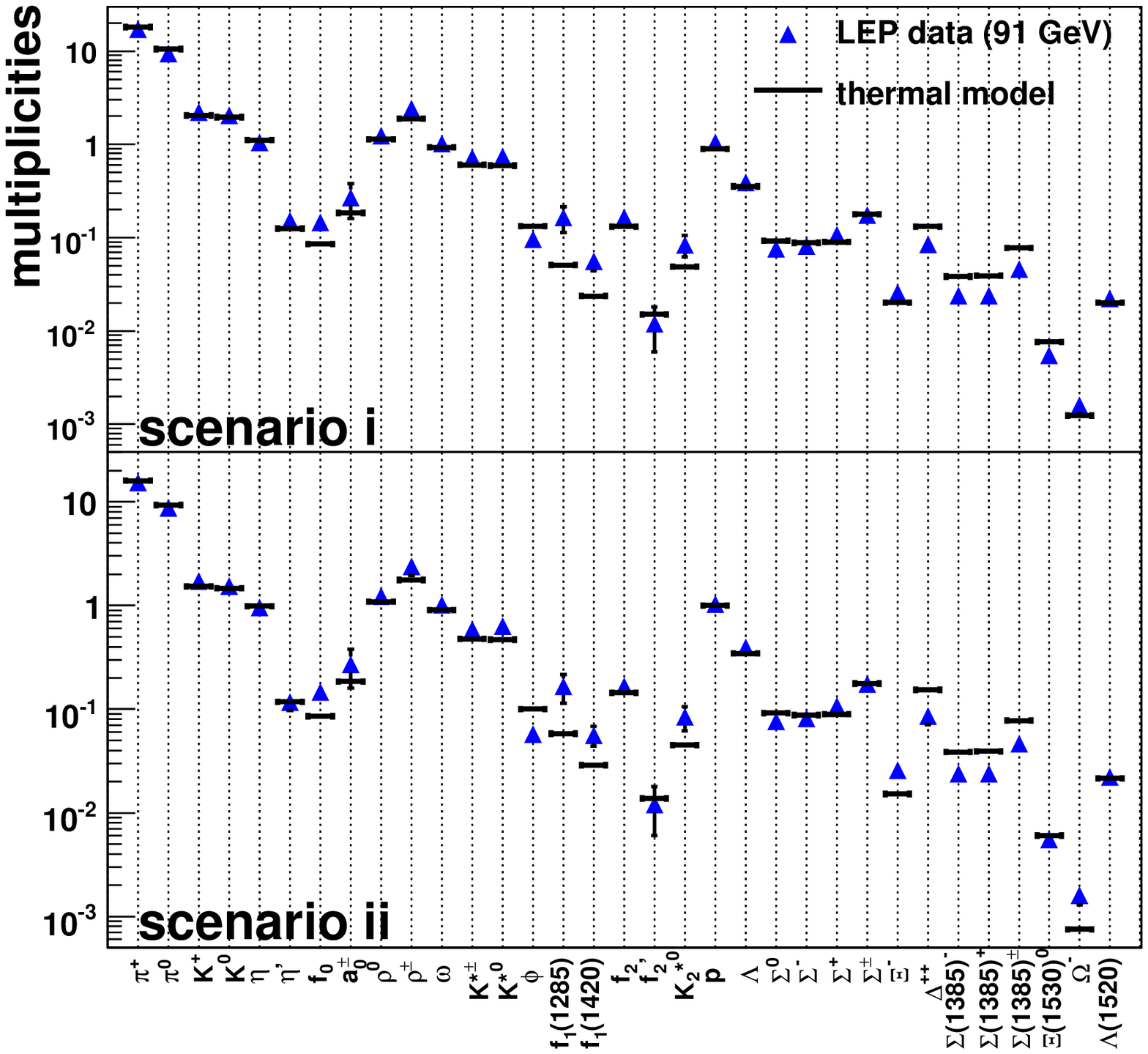}
\vspace{-0.5cm}
\caption{Comparison between the best fit Thermal Model calculations and experimental hadron multiplicities for $e^+e^-$ collisions at $\sqrt{s}$=91 GeV. The upper panel shows the case including in the data the contribution from heavy quarks, the lower panel is after subtraction of this contribution.}
\label{fig:91GeV}
\end{figure}
\begin{figure}[]
\begin{center}
\includegraphics[width=1.1\textwidth]{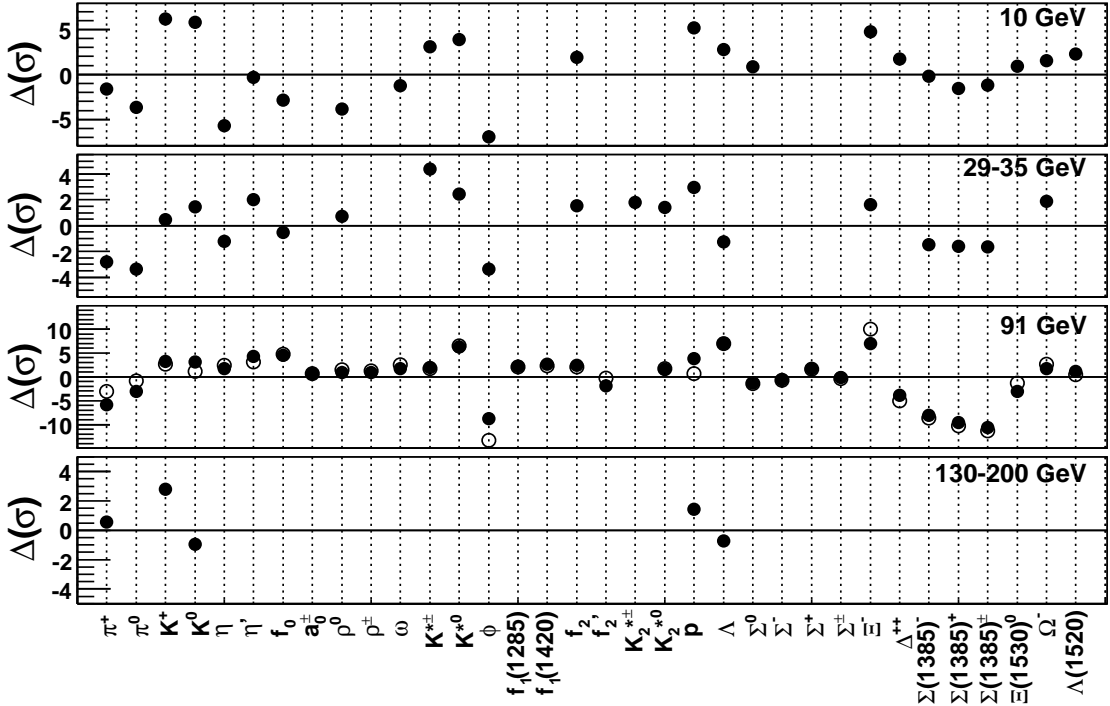}
\vspace{-1.5cm}
\caption{Difference (in units of experimental error) between experimental data and thermal model fits at four energies. 
At $\sqrt{s}$=91 GeV both scenarios are shown, the uncorrelated scheme (black point) and the results of the fit to data after subtraction of the heavy quark decay contribution (open circles).}
\label{fig:sigma}
\end{center}
\end{figure}
\begin{figure}[]
\begin{tabular}{cc} 
\begin{minipage}{.49\textwidth}
\centering\includegraphics[width=1.05\textwidth]{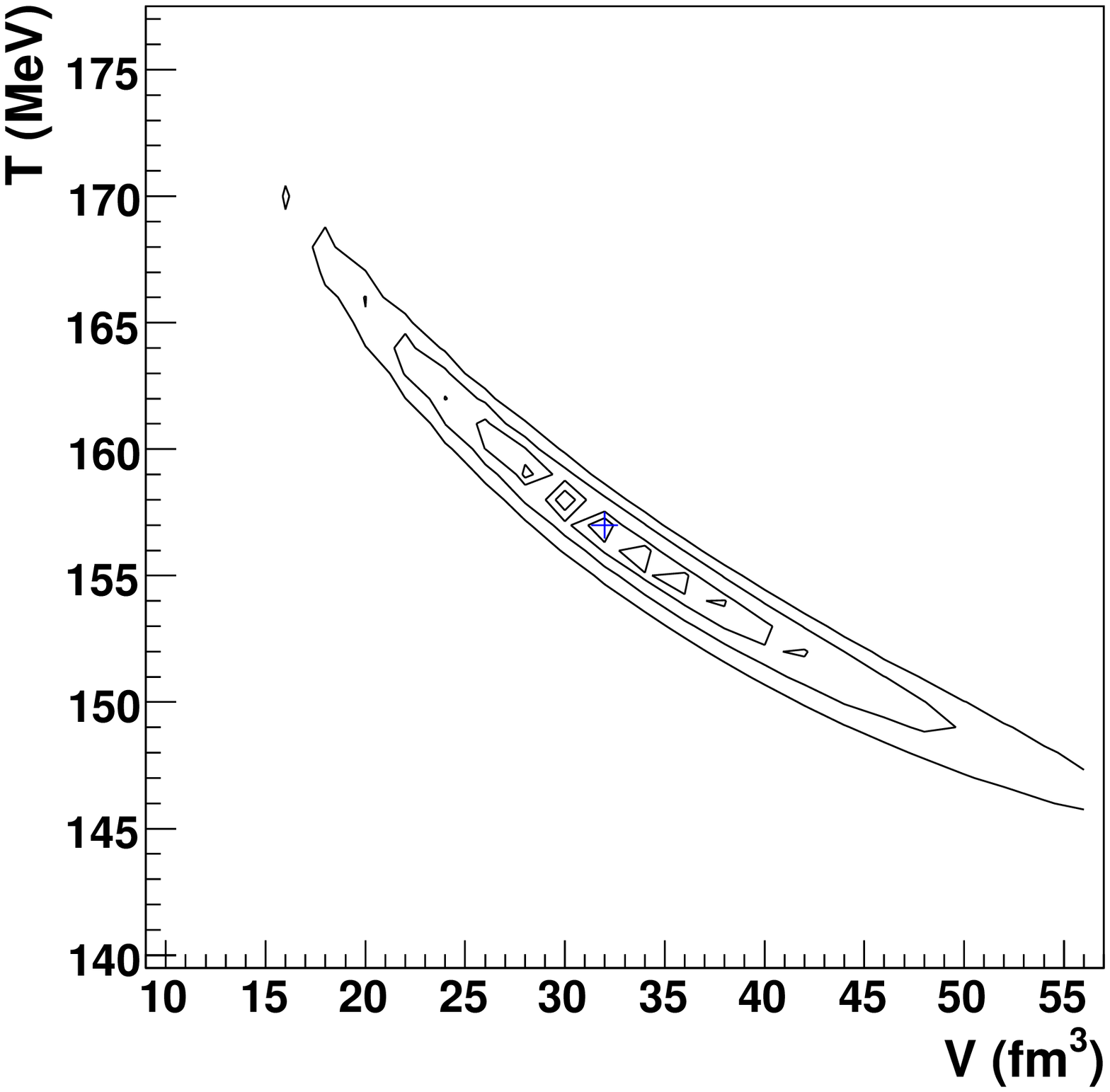}
\end{minipage} & \begin{minipage}{.49\textwidth}
\centering\includegraphics[width=1.05\textwidth]{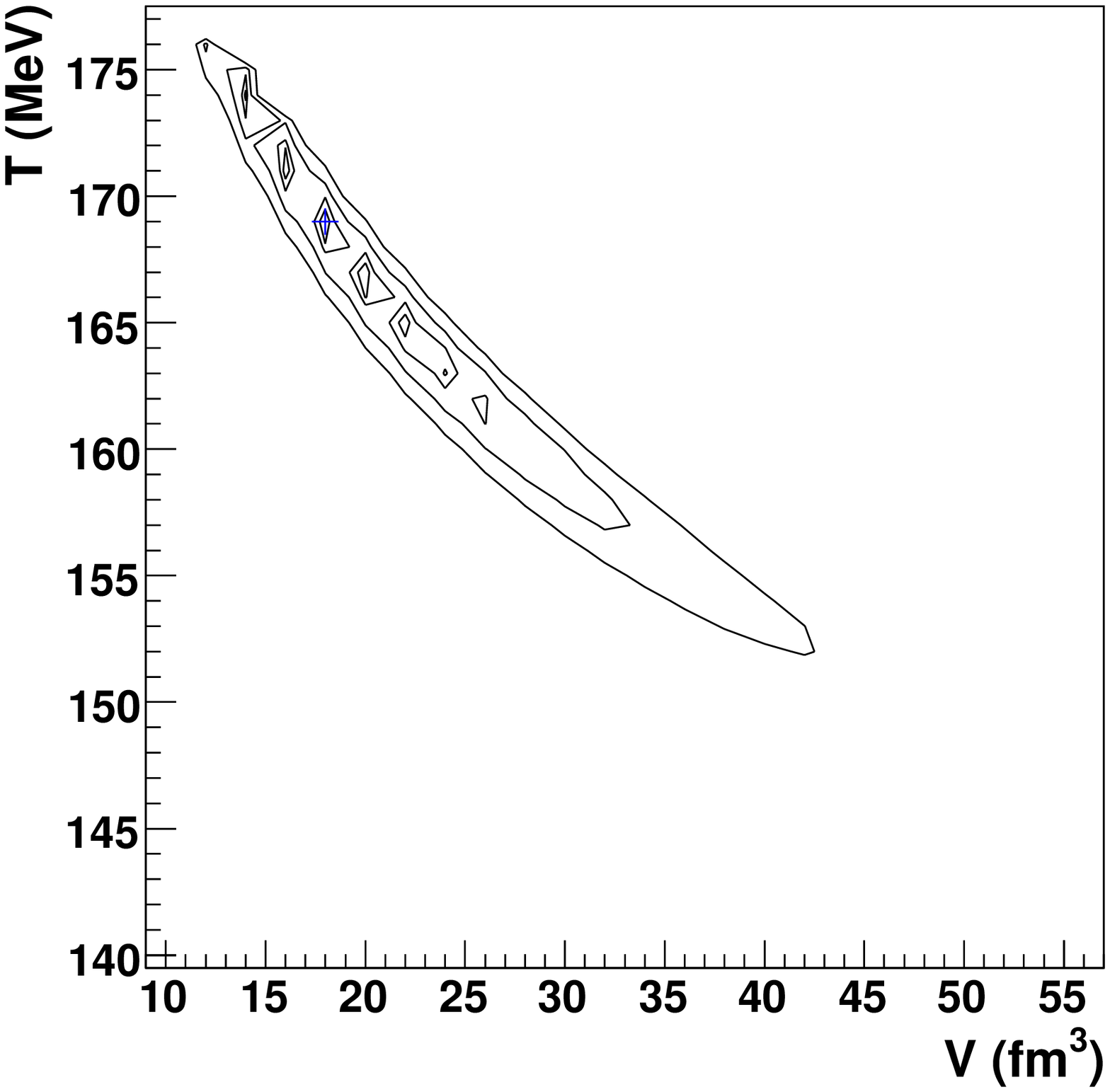}
\end{minipage}
\end{tabular}
\caption{$\chi^2$ contour lines in temperature and volume space for the overall fit (left panel) and after subtraction of charm and bottom decays (right panel) for $\sqrt{s}$=91 GeV. The contours correspond to $\chi^2_{min}$+10\%, +20\%, +50\%, +100\%. 
The best fit values are indicated by the blue crosses.}
\label{fig:91GeVcont}
\end{figure}

The resulting best fit to the data at the energy $\sqrt{s}$=91 GeV is shown in Fig.~\ref{fig:91GeV}. The two cases of the fit, without and with the subtraction of the contribution from heavy quarks (the magnitude of this contribution is listed in Table~\ref{tab:comp1}), are shown. The overall behavior of the data follows an approximately exponential decrease of particle yield with increasing particle mass. Such a behavior is expected  in the hadron resonance gas model due to the  Boltzmann factors, thus indicating the presence of statistical features of hadron production in elementary collisions.\\
The quantitative description of the data with the Statistical Model is, however, rather poor and certainly no improvement is visible for the case of subtracting charm and bottom contributions (scenario ii). The poor fit quality which is already visible in Fig.~\ref{fig:91GeV} becomes striking when investigating Fig.~\ref{fig:sigma}, where for the four energies the difference $\Delta$ (in units of the experimental error) between the experimental data and the Statistical Model calculations for the best fit values are shown.\\
A summary of the fit parameters obtained for the various data sets is presented in Table~\ref{tabl:param}. The errors of the thermal parameters listed in Table~\ref{tabl:param} are the result of a minimization performed with MINUIT \cite{minuit}, interfaced with the used code. For the true size of the errors, see the discussion in connection to Fig.~\ref{fig:91GeVcont} later in this chapter.\\
\begin{table}[]
\begin{center}
   \begin{tabular}{ccccc}
      \hline
      \hline
      $\sqrt{s}$ [GeV] & $T$ [MeV] & $V$ [fm$^3$] & $\gamma_s$ & $\chi^2$/dof\\
      \hline      
      10 & 159$\pm$1.7 & 14$\pm$1.5 & 0.80$\pm$0.02 & 325/21 \\
      29-35 & 160$\pm$1.7 & 18$\pm$1.4 & 0.96$\pm$0.03 & 105/18 \\
      91 (all) & 157$\pm$0.50 & 32$\pm$1 & 0.80$\pm$0.007 & 659/30 \\
      91 (-c,b) & 169$\pm$0.50 & 18$\pm$1 & 0.68$\pm$0.01 & 781/30 \\
      130-200 & 155$\pm$2.8 & 40$\pm$4.3 & 0.78$\pm$0.03 & 12.5/2 \\
      \hline
      \hline
   \end{tabular}
   \caption{Values of fit parameters and $\chi^2$ values per degree of freedom 
   for different energies.} 
   \label{tabl:param}  
\end{center}
\end{table}Typical $\chi^2$ values per degree of freedom lie between 5 and 20 and discrepancies between single data points and fit values larger than 5 standard deviations are not rare. Furthermore, there is no clear pattern observed: the fits are comparably poor for baryons and mesons, as well as for non-strange and strange hadrons. In particular, for all energies the yields of $\phi$ mesons and $\Lambda$ and $\Sigma$ hyperons are poorly reproduced. Large deviations are seen also for kaons.\\
The case of the fit after subtraction of the decay contribution of charmed and bottom hadrons (scenario ii), explored at $\sqrt{s}$=91 GeV, is characterized by a larger $\chi^2$ value compared to the overall fit (scenario i). The extracted fit parameters differ in the two cases. In particular, $T$ is higher and $V$ is smaller for scenario ii) compared to scenario i), following the ($T,V$) correlation shown in Fig.~\ref{fig:91GeVcont}.\\
It should be mentioned that while the general agreement among the four LEP experiments is excellent, the measured yields of $\Sigma^*$ hyperons differ by more than 70\%.  Excluding the $\Sigma^*$ hyperons would cause an increase of $\gamma_s$ at $\sqrt{s}$=91 GeV, because the $\Sigma^*$ yields calculated in the model are overestimated (see Fig.~\ref{fig:91GeV}). This would also slightly improve the situation for the $\Omega$ and $\Lambda$ multiplicities which are much higher than those predicted by the model. However, excluding those particles from the fit does not result in a significant improvement of the $\chi^2$ values, as is discussed below.\\
\begin{table}[]
\begin{center}
\begin{tabular}{cccc|c}
\hline
\hline
particle & uncorrelated & correlated ($u$,$d$,$s$) & correlated ($u$,$d$,$s$) & contribution \\
~ & ~ & $N$=$Q$=0 & fractional $N$, $Q$ & from $c$, $b$ (in \%)\\
\hline
$\pi^{+}$ & 16.01 & 16.23 & 16.48 & 11.7\\
$\pi^0$  & 9.29 & 9.40 & 9.52 & 8.4\\
K$^+$ & 1.53 & 1.62 & 1.65 & 30.0\\
$\rho^0$(770) & 1.09 & 1.09 & 1.10 & 1.4\\
K$^{*0}$(892) & 0.469 & 0.507 & 0.514 & 16.5\\
p & 0.999 & 1.02 & 1.13 & 1.7\\
$\phi$(1020) & 0.100 & 0.100 & 0.100 & 68.0\\
$\Lambda$ & 0.343 & 0.363 & 0.383 & 1.5 \\
$\Sigma^{+}$(1385) & 0.0392 & 0.0416 & 0.0438 & $\sim$0\\
$\Xi^-$ & 0.0152 & 0.0172 & 0.0177 & $\sim$0 \\
$\Omega$ & 0.000751 & 0.000936 & 0.000960 & $\sim$0\\
\hline
$\chi^2$/dof & 781/30 & 762/30 & 716/30 & -\\
\hline
\hline
\end{tabular}
\caption{Calculated yields (corresponding to the fit of the subtracted data at 91 GeV, $T$=169 MeV, $V$=18 fm$^3$ and $\gamma_s$=0.68) assuming vanishing or fractional baryon and charge quantum numbers. Also $\chi^2$/dof of the fits for the three cases is shown. In the rightmost column the percentage increase of the yields due to the contribution of charm and bottom events is shown, as calculated from the experimental data.
}
\label{tab:comp1}
\end{center}
\end{table}Now the difference between calculations with uncorrelated and correlated quark-antiquark ($u$,$d$,$s$) schemes will be discussed. In this context one should note that in the correlated jet scheme there is an open issue concerning the treatment of electric charge and baryon numbers of the quarks. A calculation for vanishing charge and baryon number and for fractional quark quantum numbers is shown in Table~\ref{tab:comp1}. The results in these 2 cases differ little for all hadrons, with the exception of protons, for which a difference of 11\% is observed. However, the quality of the fit to the data (charm-subtracted data at 91 GeV) is not significantly improved.\\
A further difficulty is visible if one inspects $\chi^2$ contour lines as shown in Fig.~\ref{fig:91GeVcont} in ($T$,$V$) space for both fit scenarios at 91 GeV. One notices in this Figure a strong anti-correlation between the fit parameters which is also present in the (T,$\gamma_s$) space (shown in appendix \ref{ap:plots}). Closer inspection reveals, in addition, a series of local minima which indicates the difficulty in the determination of the fit parameters. Such local minima are typical for poor fits and imply that the true uncertainties in the fit parameters are likely much larger than the values obtained from the standard fit procedure \cite{minuit} employed here.\\
Despite these caveats about fit quality and uncertainties it is noteworthy that the temperature parameters obtained from most data sets are close to 160 MeV and nearly independent of energy, similar to results of previous investigations. In contrast, the volume increases with the center of mass energy, reaching a value at 91 GeV which is larger by almost a factor of 2, compared to results reported in \cite{Becattini:1995if}, but somewhat smaller than the value obtained in \cite{Becattini:1997uf,Becattini:1996gy}. These differences may also be connected with the local minima observed in the $\chi^2$ surface discussed above. The values obtained for the strangeness under-saturation parameter $\gamma_s$ range between 0.96 and 0.68 and exhibit no clear trend with energy.\\
In the following the 91 GeV fit will be investigated in detail, as the data set at that energy contains the largest number of measured hadron yields. Since there is no statistically significant difference between the two fit scenarios, scenario i) is considered, that is, the fit to data without subtraction the charm and bottom decay contributions. To check whether the high $\chi^2$/dof values are caused by discrepancies for a few particular particles, the four worst hadron species (the three $\Sigma^*$'s and $\phi$) are excluded. Still a high $\chi^2$/dof=206/26 (for $T$=159 MeV, $V$=28 fm$^3$, $\gamma_s$=0.86) remains.\\
The data from LEP comprise many particle species and their yields typically are measured with accuracies of a few percent. Clearly the precision of the experimental data set provides a stringent test of the Statistical Model. To provide a quantitative estimate at which accuracy level the Statistical Model breaks down and to allow comparison with the situation encountered for RHIC data a fit at $\sqrt{s}$=91 GeV is performed to those hadron yields which were used for the fit of central nucleus-nucleus collision data at $\sqrt{s_{NN}}$=200 GeV~\cite{Andronic:2005yp}, namely $\pi^+$, $\pi^0$, $K^+$, $K^0$, $K^{*+}$, $K^{*0}$, $p$, $\Lambda$, $\Xi^-$, $\Omega^-$, and $\phi$.\\
The resulting thermal parameters are: $T$=162$\pm$2 MeV $V$=26$\pm$2 fm$^3$, and $\gamma_s$=0.78$\pm$0.02, with a still poor $\chi^2$/dof of 263/8. Excluding the $\phi$-meson from the fit yields $T$=160$\pm$2 MeV, $V$=26$\pm$2 fm$^3$ and $\gamma_s$=0.90$\pm$0.02, with  $\chi^2$/dof=57/7. A more direct comparison between thermal fits for heavy ion and $e^+e^-$ data can be obtained by assigning the uncertainties of the RHIC data to the corresponding LEP yields (namely uncertainties of the order of 13\%). Constraining, as for the RHIC data, the fit parameters to $T$ and $V$, i.e. setting $\gamma_s$=1 yields then $T$=157$\pm$2 MeV and $V$=26$\pm$2 fm$^3$, with $\chi^2$/dof=38/9. A reasonable fit can only be obtained by also letting $\gamma_s$ vary freely, with resulting parameters $T$=166$\pm$2 MeV, $V$=22$\pm$2fm$^3$ and $\gamma_s$=0.76$\pm$0.02, with $\chi^2$/dof=11.8/8. These exercises demonstrate that a Statistical Model description of $e^+e^-$ data fails badly without the introduction of the non-equilibrium parameter $\gamma_s$. Even using $\gamma_s$ the statistical description breaks down completely at an accuracy level for the data of below 10\%.

\section{Energy content}
\label{sec:energy}

Another noteworthy difference between fireballs in $e^+e^-$ and nucleus-nucleus collisions is their energy content. For its determination the energy density $\epsilon$ in the hadronic gas is computed to yield the thermal energy content $E$=$\epsilon V$ of one jet at chemical decoupling.\\
The energy can be derived by applying eq. (\ref{eq:stat5}) to eq. (\ref{eq:partition19.3})
\begin{equation}
\begin{split}
E &= -\frac{\partial\ln \mathcal{Z}}{\partial\beta}\\
&= \sum_j \frac{\mathcal{Z}(\vec{X} - \vec{x}_j)}{\mathcal{Z}(\vec{X})}E^1_j + \sum_b \sum^{\infty}_{k=2}\frac{\mathcal{Z}(\vec{X} - k\vec{x}_b)}{\mathcal{Z}(\vec{X})}E_b^{k}
\end{split}
\end{equation}
with
\begin{equation}
E_j^k = \frac{g_j V}{(2\pi\hbar)^3} \int d^3p\; \epsilon_je^{-\beta \epsilon_j k}
\end{equation}
where $j$ runs over all particles and $b$ runs over the light bosons. The equation shows, that the canonical factor will decrease the energy as in the case of the particle yields.\\
The results for the parameters reported in Table~\ref{tabl:param} are shown in Table \ref{tabl:energy}. The size of the compression because of the canonical factor decreases with increasing energy, like one would expect. It is also possible to see the small but visible effect caused by the underestimation of Boltzmann approximation in comparison to the correct quantum statistic calculation. Note that the thermal energy within each jet is only a small fraction of $\sqrt{s}/2$, (e.g. about 19\% at 91 GeV). Apparently, in the thermal interpretation of $e^+e^-$ collisions, most of the c.m. energy is not available for particle production. This is in strong contrast with results for nucleus-nucleus collisions. For this purpose central collision events for 20 and 40 GeV/nucleon Pb-Pb collisions \cite{Andronic:2005yp} are analyzed, where it makes sense to consider data integrated over the full phase space. In these cases the energy content in the fireball amounts to 61\% and 63\% of the total c.m. energy at 20 and 40 AGeV, respectively, implying that most of the total c.m. energy in a nucleus-nucleus collision is thermal, with the remaining non-thermal fraction likely to be due to collective flow.\\
These differences are not consistent with the finding that particle production in $e^+e^-$, pp and nucleus-nucleus collisions is universally governed by the available c.m. energy \cite{phobos,cley2,basile}. 
\begin{table}[]
   \begin{center}
   \begin{tabular}{cccc}
      \hline
      \hline
      & Quantum & Boltzmann & Grand\\
      $\sqrt{s}$ [GeV] & statistics [MeV] & statistics [MeV] & canonical [MeV]\\
      \hline      
      10 & 3183.07 (60.1\%) & 3169.46 (59.8\%) & 5298.64\\
      29-35 & 5380.70 (69.9\%) & 5354.05 (69.6\%) & 7697.03\\
      91 & 8773.02 (80.1\%) & 8713.98 (79.6\%) & 10951.16\\
      130-200 & 10159.63 (82.9\%) & 10085.10 (82.3\%) & 12256.37\\
      \hline
      \hline
   \end{tabular}
   \caption{Energy content of one fireball for the four different energies, calculated in quantum statistics, Boltzmann statistics and grand-canonical. The comparison to the grand-canonical result are given in \%.} 
   \label{tabl:energy}  
   \end{center}
\end{table}

\section{Heavy flavored particles}
\label{sec:heavy}

\begin{figure}[t]
\begin{center}
\includegraphics[width=11cm,height=8cm]{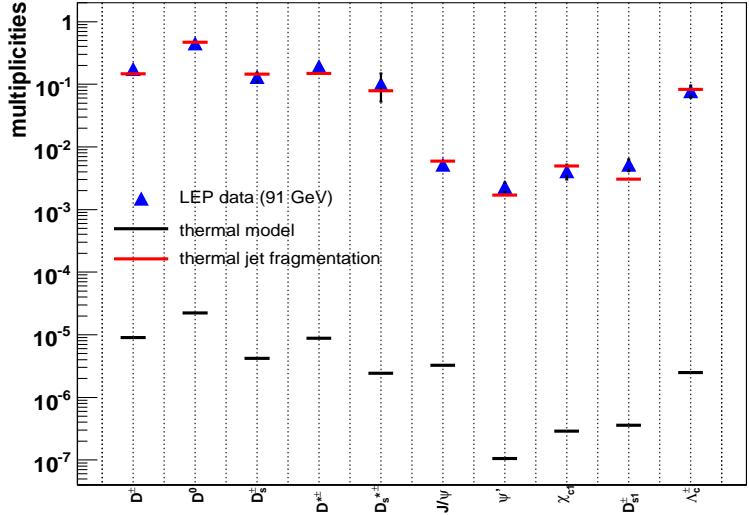}
\caption{Model predicted values for charged particles, compared to the LEP measurements at 91 GeV. The black lines corresponds to the uncorrelated jet scheme and the red lines (thermal jet fragmentation) corresponds to the correlated jet scheme.}
\label{fig:charm}
\end{center}
\end{figure}
\begin{figure}[t]
\begin{center}
\includegraphics[width=11cm,height=8cm]{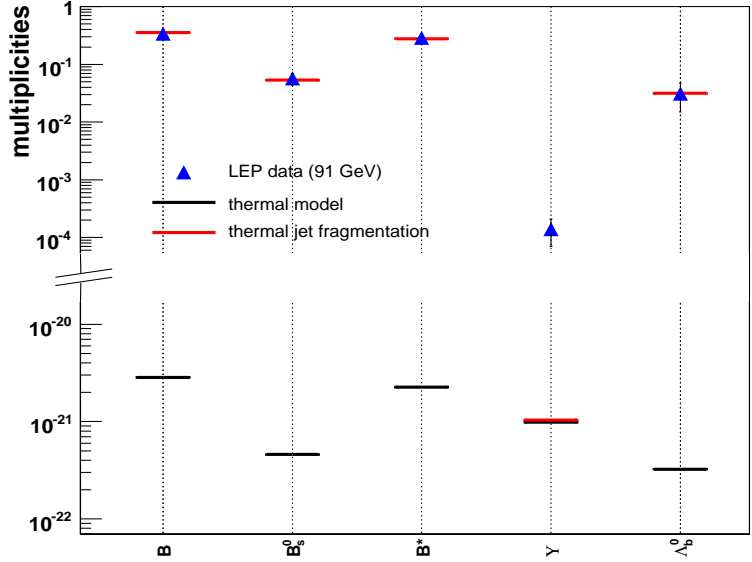}
\caption{Model predicted values for bottom particles, compared to the LEP measurements at 91 GeV. The thermal model corresponds to the uncorrelated jet scheme, thermal jet fragmentation is the results of the correlated jet scheme.}
\label{fig:bottom}
\end{center}
\end{figure}

\begin{figure}[]
\begin{center}
\includegraphics[width=9cm,height=9cm]{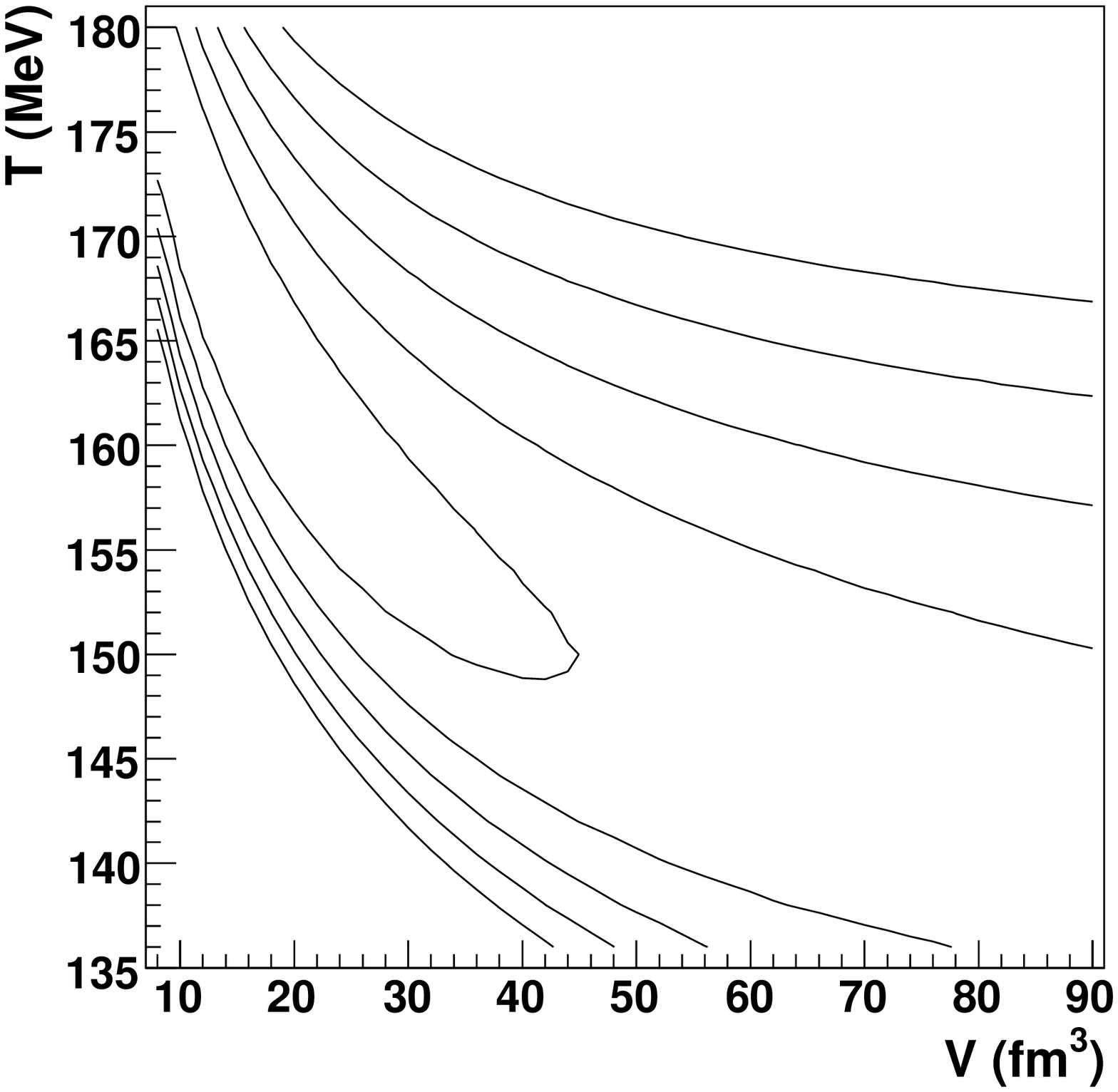}
\vspace{-0.6cm}
\caption{Contours plot obtained by a fit to the four measured bottom particle yields ($\Upsilon$ excluded).}
\label{fig:bottom2}
\end{center}
\end{figure}

The thermal production of particles containing charm or bottom quarks is very small, but as already shown in section \ref{sec:heavy2} there is a strong enhancement of charm particles in c-jets and bottom particles in b-jets. The question is, whether a correlated jet scheme, with the appropriate treatment of quantum numbers in the corresponding jets is able to describe the measured heavy particle yields.\\
Figure~\ref{fig:charm} and \ref{fig:bottom} show the measured charm and bottom yields together with the calculated values in a correlated (thermal jet fragmentation) and uncorrelated (thermal model) jet scheme. The uncorrelated scheme corresponds to the pure thermal production. The parameters are $T$=169 MeV, $V$=18 fm$^3$ and $\gamma_s$=0.68 as found for the c, b subtracted fit to the light particles. In contrary to the uncorrelated scheme (black lines) the correlated jet scheme (red lines) is able to describe the measured data with the exception of the $\Upsilon$ yield in Figure \ref{fig:bottom}, which can only be produced thermally even in the correlated jet scheme, because it has no net bottomness. The $J/\Psi$ yield can be described because the main contribution of it comes from bottom decays. The $\chi^2$/dof of the four bottom particles (without the $\Upsilon$) is 0.2/4 and of the charm particles it is 76/10. Nevertheless it seems, that it is almost independent of the parameters what can be seen in Figure~\ref{fig:bottom2}.\\
It was already discussed, that there is the necessity of an additional parameter $\gamma_s$ for particles carrying strangeness caused by an incomplete equilibrium. The reason is the high mass of the strange quark and therewith the Boltzmann suppression. The mass of the charm quark is much higher than the strange quark mass, thus there is no expectation of an equilibrium of charm particles. So why is it possible to describe the charm and bottom particle yields without any under-saturation factor as used for strange particles? \\
The thermal production of heavy quark is very low (black lines in Figure \ref{fig:charm} and \ref{fig:bottom}). The charmness and bottomness production takes place in hard collisions. This is taken into account by the parameters $R_q$. The sum over all initial charm particles gives exactly $R_c$=0.17 (or $R_b$=0.22 for bottom particles). This can be seen by investigating the charm particle yields using equation (\ref{eq:partition18}) and (\ref{eq:partition3.1}) which is in the case of a $c$-jet
\begin{equation}
\langle n^{charm}_j\rangle = \frac{z^1_j\mathcal{I}_{N-N_j,S-S_j,Q-Q_j}}{\sum_{j_c} z^1_{j_c} \mathcal{I}_{N-N_{j_c},S-S_{j_c},Q-Q_{j_c}}}
\end{equation}
(and similar for bottom particles).\\
This is not true for $u$, $d$ and $s$ quark, because the thermal production of this particles is much bigger than the production in hard collisions (In the case of strange particles there is indeed a conflict between the parameter $\gamma_s$ and $R_s$). The production of strange and charm particles in the different jets is shown in Table \ref{tab:jets}.
\begin{table}[]
   \begin{center}
   \begin{tabular}{cllll}
\hline
\hline
\multicolumn{1}{l}{\textbf{jet}} & \multicolumn{2}{c}{\textbf{strangeness} (S=-1)} & \multicolumn{2}{c}{\textbf{charmness} (C=1)}\\
\hline
$u$-jet: & & 0.126 & & 2.184$\cdot 10^{-6}$\\
$\bar{u}$-jet: & & 0.127 & & 2.184$\cdot 10^{-6}$\\
$d$-jet: & & 0.163 & & 2.809$\cdot 10^{-6}$\\
$\bar{d}$-jet: & & 0.162 & & 2.809$\cdot 10^{-6}$\\
$s$-jet: & & 0.322 & & 3.06$\cdot 10^{-6}$\\
$\bar{s}$-jet: & & 0.108 & & 3.06$\cdot 10^{-6}$\\
$c$-jet: & & 0.144 & & 0.17\\
$\bar{c}$-jet: & & 0.145 & & 6.776$\cdot 10^{-30}$\\
$b$-jet: & & 0.191 & & 3.036$\cdot 10^{-6}$\\
$\bar{b}$-jet: & & 0.189 & & 3.036$\cdot 10^{-6}$\\
\hline
\hline
   \end{tabular}
\caption{Sum over all initial strange/charm particle yields in the different jets. The values are already multiplied wit $R_q$ (see eq. (\ref{eq:fpf2})).}
\label{tab:jets}
   \end{center}
\end{table}It can be seen, that even in the antistrange jet there is a production of strange quarks, because strangeness is produced thermally. One could ask why is there a difference of strangeness production in e.g. the $u$ and $\bar{u}$ jets. This differences are caused by the electric charge. The values in Table~\ref{tab:jets} are the sums over all initial produced strange ($S$=-1) particle yields. This is e.g. in case of kaons the $K^-$ and the $\overline{\text{K}}{}^0$. and it is clear that there are more $K^-$ in an $\bar{u}$-jet than in an $u$-jet. Similar for the other jets, whereas the difference in the $s$-jets is obviously caused by the strangeness quantum number. It is again visible, that the strangeness quantum number plays a much bigger role than the electric charge quantum number, what was already discussed in chapter \ref{sec:about} and shown in Figure \ref{fig:Zvsall}.\\
In contrary to strange particles thermal production of charm particles is very small and the final charm yields are dominated by the production in c-jets. The overall charmness in such a jet is defined by the parameter $R_c$.

\chapter{Summary and Conclusion}
\label{sec:summary}

Within this thesis a numerically calculable quantum statistical canonical partition function taking into account the conservation of four and five quantum numbers was derived~\cite{Andronic:2009sv}. Based on these functions a computer program was implemented to calculate the hadron production in particle collisions~\cite{Beutler}. Furthermore the canonical factor with its special features was investigated and explained in detail. The comprehensive new set of measured yields of hadrons with light quarks ($u$, $d$, $s$) in $e^+e^-$ collisions was analyzed in a range of collision energies from 10 GeV up to 200 GeV~\cite{Andronic:2008ev}. At $\sqrt{s}$=91 GeV, two scenarios were considered, namely fitting data without and with subtraction of the decay products of charmed and bottom hadrons. The two thermodynamical parameters temperature $T$ and volume $V$, and a strangeness under-saturation factor $\gamma_s$ were obtained from a $\chi^2$ minimization procedure. While, as in previous investigations~\cite{Becattini:1995if,Becattini:1997uf,Becattini:1996gy}, the resulting temperature value was close to 160 MeV, independent of energy, the overall description of the high-precision LEP data is rather poor, independent of whether heavy quark contributions are subtracted or not. The $\chi^2$/dof values are larger than 5 for all fits and call into question the validity of the thermodynamical approach for these data. This conclusion still holds even if the analysis is restricted to the same set of hadrons which were analyzed in the context of thermal model fits to Au-Au collision data from the RHIC accelerator.\\
The apparent statistical fingerprint visible in the LEP data and first observed in \cite{Becattini:1995if} breaks down at an accuracy level of about 10\%. Even at that level the $e^+e^-$ data cannot at all be described without the explicit assumption of strangeness under-saturation, implying that hadron production in $e^+e^-$ originates from a state which is quite far from true
thermodynamic equilibrium. This conclusion is further supported by the observation that the corresponding fireball volume contains only a small fraction of the overall center of mass energy, in strong contrast to the situation in nucleus-nucleus collisions, where the fireball volume contains the majority fraction of the available energy.\\
Furthermore it was observed, that the implementation of fireballs with charm and bottom quantum numbers can be used to describe the heavy particle production in $e^+e^-$ collisions \cite{Andronic:2009sv} determined by the new input values $R_q$ (except of the $\Upsilon$ yield). This agrees with the observation of \cite{Becattini:1995if,Becattini:2008tx}. However, the contours plot showed that a wide parameter range of volume and temperature is able to explain the measured data. Nevertheless it seems that the production of this particles is indeed controlled by the particle phase space and the canonical factor.

\newpage

\renewcommand{\appendixtocname}{Appendix}

\begin{appendix}
\addappheadtotoc

\chapter{Mathematical and numerical approximations}

The following integral has to be solved (\ref{eq:partition19.4})
\begin{equation}
z^k_j = \frac{g_jV}{k(2\pi\hbar)^3}\int d^3p\; e^{-\frac{\sqrt{p^2+m_j^2}}{T}k}
\label{eq:appendix1}
\end{equation}
because the integral depends only on the absolute value of $p$, it simplifies by changing into spherical coordinates
\begin{equation}
z^k_j = \frac{g_jV}{k2\pi^2\hbar^3}\int^r_0 dr\; r^2 e^{-\frac{\sqrt{r^2+m_j^2}}{T}k}
\label{eq:appendix2}
\end{equation}
This integral can be solved numerically.\\
It is also possible to use the following relation
\begin{equation}
\frac{1}{T^3}\int dr\; r^2 e^{-\frac{\sqrt{r^2+m_j^2}}{T}} = \left(\frac{m}{T}\right)^2K_2\left(\frac{m}{T}\right)
\end{equation}
with the modified Bessel-function of order 2. This relation obeys the following limiting cases
\begin{eqnarray}
\left(\frac{m}{T}\right)^2K_2\left(\frac{m}{T}\right) &\rightarrow& 
\begin{cases} 2 & \text{for} \quad m\rightarrow 0\cr
\sqrt{\frac{\pi m^3}{2T^3}}e^{-\frac{m}{T}} & \text{for}\quad m\gg T
\end{cases}
\end{eqnarray}
and equation (\ref{eq:appendix2}) becomes
\begin{equation}
z^k_j = \frac{g_jVT}{2k^2\pi^2\hbar^3}m_j^2K_2\left(\frac{km_j}{T}\right)
\label{eq:appendix4}
\end{equation}

\section{Calculation comparison}
\label{ap:compare}

\begin{table}[H]
\begin{center}
\begin{tabular}{ccc}
\hline
\hline
particle & Bessel-way  & Integral-way \\
& (section \ref{sec:extent}) & (section \ref{sec:second})\\
\hline
$\pi^{\pm}$ & 3.186 & 3.185\\
$\pi^0$ & 3.629 & 3.634\\
$\eta$ & 0.510 & 0.510\\
$\rho^+$(770) & 0.779 & 0.779\\
$\phi$(1020) & 0.1243 & 0.1243\\
p & 0.1288 & 0.1285\\
$\Lambda$ & 0.03602 & 0.03593\\
$\Sigma^{+}$ & 0.02230 & 0.02225\\
$\Xi^-$ & 0.005667 & 0.005690\\
$\Omega^-$ & 0.0005432 & 0.0005494\\
\hline
\hline
\end{tabular}
\setlength{\captionwidth}{0.96\textwidth}
\caption{Comparison of the two ways of calculation explained in chapter \ref{sec:themodel}. The shown yields are the initial ones, what means that there is no treatment of any decay. The deviations are caused by numerical approximations (uncorrelated scheme).}
\label{tabl:compare}
\end{center}
\end{table}

\chapter{Quantum statistical calculation (for bosons and fermions)}
\label{ap:fermions}

The quantum statistical canonical partition function is (\ref{eq:partition19})
\begin{equation}
\begin{split}
\mathcal{Z}_{N,S,Q,C,B}(\vec{X}) &= \frac{1}{(2\pi)^5}\int d^5\vec{\phi}\; e^{i\vec{X}\vec{\phi}}\\
&\quad \exp\left\{\frac{g_jV}{(2\pi\hbar)^3}\int d^3p \ln(1\pm e^{-\frac{\sqrt{\vec{p}^2+m_j^2}}{T}-i\vec{x}_j\vec{\phi}})^{\pm1}\right\}\notag
\end{split}
\end{equation}
as already shown in section \ref{sec:extent}, where the correct quantum statistical calculation was implemented for bosons only (\ref{eq:partition19.2})
\begin{equation}
\ln(1-x)^{-1} = \sum^{\infty}_{k=1}\frac{x^{k}}{k}\notag
\end{equation}
For fermions the following expansion has to be used
\begin{equation}
\ln(1+x) = \sum^{\infty}_{k=1}(-1)^{k+1}\frac{x^{k}}{k}
\label{ap:fermions1}
\end{equation}
By including the whole series for both, fermions and bosons, to the partition function one gets
\begin{align}\label{ap:fermions2}
\mathcal{Z}_{N,S,Q,C,B}(\vec{X}) &= \frac{Z_{0}}{(2\pi)^5}\int d^5\vec{\phi}\; e^{i\vec{X}\vec{\phi}} \exp\left\{\sum_j z^1_je^{-i\vec{x}_j\vec{\phi}} + \sum_{j}\sum_{k=2}^{\infty}(\pm)^{k+1}z^k_je^{-ik\vec{x}_j\vec{\phi}}\right\}\notag\\
 &=\frac{Z_{0}}{(2\pi)^5}\int d^5\vec{\phi}\; e^{i\vec{X}\vec{\phi}} \exp\left\{\sum_j \sum_{k=1}^{\infty}(\pm)^{k+1}z^k_je^{-ik\vec{x}_j\vec{\phi}}\right\}
\end{align}
where the $+$ is for bosons and the $-$ for fermions. The particle partition functions are defined like in (\ref{eq:partition19.4}).\\
The particle multiplicities are
\begin{align}
\langle n_{j}\rangle &= \left.\frac{\partial\ln \mathcal{Z}}{\partial \lambda_{j}}\right|_{\lambda_{j}=1}\notag\\
&= \sum^{\infty}_{k=1}(\pm)^{k+1}kz^k_{j}\frac{\mathcal{Z}_{N,S,Q,C,B}(\vec{X}-k\vec{x}_{j})}{\mathcal{Z}_{N,S,Q,C,B}(\vec{X})}
\end{align}
To get a numerically calculable term for the integrals in (\ref{ap:fermions1}) one has to apply the same manipulations as discussed in section \ref{sec:extent}. The result will be
\begin{flalign*}
\mathcal{I}_{N,S,Q} & = \left[\prod^{10}_{j=1}\sum^{\infty}_{n_j={-\infty}}\right]I_{n1}(2Z_{p})I_{n2}(2Z_{\Delta^{\mp}})I_{n3}(2Z_{\Delta^{++}})I_{n4}(2Z_{K^{\pm}})\\
&\quad I_{n5}(2Z_{\Lambda})I_{n6}(2Z_{\Sigma^+})I_{n7}(2Z_{\Sigma^-})I_{n8}(2Z_{\Xi^0})I_{n9}(2Z_{\Xi^{\mp}})I_{n10}(2Z_{\Omega^{\mp}})\\
&\quad\left[\prod^{\infty}_{k=2}\sum^{\infty}_{n_k=-\infty}\right]I_{n_k}(2Z^{k}_{\pi^{\pm}})\left[\prod^{\infty}_{h=2}\sum^{\infty}_{n_h=-\infty}\right]I_{n_h}(2Z^{h}_{K^0})\left[\prod^{\infty}_{l=2}\sum^{\infty}_{n_l=-\infty}\right]I_{n_l}(2Z^{l}_{K^{\pm}})\\
&\quad\left[\prod^{\infty}_{m=2}\sum^{\infty}_{n_m=-\infty}\right]I_{n_m}(2Z^{m}_{p})\left[\prod^{\infty}_{n=2}\sum^{\infty}_{n_n=-\infty}\right]I_{n_n}(2Z^{n}_{\Delta^{\mp}})\left[\prod^{\infty}_{o=2}\sum^{\infty}_{n_o=-\infty}\right]I_{n_o}(2Z^{o}_{\Delta^{++}})\\
&\quad\left[\prod^{\infty}_{p=2}\sum^{\infty}_{n_p=-\infty}\right]I_{n_p}(2Z^{p}_{\Lambda})\left[\prod^{\infty}_{q=2}\sum^{\infty}_{n_q=-\infty}\right]I_{n_q}(2Z^{q}_{\Sigma^+})\left[\prod^{\infty}_{r=2}\sum^{\infty}_{n_r=-\infty}\right]I_{n_r}(2Z^{r}_{\Sigma^-})\\
&\quad\left[\prod^{\infty}_{s=2}\sum^{\infty}_{n_s=-\infty}\right]I_{n_s}(2Z^{s}_{\Xi^0})\left[\prod^{\infty}_{t=2}\sum^{\infty}_{n_t=-\infty}\right]I_{n_t}(2Z^{t}_{\Xi^{\mp}})\left[\prod^{\infty}_{u=2}\sum^{\infty}_{n_u=-\infty}\right]I_{n_u}(2Z^{u}_{\Omega^{\mp}})\\
&\quad I_{-\nu_1}(2Z_n)I_{-\nu_1}(2Z_{K^0}) I_{-\nu_2}(2Z_{\pi^{\pm}}) 
\refstepcounter{equation}\tag{\theequation}
\end{flalign*}
with
\begin{align}
\nu_1 &= N+n1+n2+n3+n5+n6+n7+n8+n9+n10\notag\\
&\quad+\sum mn_m+\sum nn_n+\sum on_o+\sum qn_q\notag\\
&\quad+\sum rn_r+\sum sn_s+\sum tn_t+\sum un_u\\
\nu_2 &= S+n4-n5-n6-n7-2n8-2n9-3n10\notag\\
&\quad+\sum hn_h+\sum ln_l-\sum pn_p-\sum qn_q -\sum rn_r\notag\\
&\quad-\sum 2sn_s-\sum 2tn_t-\sum 3un_u\\
\nu_3 &= Q+n1-n2+2n3+n4+n6-n7-n9-n10\notag\\
&\quad+\sum kn_k+\sum ln_l+\sum mn_m-\sum nn_n\notag\\
&\quad+\sum 2on_o+\sum qn_q-\sum rn_r-\sum tn_t-\sum un_u
\end{align}
Inserting this integral in equation (\ref{eq:partition23}) gives the final partition function.
\begin{equation}
\begin{split}
\mathcal{Z}_{N,S,Q,C,B}(\vec{X}) & \approx Z_{0}\bigg(\mathcal{I}_{N,S,Q}\;\delta_{C,0}\delta_{B,0}\\
&\quad +  \sum_{j_{c}}z^1_{j_{c}}\mathcal{I}_{N+N_{j_c},S+S_{j_c},Q+Q_{j_c}} \delta_{C,C_{j_{c}}}\delta_{B,B_{j_c}}\\
&\quad +  \sum_{j_{b} \;\;\&\atop C_{j_b} = 0}z^1_{j_{b}}\mathcal{I}_{N+N_{j_b},S+S_{j_b},Q+Q_{j_b}} \delta_{C,0}\delta_{B,B_{j_{b}}}\\
&\quad +  \sum_{j_{c}}\sum_{j_{b}\;\;\&\atop C_{j_b} = 0}z^1_{j_{c}}z^1_{j_{b}}\mathcal{I}_{N+N_{j_c}+N_{j_b},S+S_{j_c}+S_{j_b},Q+Q_{j_c}+Q_{j_b}} \delta_{C,C_{j_{c}}}\delta_{B,B_{j_{c}}+B_{j_b}}\bigg)
\end{split}
\end{equation}

\chapter{Fit results}
\label{ap:results}

\section{Particle yields}
\label{ap:yields}

\begin{table}[H]
\begin{center}
\begin{tabular}{ccccccc}
\hline
\hline
particle & $\sqrt{s}$=10 GeV & $\sqrt{s}$=29-35 GeV & $\sqrt{s}$=91 GeV & $\sqrt{s}$=130-200 GeV\\ 
\hline
$\pi^{\pm}$& 6.91 & 10.97 & 18.08 & 21.07\\
$\pi^0$ & 4.32 & 6.60 & 10.47 & 12.10\\
$K^{\pm}$ & 0.627 & 1.36 & 2.04 & 2.36\\
$K^0$ & 0.593 & 1.30 & 1.96 & 2.28\\
$\eta$ & 0.501 & 0.796 & 1.11 & 1.25\\
$\rho^{\pm}$(770) & 0.739 & 1.13 & 1.89 & 2.18\\
$\rho^0$(770) & 0.518 & 0.726 & 1.13 & 1.29\\
$K^{*\pm}$(892) & 0.175 & 0.402 & 0.604 & 0.694\\
$K^{*0}$(892) & 0.172 & 0.395 & 0.594 & 0.682\\
$p$ & 0.192 & 0.459 & 0.895 & 1.05\\
$\phi$(1020) & 0.0646 & 0.125 & 0.133 & 0.142\\
$\Lambda$ & 0.0685 & 0.213 & 0.354 & 0.409\\
$\Sigma^{+}$ & 0.0174 & 0.0529 & 0.0901 & 0.104\\
$\Sigma^{+}$(1385) & 0.00746 & 0.0228 & 0.0391 & 0.0452\\
$\Xi^-$ & 0.00277 & 0.0133 & 0.0201 & 0.0235\\
$\Xi^0$(1530) & 0.00106 & 0.00509 & 0.00762 & 0.00882\\
$\Omega^-$ & 0.000105 & 0.000879 & 0.00124 & 0.00147\\
\hline
\hline
\end{tabular}
\caption{A random set of calculated particle yields (particle $+$ antiparticle in two jets) for the best fit values (see Table \ref{tabl:param}) at the four different energies used in the analysis. The uncorrelated jet scheme is employed.}
\label{tabl:yields}  
\end{center}
\end{table}

\newpage

\section{Plots}
\label{ap:plots}

\subsection{10 GeV}

\begin{figure}[H]
\vspace{-0.5cm}
\begin{center}
\label{fig:10GeV}
\includegraphics[width=0.8\textwidth,height=9cm]{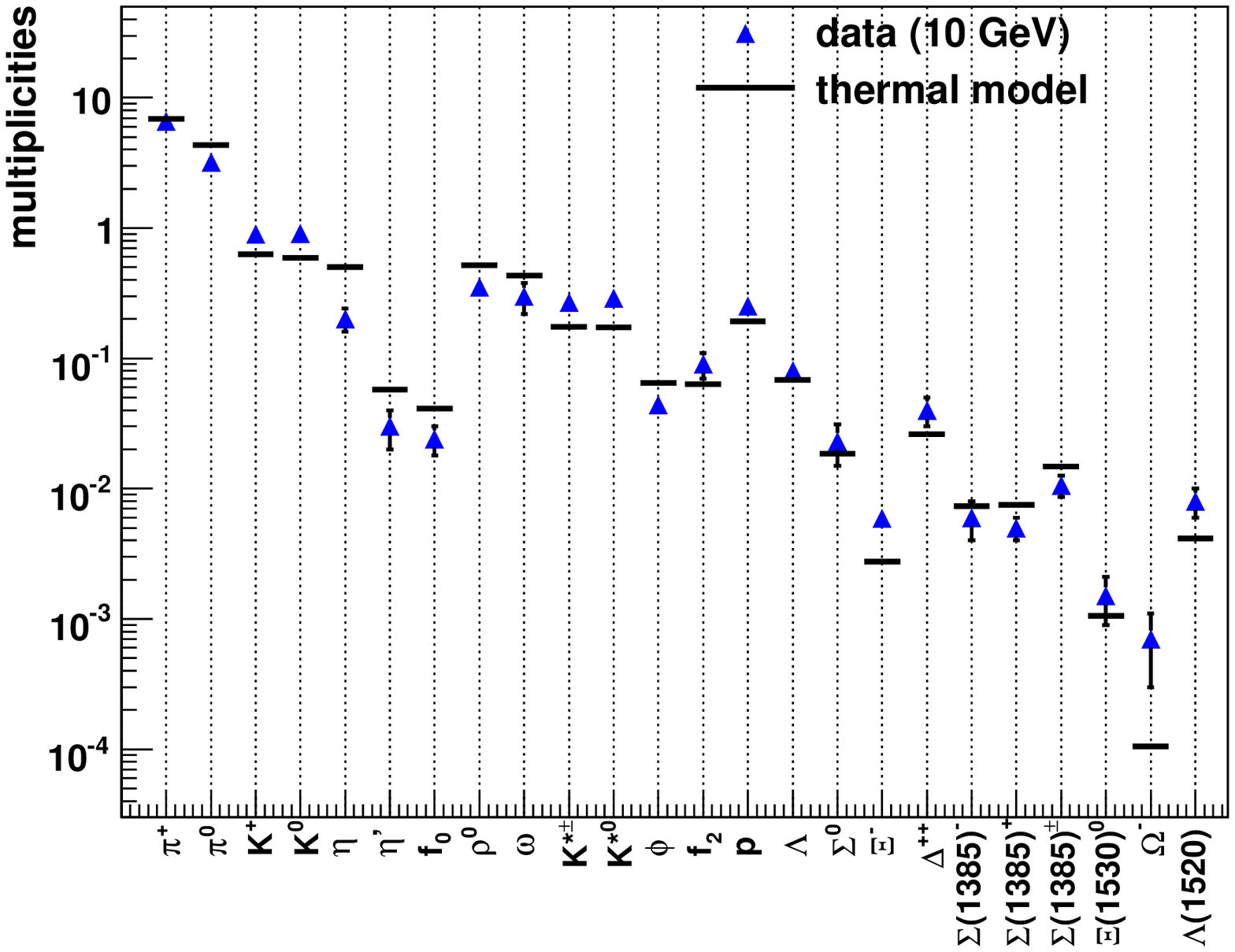}
\vspace{-0.5cm}
\caption{Comparison between thermal model predictions and experimental particle multiplicities for $e^+e^-$ collisions at $\sqrt{s}$=10 GeV.}
\end{center}
\end{figure}
\vspace{-1.3cm}
\begin{figure}[H]
\begin{tabular}{cc} 
\begin{minipage}{.49\textwidth}
\centering\includegraphics[width=1.05\textwidth]{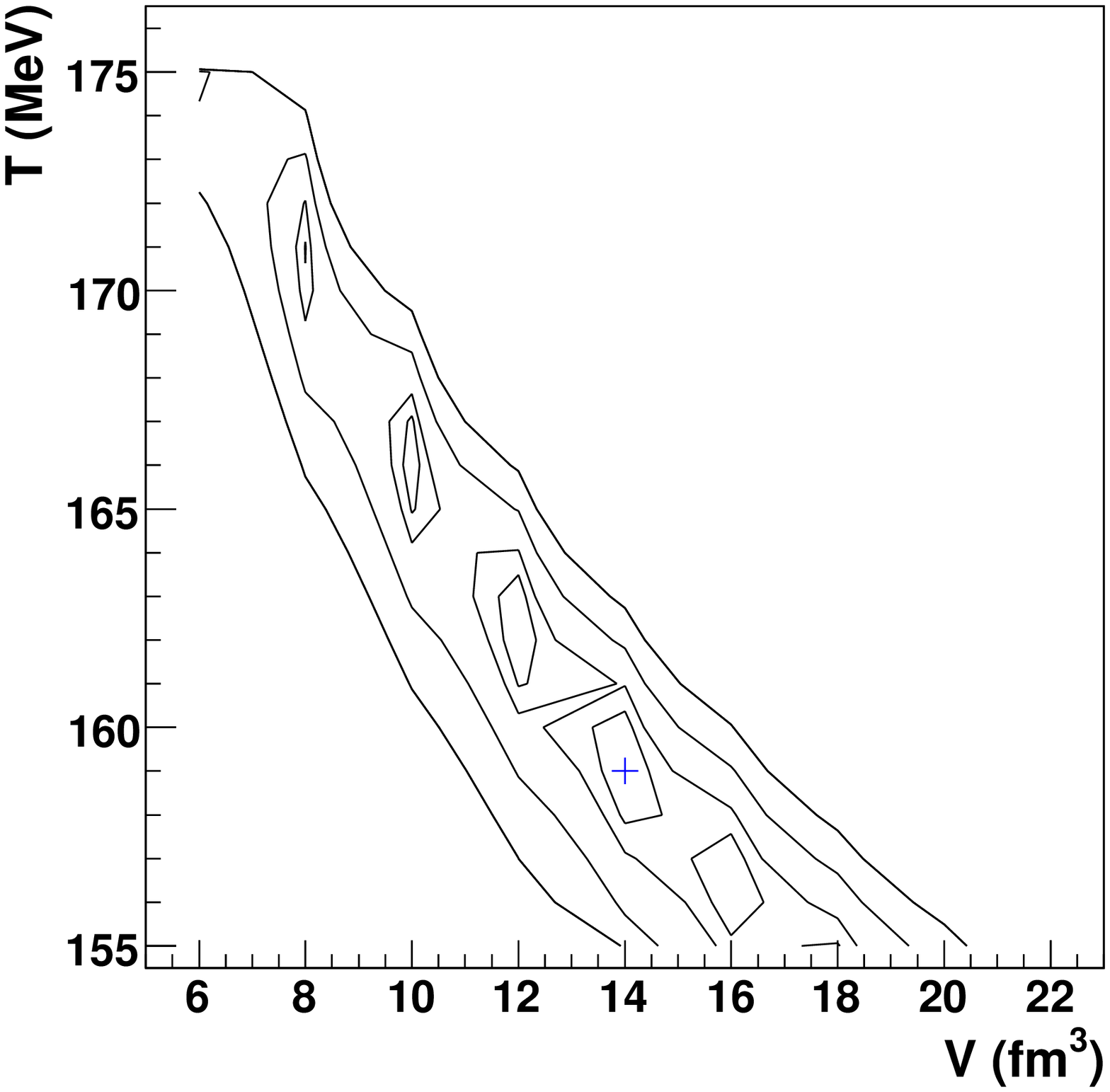}
\end{minipage} & \begin{minipage}{.49\textwidth}
\centering\includegraphics[width=1.05\textwidth]{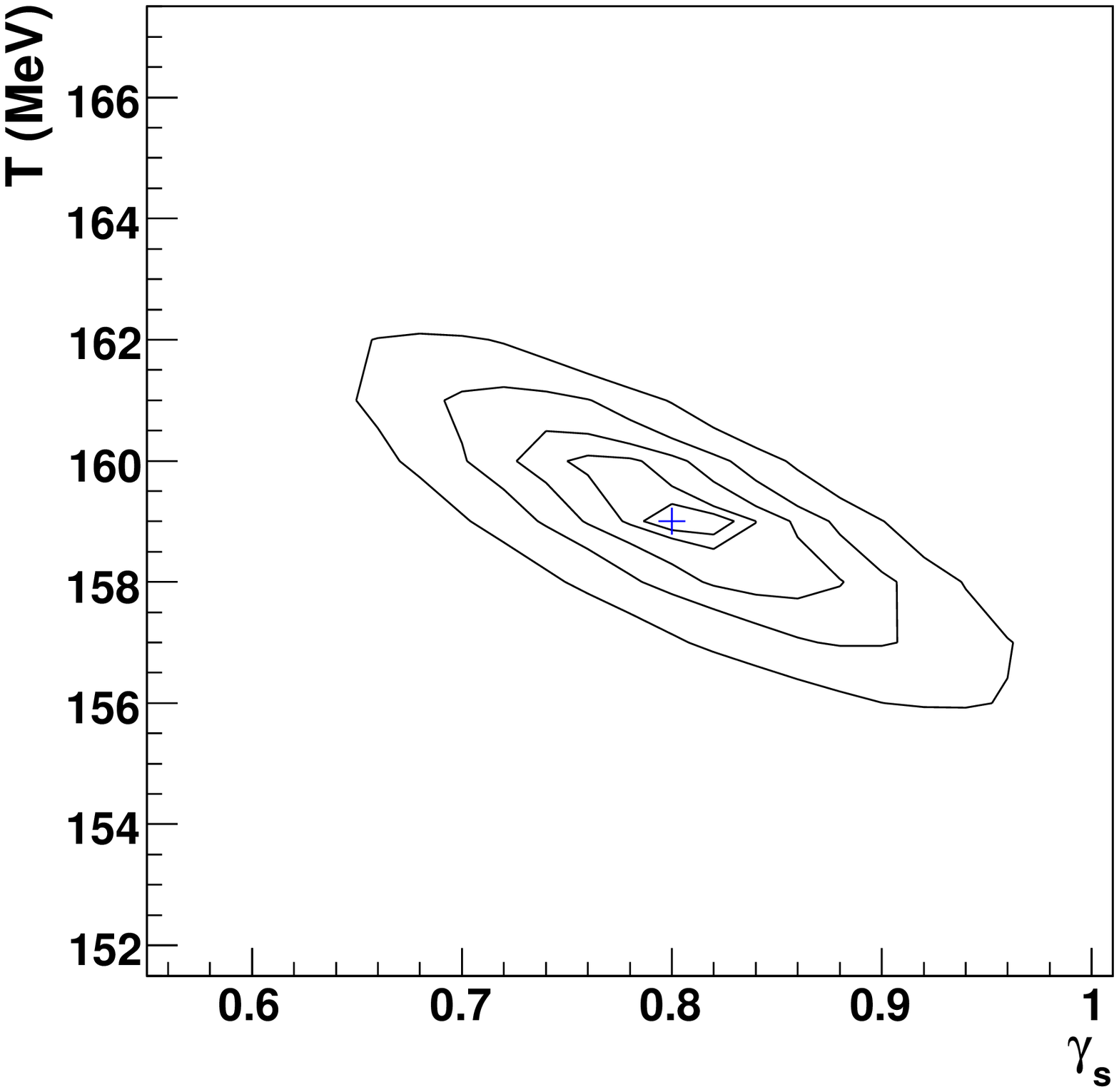}
\end{minipage}
\end{tabular}
\vspace{-0.9cm}
\caption{$\chi^2$ contour lines in temperature and volume (left panel) and temperature and $\gamma_s$ (right panel) space for $\sqrt{s}$=10 GeV. The contours correspond to $\chi^2_{min}$ left: $+10\%$, $+20\%$, $+50\%$, $+100\%$ right: $+1\%$, $+2\%$, $+5\%$, $+10\%$, $+20\%$. The best fit values are indicated by the blue crosses.}
\label{fig:10GeVcont1}
\end{figure}

\begin{figure}[H]
\begin{tabular}{cc} 
\begin{minipage}{.49\textwidth}
\centering\includegraphics[width=1.05\textwidth]{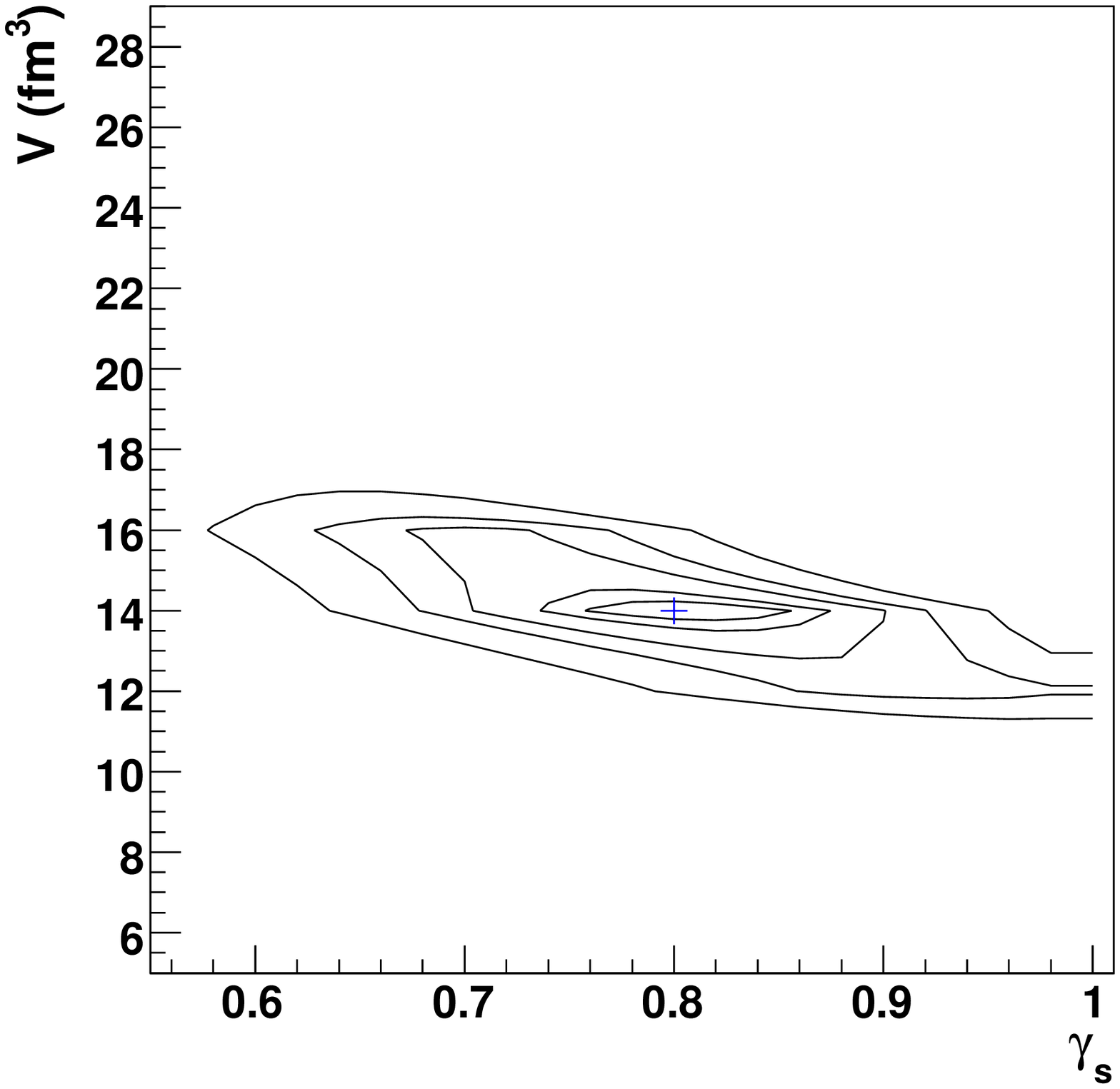}
\end{minipage} & \begin{minipage}{.49\textwidth}
\end{minipage}
\end{tabular}
\vspace{-0.5cm}
\caption{$\chi^2$ contour lines in volume and $\gamma_s$ space for $\sqrt{s}$=10 GeV. The contours correspond to $\chi^2_{min} +5\%$, $+10\%$, $+20\%$, $+30\%$, $+50\%$. The best fit value is indicated by the blue cross.}
\label{fig:10GeVcont2}
\end{figure}

\subsection{29-35 GeV}
\vspace{-0.5cm}
\begin{figure}[H]
\begin{center}
\label{fig:35GeV}
\includegraphics[width=0.8\textwidth,height=9cm]{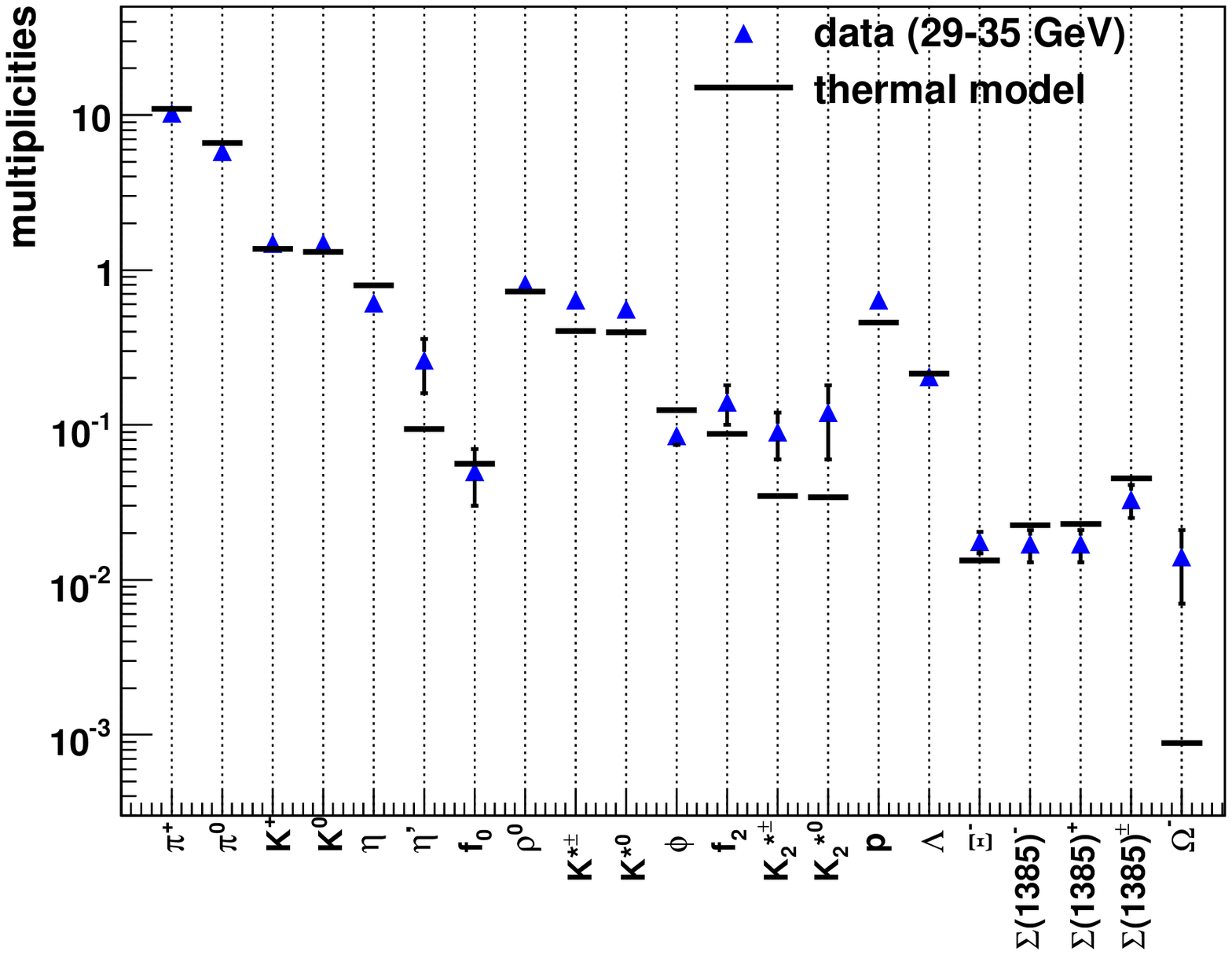} 
\vspace{-0.3cm}
\caption{Comparison between thermal model predictions and experimental particle multiplicities for $e^+e^-$ collisions at $\sqrt{s}$=29-35 GeV.}
\end{center}
\end{figure}

\begin{figure}[H]
\begin{tabular}{cc} 
\begin{minipage}{.49\textwidth}
\centering\includegraphics[width=1.05\textwidth]{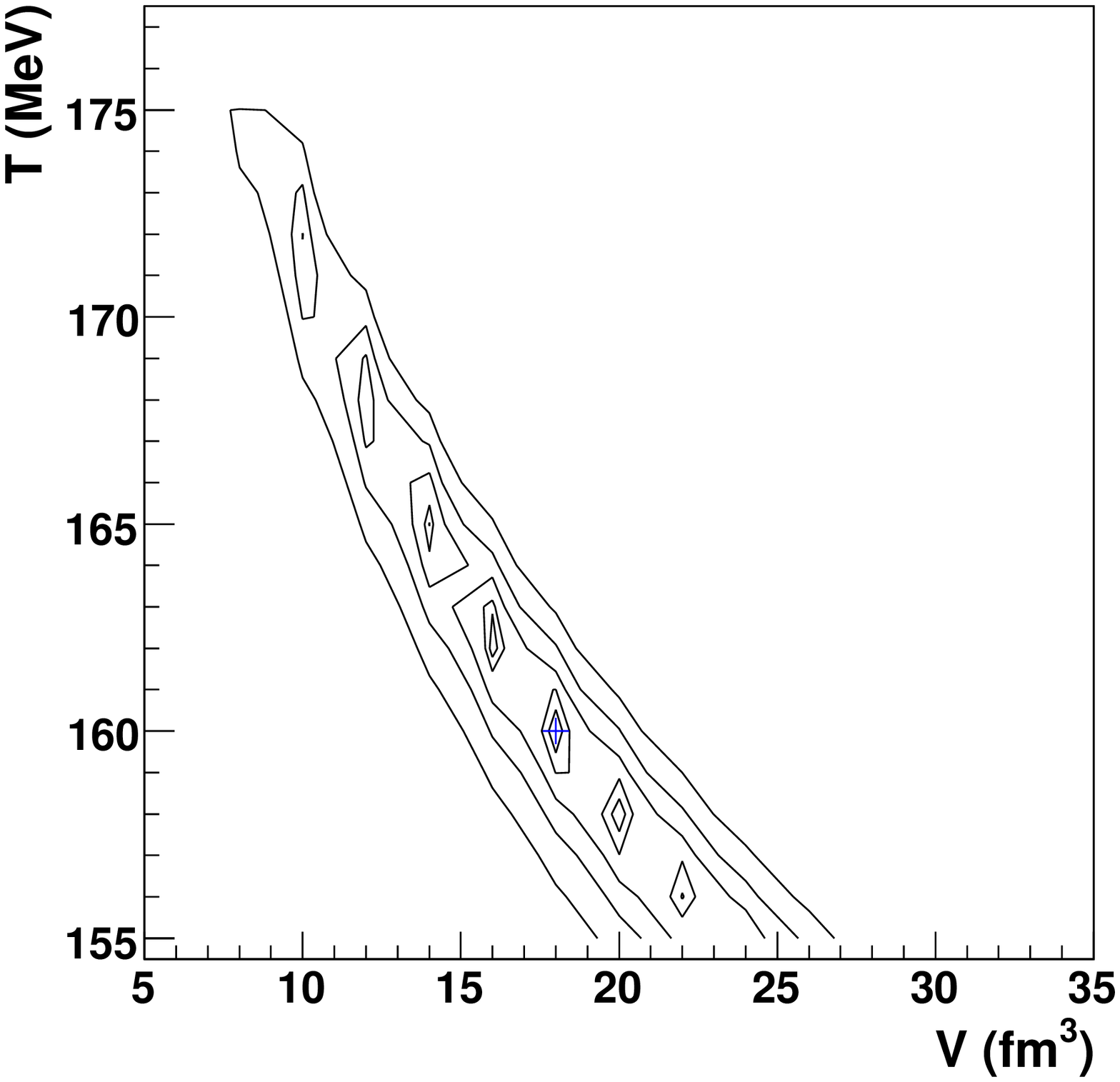}
\end{minipage} & \begin{minipage}{.49\textwidth}
\centering\includegraphics[width=1.05\textwidth]{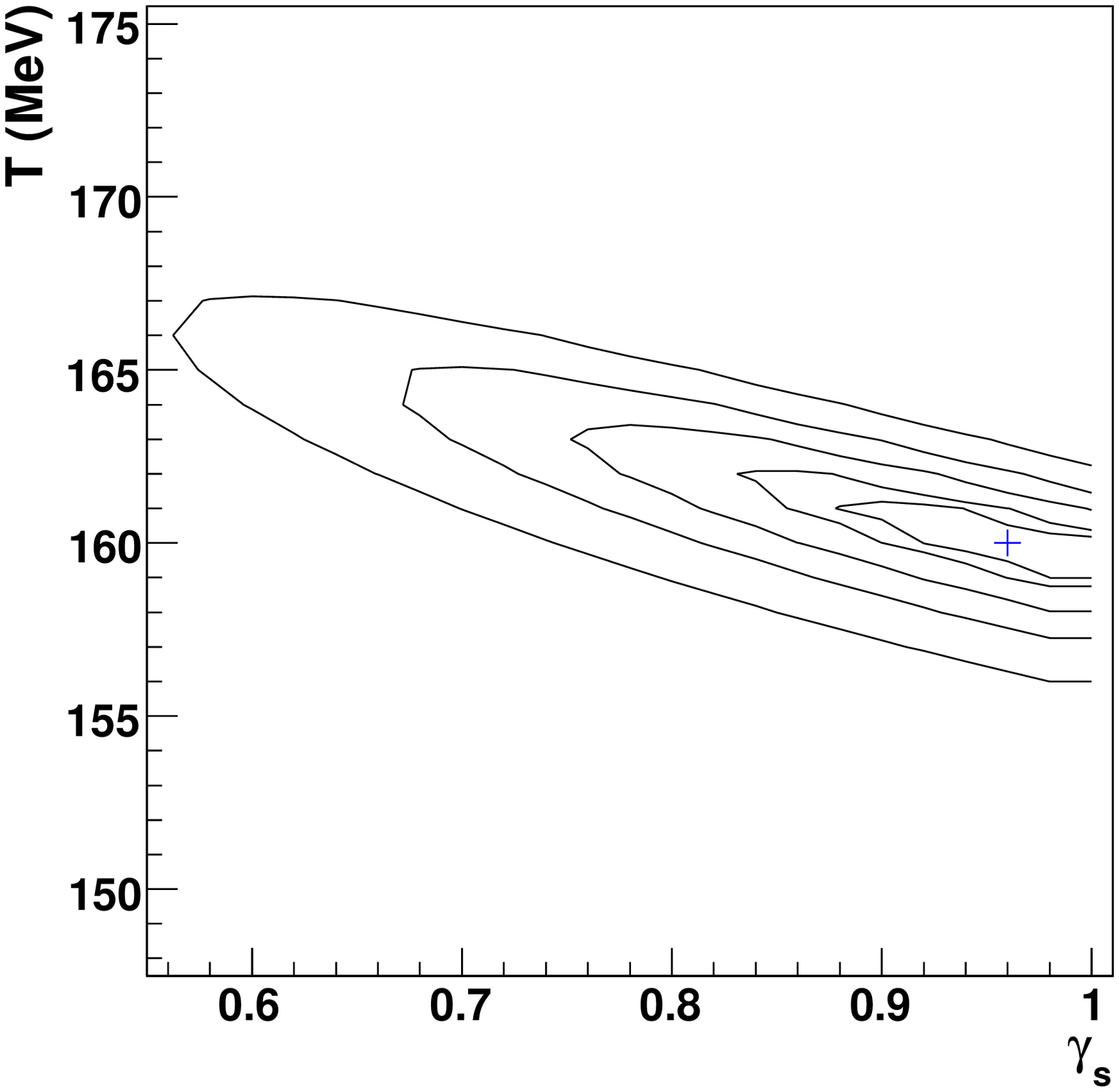}
\end{minipage}
\end{tabular}
\caption{$\chi^2$ contour lines in temperature and volume (left panel) and temperature and $\gamma_s$ (right panel) space for $\sqrt{s}$=29-35 GeV. The contours correspond to $\chi^2_{min}+10\%$, $+20\%$, $+50\%$, $+100\%$, $+200\%$. The best fit values are indicated by the blue crosses.}
\label{fig:35GeVcont1}
\end{figure}
\begin{figure}[H]
\begin{tabular}{cc} 
\begin{minipage}{.49\textwidth}
\centering\includegraphics[width=1.05\textwidth]{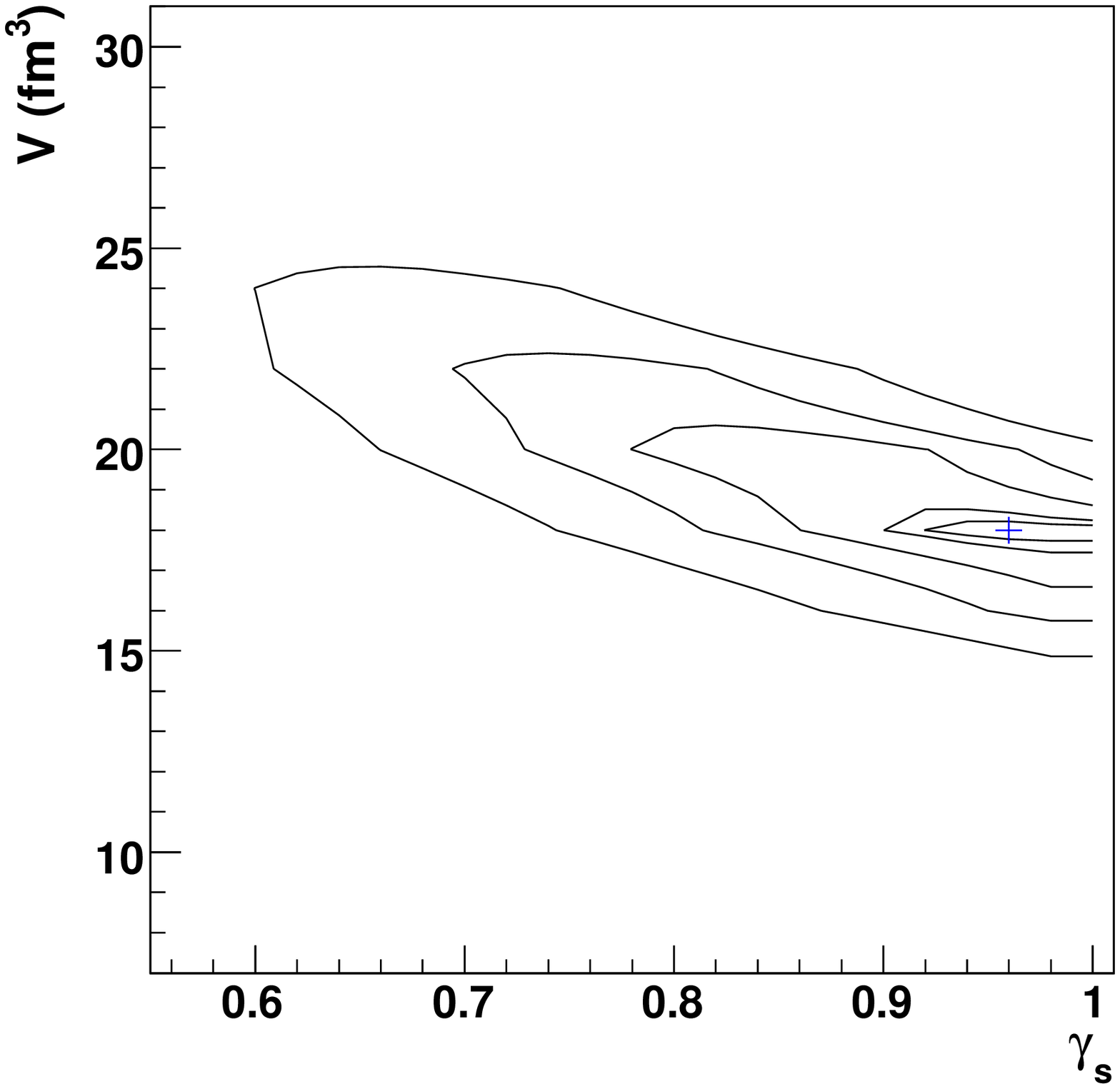}
\end{minipage} & \begin{minipage}{.49\textwidth}
\end{minipage}
\end{tabular}
\caption{$\chi^2$ contour lines in volume and $\gamma_s$ space for $\sqrt{s}$=29-35 GeV. The contours correspond to $\chi^2_{min}+10\%$, $+20\%$, $+50\%$, $+100\%$, $+200\%$. The best fit value is indicated by the blue cross.}
\label{fig:35GeVcont2}
\end{figure}

\subsection{91 GeV (uncorrected)}

\begin{figure}[H]
\begin{center}
\vspace{-1.3cm}
\includegraphics[width=0.8\textwidth,height=9cm]{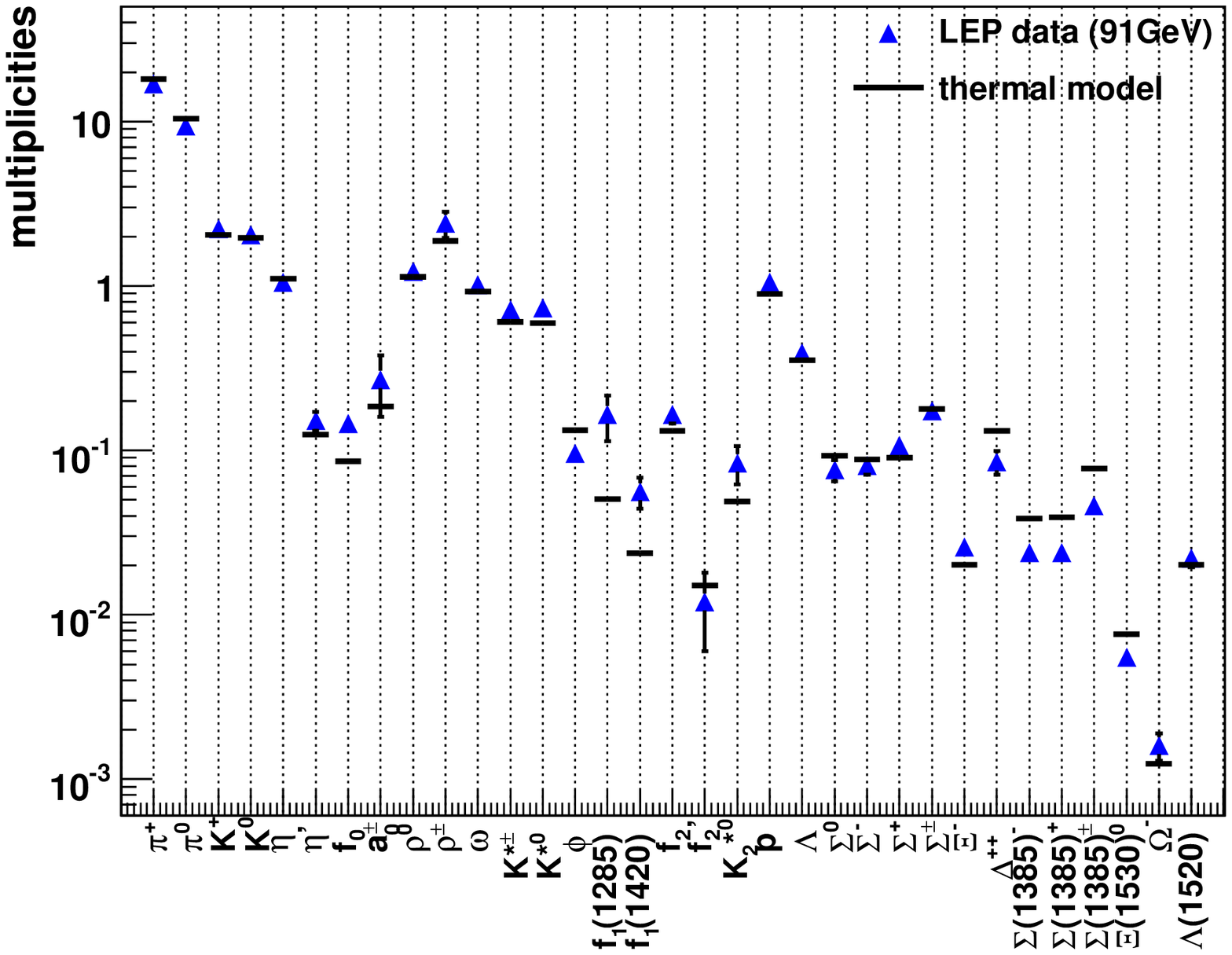}
\vspace{-0.5cm}
\caption{Comparison between thermal model predictions and experimental particle multiplicities for $e^+e^-$ collisions at $\sqrt{s}$=91 GeV.}
\end{center}
\end{figure}
\vspace{-2cm}
\begin{figure}[H]
\begin{tabular}{cc} 
\begin{minipage}{.49\textwidth}
\centering\includegraphics[width=1.05\textwidth]{chi2_tv91}
\end{minipage} & \begin{minipage}{.49\textwidth}
\centering\includegraphics[width=1.05\textwidth]{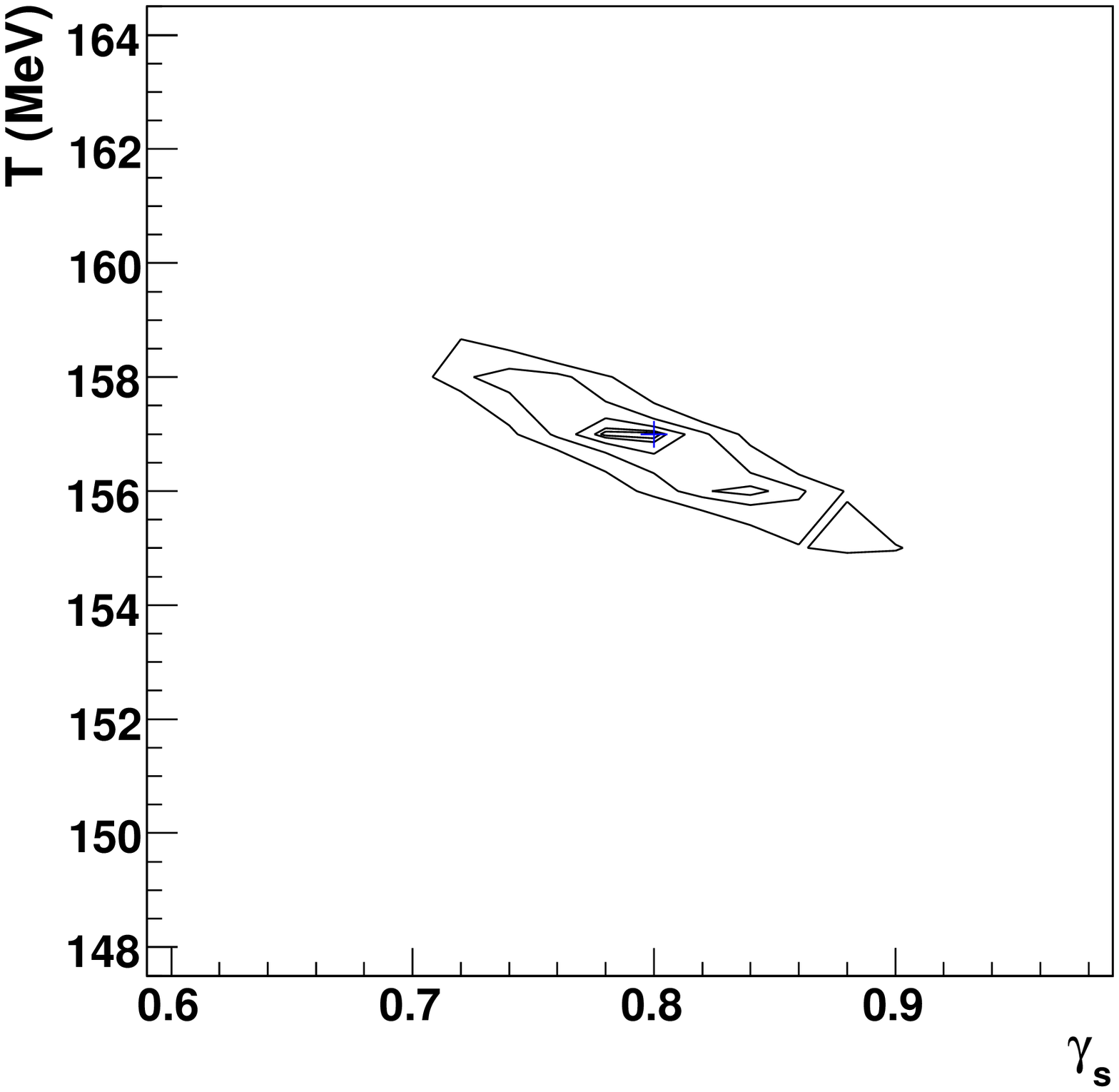}
\end{minipage}
\end{tabular}
\vspace{-0.5cm}
\caption{$\chi^2$ contour lines in temperature and volume (left panel) and temperature and $\gamma_s$ (right panel) space for $\sqrt{s}$=91 GeV. The contours correspond to left $\chi^2_{min}$ left: $+10\%$, $+20\%$, $+50\%$, $+100\%$, right: $+1\%$, $+2\%$, $+5\%$, $+10\%$, $+20\%$. The best fit values are indicated by the blue crosses.}
\label{fig:91GeVcont1}
\end{figure}

\begin{figure}[H]
\begin{tabular}{cc} 
\begin{minipage}{.49\textwidth}
\centering\includegraphics[width=1.05\textwidth]{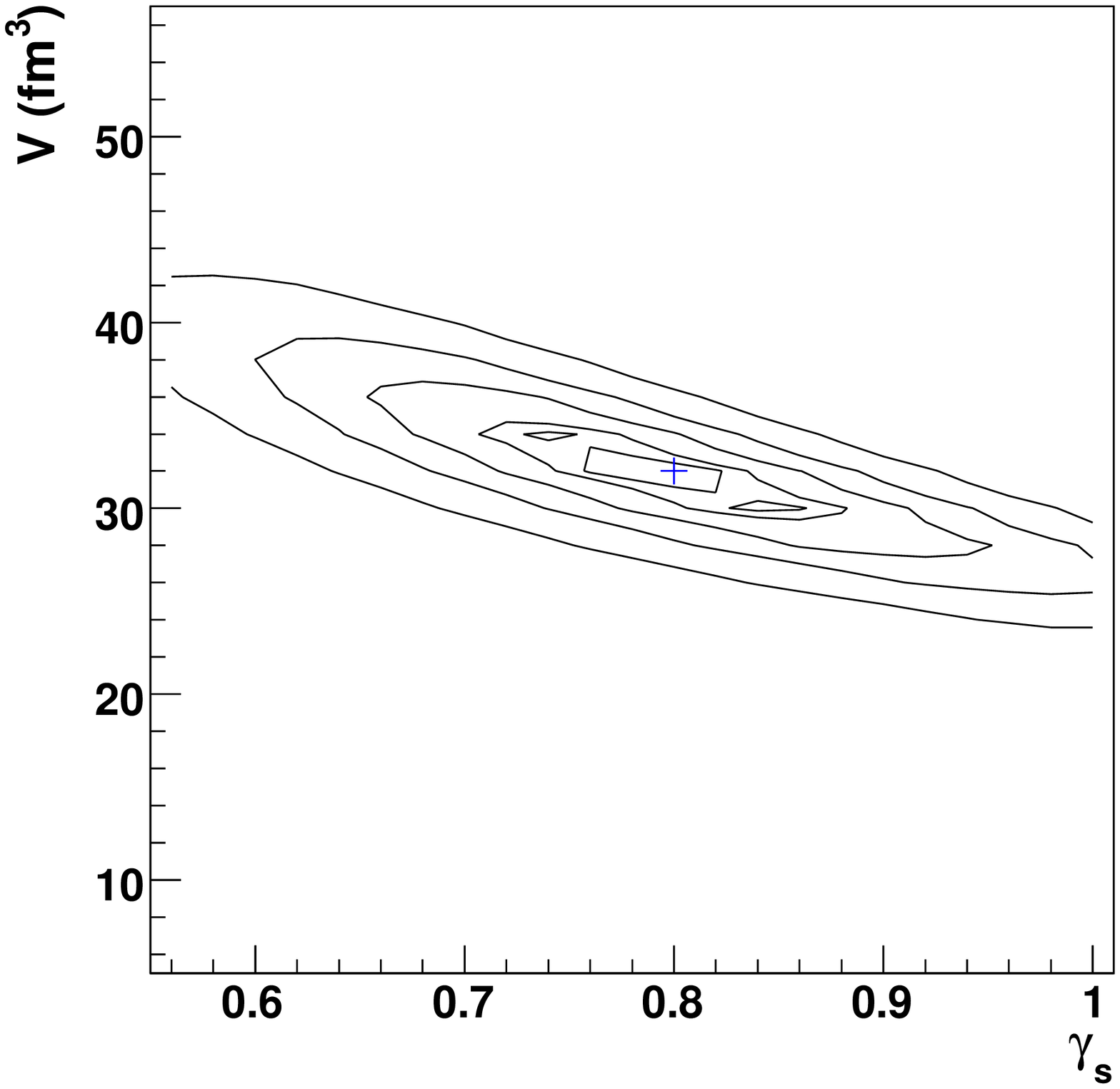}
\end{minipage} & \begin{minipage}{.49\textwidth}
\end{minipage}
\end{tabular}
\vspace{-0.5cm}
\caption{$\chi^2$ contour lines in volume and $\gamma_s$ space for $\sqrt{s}$=91 GeV. The contours correspond to $\chi^2_{min} +5\%$, $+10\%$, $+20\%$, $+30\%$, $+50\%$. The best fit value is indicated by the blue cross.}
\label{fig:91GeVcont2}
\end{figure}

\subsection{91 GeV (c, b corrected)}

\begin{figure}[H]
\begin{center}
\vspace{-0.5cm}
\includegraphics[width=0.8\textwidth,height=9cm]{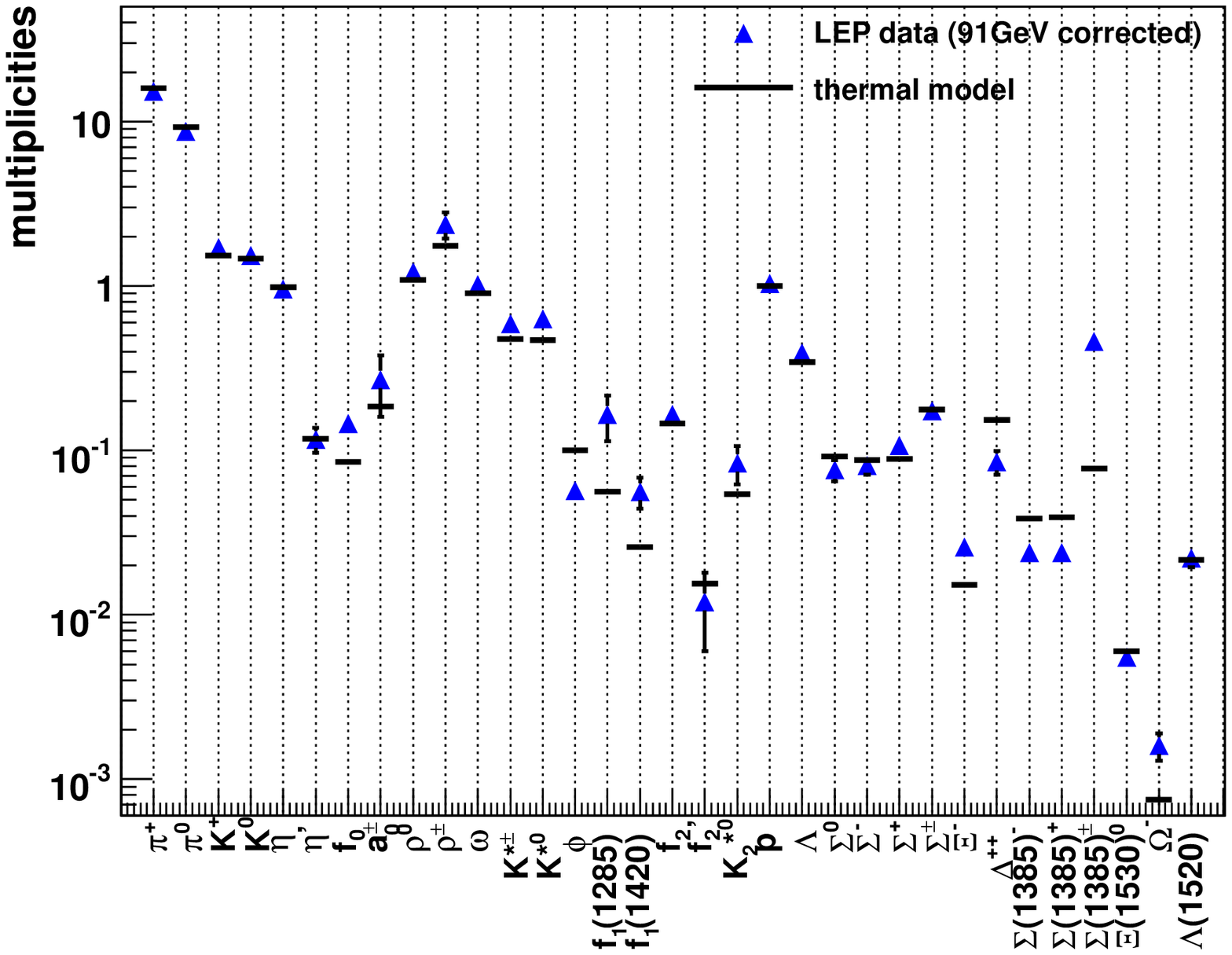}
\vspace{-0.5cm}
\caption{Comparison between thermal model predictions and experimental particle multiplicities for $e^+e^-$ collisions at $\sqrt{s}$=91 GeV.}
\end{center}
\end{figure}

\begin{figure}[H]
\begin{tabular}{cc} 
\begin{minipage}{.49\textwidth}
\centering\includegraphics[width=1.05\textwidth]{chi2_tv91GeVc}
\end{minipage} & \begin{minipage}{.49\textwidth}
\centering\includegraphics[width=1.05\textwidth]{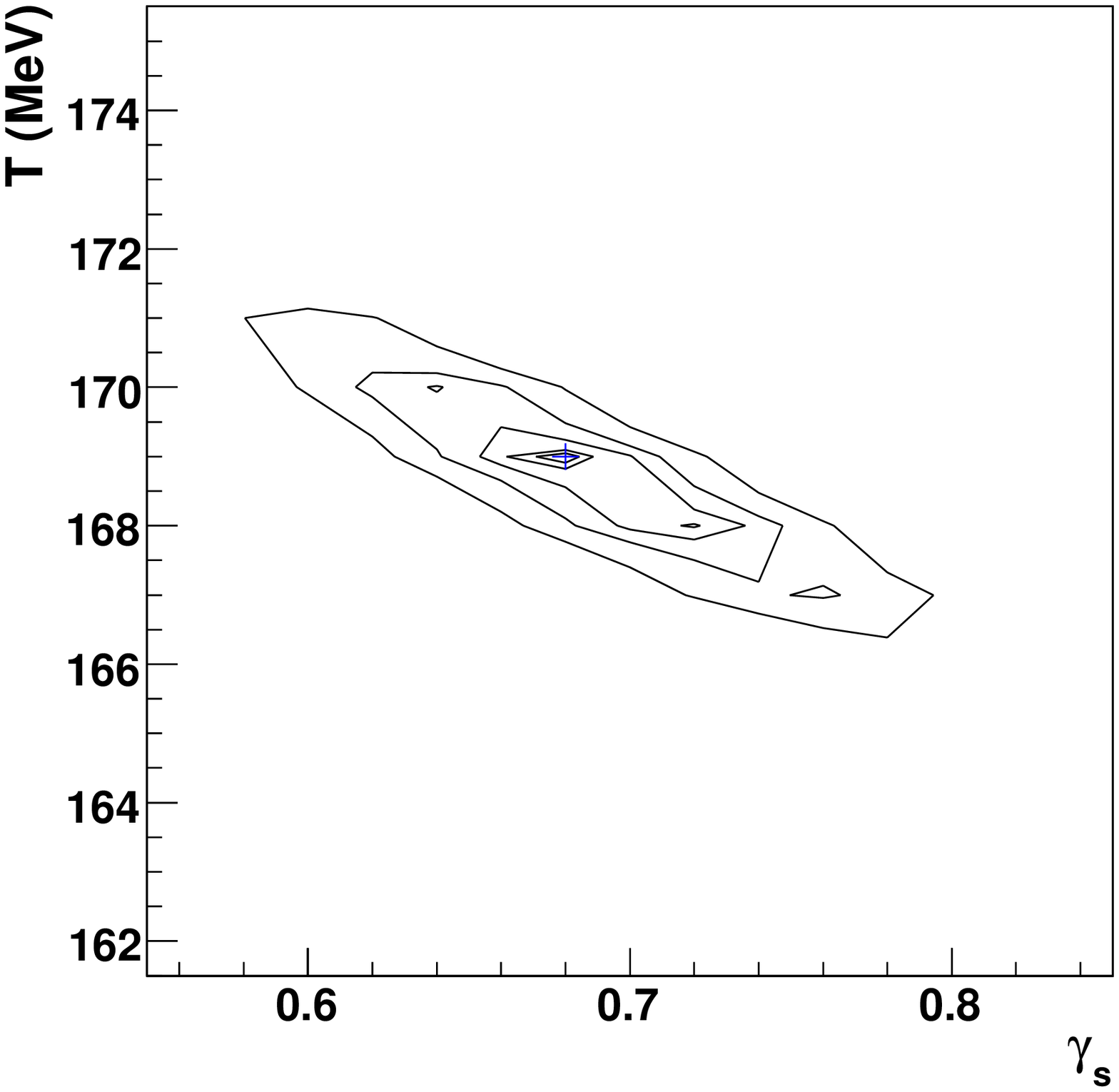}
\end{minipage}
\end{tabular}
\vspace{-0.5cm}
\caption{$\chi^2$ contour lines in temperature and volume (left panel) and temperature and $\gamma_s$ (right panel) space for $\sqrt{s}$=91 GeV. The contours correspond to $\chi^2_{min} +10\%$, $+20\%$, $+50\%$, $+100\%$, $+200\%$. The best fit values are indicated by the blue crosses.}
\label{fig:91GeVcorcont1}
\end{figure}

\begin{figure}[H]
\begin{tabular}{cc} 
\begin{minipage}{.49\textwidth}
\centering\includegraphics[width=1.05\textwidth]{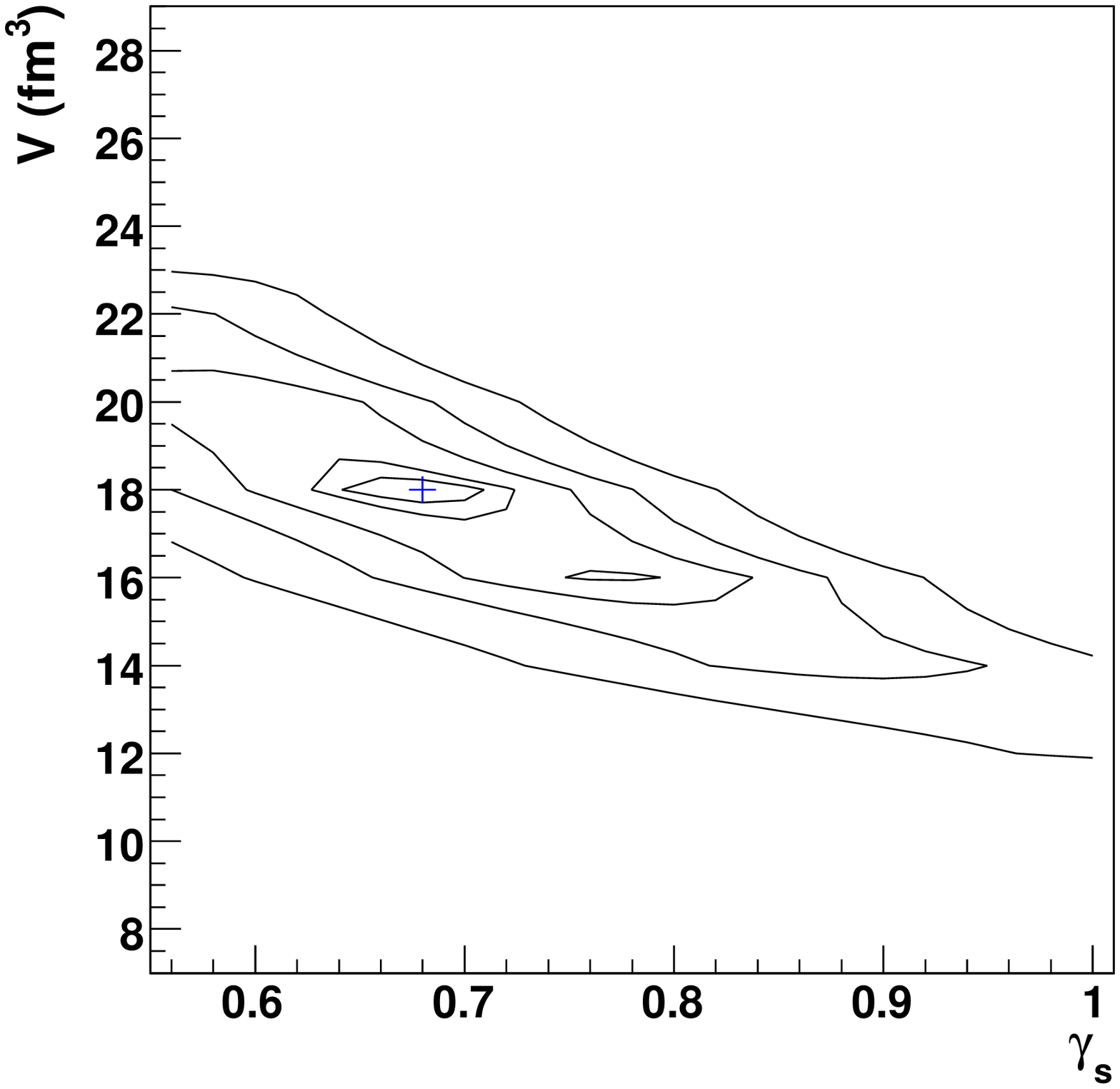}
\end{minipage} & \begin{minipage}{.49\textwidth}
\end{minipage}
\end{tabular}
\caption{$\chi^2$ contour lines in volume and $\gamma_s$ space for $\sqrt{s}$=91 GeV. The contours correspond to $\chi^2_{min} +5\%$, $+10\%$, $+20\%$, $+30\%$, $+50\%$. The best fit value is indicated by the blue cross.}
\label{fig:91GeVcorcont2}
\end{figure}

\subsection{130-200 GeV}

\begin{figure}[H]
\vspace{-0.5cm}
\begin{center}
\centering\includegraphics[width=0.8\textwidth,height=9cm]{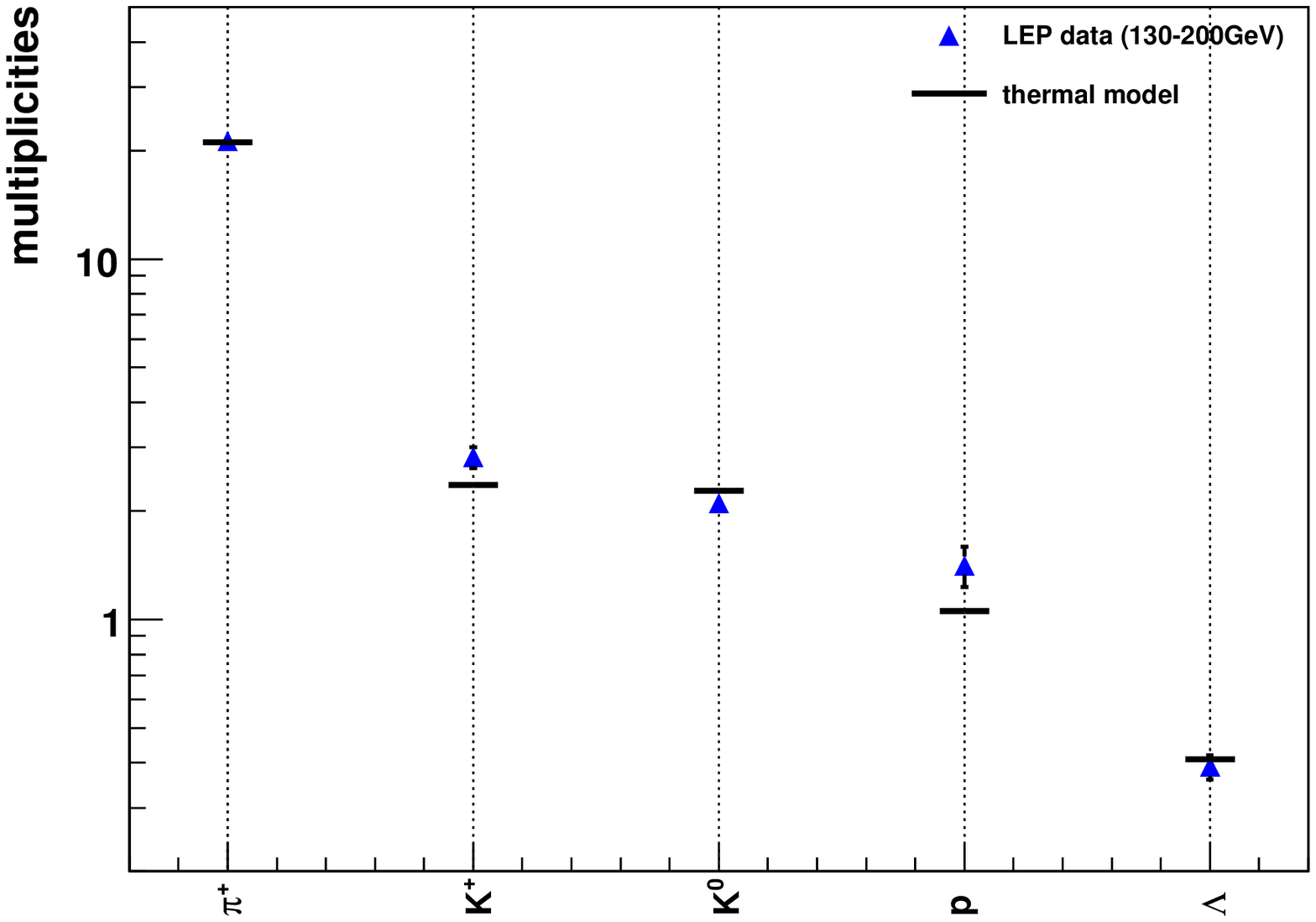}
\vspace{-1cm}
\caption{Comparison between thermal model predictions and experimental particle multiplicities for $e^+e^-$ collisions at $\sqrt{s}$=130-200 GeV.}
\label{fig:200GeV}
\end{center}
\end{figure}
\vspace{-1cm}
\begin{figure}[H]
\begin{tabular}{cc} 
\begin{minipage}{.49\textwidth}
\centering\includegraphics[width=1.05\textwidth]{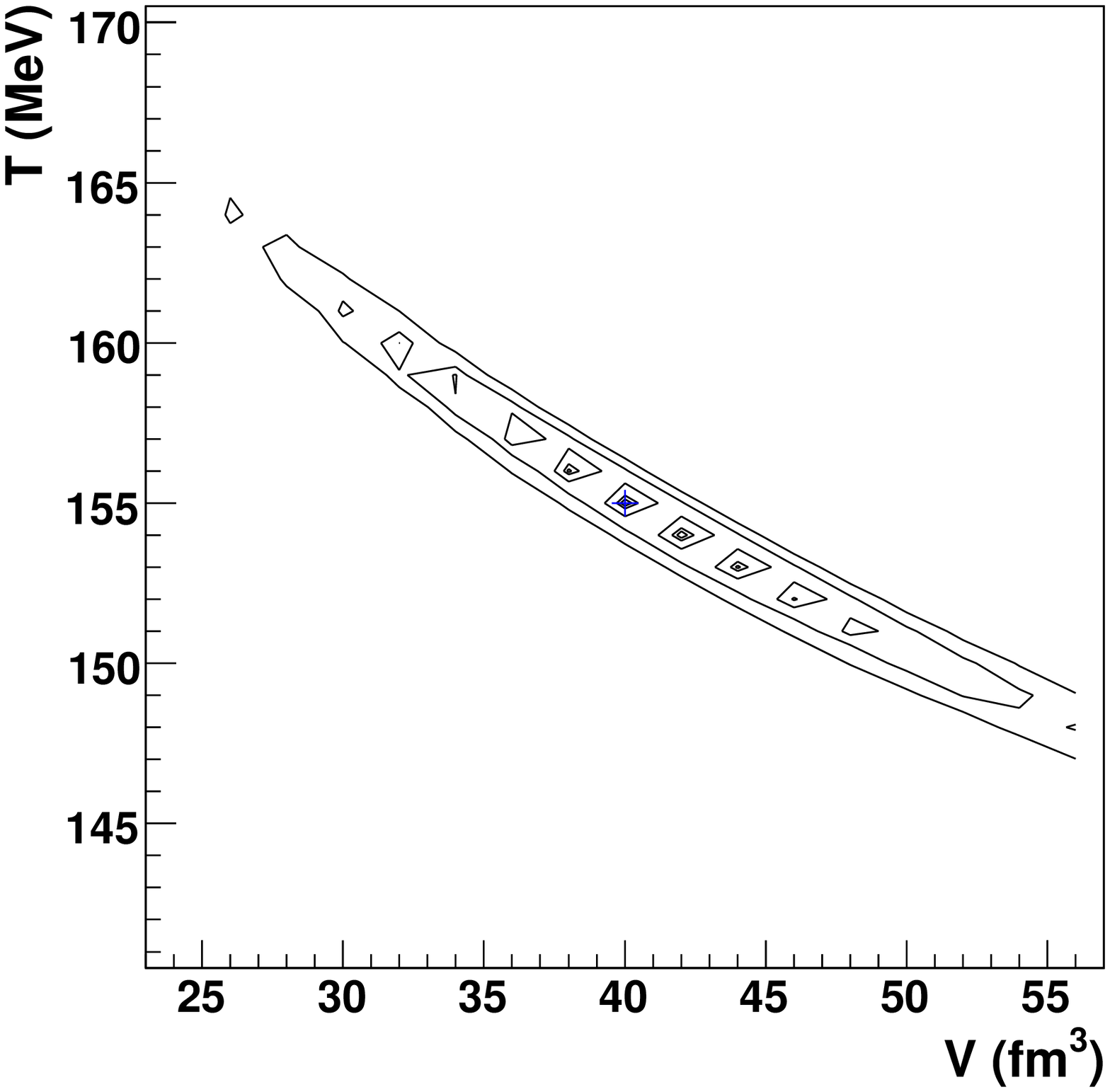}
\end{minipage} & \begin{minipage}{.49\textwidth}
\centering\includegraphics[width=1.05\textwidth]{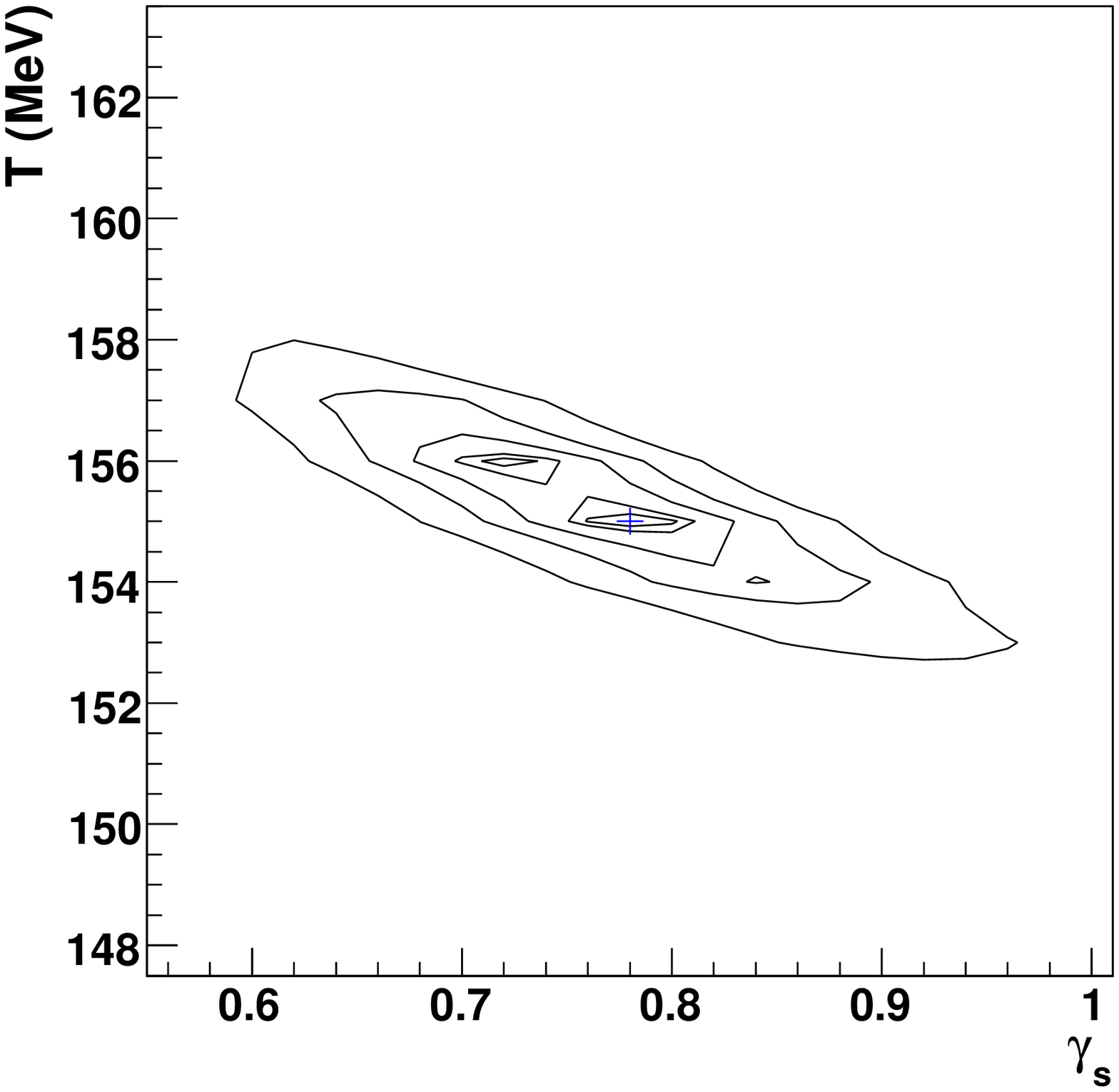}
\end{minipage}
\end{tabular}
\caption{$\chi^2$ contour lines in temperature and volume (left panel) and temperature and $\gamma_s$ (right panel) space for $\sqrt{s}$=130-200 GeV. The contours correspond to $\chi^2_{min} +10\%$, $+20\%$, $+50\%$, $+100\%$, $+200\%$. The best fit values are indicated by the blue crosses.}
\label{fig:200GeVcont1}
\end{figure}

\begin{figure}[H]
\begin{tabular}{cc} 
\begin{minipage}{.49\textwidth}
\centering\includegraphics[width=1.05\textwidth]{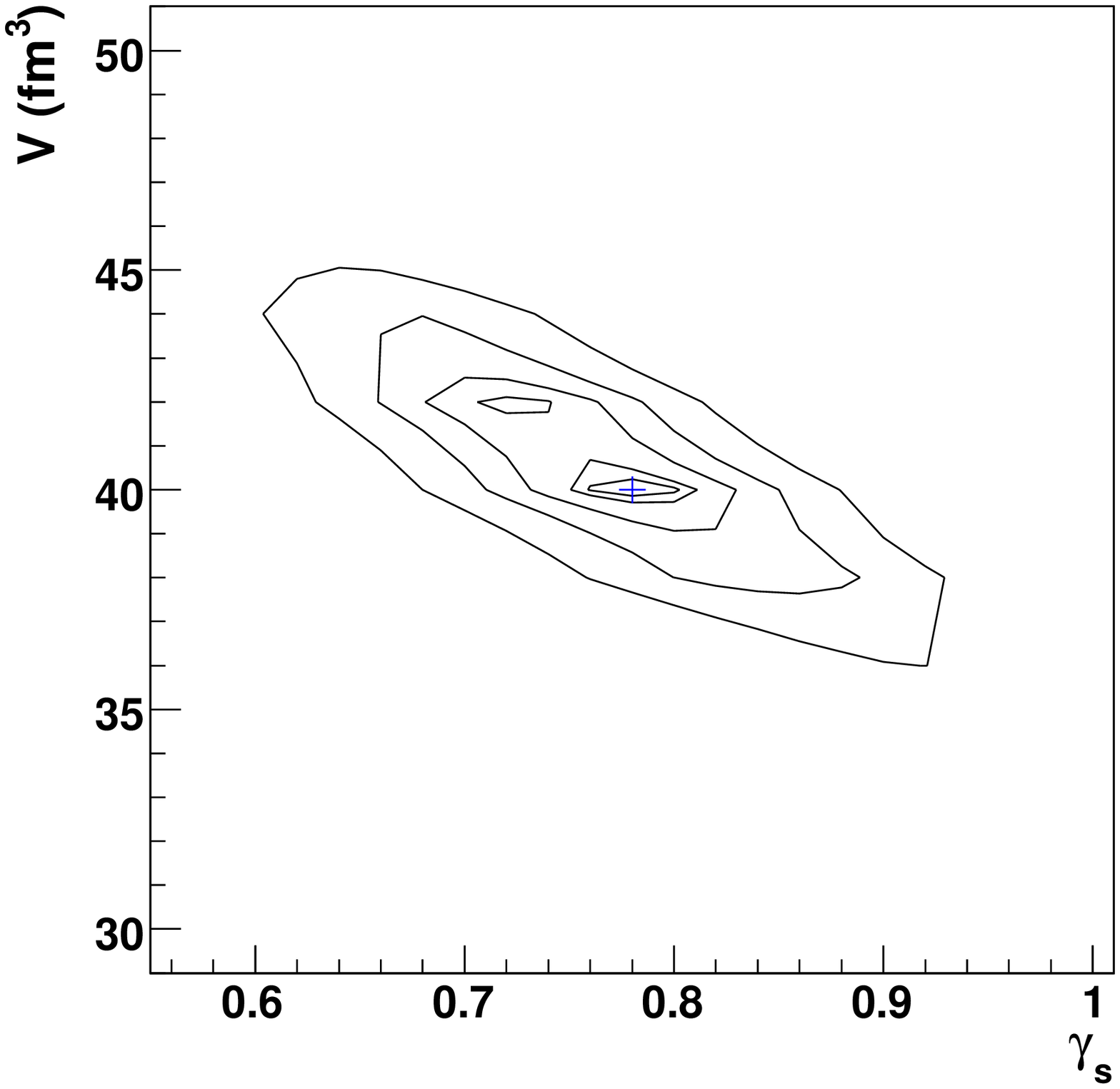}
\end{minipage} & \begin{minipage}{.49\textwidth}
\end{minipage}
\end{tabular}
\caption{$\chi^2$ contour lines in volume and $\gamma_s$ space for $\sqrt{s}$=130-200 GeV. The contours correspond to $\chi^2_{min} +10\%$, $+20\%$, $+50\%$, $+100\%$, $+200\%$. The best fit value is indicated by the blue cross.}
\label{fig:200GeVcont2}
\end{figure}

\chapter{Data source \cite{Yao:2006px}}
\label{ap:data}

\section{Heavy particles}
\vspace{-0.5cm}
\begin{footnotesize}\begin{minipage}{\textwidth}
\begin{longtable}{ccccc}
\hline
\hline
\textbf{particle} & \textbf{$\sqrt{s} \approx 10$ GeV} & \textbf{$\sqrt{s}$=29-35 GeV} & \textbf{$\sqrt{s}$=91 GeV} & \textbf{$\sqrt{s}$=130-200 GeV}\\
\hline
\endfirsthead
\endhead
\hline
\multicolumn{5}{c}{\textbf{see also next page}}\\
\hline
\endfoot
\endlastfoot
\multicolumn{5}{l}{\textbf{Pseudoscalar mesons:}}\\
$D^{+}$ & 0.194 $\pm$ 0.019 & 0.17 $\pm$ 0.03 & 0.175 $\pm$ 0.016 & ---\\
$D^{0}$ & 0.446 $\pm$ 0.032 & 0.45 $\pm$ 0.07 & 0.454 $\pm$ 0.030 & ---\\
$D^{+}_s$ & 0.063 $\pm$ 0.014 & 0.45 $\pm$ 0.20 & 0.131 $\pm$ 0.021 & ---\\
$B^{(1)}$ & --- & --- & 0.343 $\pm$ 0.032 & ---\\
$B_s^+$ & --- & --- & 0.057 $\pm$ 0.013 & ---\\
\hline
\multicolumn{5}{l}{\textbf{Vector mesons:}}\\
$D^{*}(2010)^+$ & 0.177 $\pm$ 0.022 & 0.43 $\pm$ 0.07 & 0.1937 $\pm$ 0.0057 & ---\\
$D^{*}(2007)^{0}$ & 0.168 $\pm$ 0.019 & 0.27 $\pm$ 0.11 & --- & ---\\
$D_s^{*}(2112^+)$ & 0.048 $\pm$ 0.014 & --- & 0.101 $\pm$ 0.048 & ---\\
$B^{*(2)}$ & --- & --- & 0.288 $\pm$ 0.026 & ---\\
$J/\Psi(1S)$ & 0.0005 $\pm$ 0.00005 & --- & 0.0052 $\pm$ 0.0004 & ---\\
$\Psi(2S)$ & --- & --- & 0.0023 $\pm$ 0.0004 & ---\\
$\Upsilon(1S)$ & --- & --- & 0.00014 $\pm$ 0.00007 & ---\\
\hline
\multicolumn{5}{l}{\textbf{Pseudovector mesons:}}\\
$\chi_{c1}(3510)$ & --- & --- & 0.0041 $\pm$ 0.0011 & ---\\
\hline
\multicolumn{5}{l}{\textbf{Tensor mesons:}}\\
$B^{**(2)}$ & --- & --- & 0.118 $\pm$ 0.024 & ---\\
$D_{s1}^{\pm}$ & --- & --- & 0.0052 $\pm$ 0.0011 & ---\\
$D_{s2}^{*\pm}$ & --- & --- & 0.0083 $\pm$ 0.0031 & ---\\
\hline
\multicolumn{5}{l}{\textbf{Baryons:}}\\
$\Lambda^+_c$ & 0.074 $\pm$ 0.031 & 0.11 $\pm$ 0.05 & 0.078 $\pm$ 0.017 & ---\\
$\Lambda^0_b$ & --- & --- & 0.031 $\pm$ 0.016 & ---\\
$\Sigma^{++}_c,\Sigma_c^0$ & 0.014 $\pm$ 0.007 & --- & --- & ---\\
\hline
\hline
\caption{Average heavy particle multiplicities per hadronic $e^{+}e^{-}$ annihilation event at $\sqrt{s} \approx$ 10, 29-35, 91, and 130-200 GeV. The rates given include decay products from resonances with $c\tau < 10$ cm, and include the corresponding anti-particle state \cite{Yao:2006px}. (1) Contains $B^+$ and $B^0$, (2) Any charged state (i.e. $B^{*+}, B^{*0}$ and $B^*_s$)}
\end{longtable}
\end{minipage}\end{footnotesize}

\section{Light particles}

\begin{footnotesize}\begin{minipage}{\textwidth}
\begin{longtable}{ccccc}
\hline
\hline
\textbf{particle} & \textbf{$\sqrt{s} \approx 10$ GeV} & \textbf{$\sqrt{s}$=29-35 GeV} & \textbf{$\sqrt{s}$=91 GeV} & \textbf{$\sqrt{s}$=130-200 GeV}\\
\hline
\endfirsthead
\endhead
\hline
\multicolumn{5}{c}{\textbf{see also next page}}\\
\hline
\endfoot
\endlastfoot
\multicolumn{5}{l}{\textbf{Pseudoscalar mesons:}}\\
$\pi^{\pm}$ & 6.6 $\pm$ 0.2 & 10.3 $\pm$ 0.4 & 17.02 $\pm$ 0.19 & 21.24 $\pm$ 0.39\\
$\pi^{0}$ & 3.2 $\pm$ 0.3 & 5.83 $\pm$ 0.28 & 9.42 $\pm$ 0.32 & ---\\
$K^{\pm}$ & 0.90 $\pm$ 0.04 & 1.48 $\pm$ 0.09 & 2.228 $\pm$ 0.059 & 2.82 $\pm$ 0.19\\
$K^{0}$ & 0.91 $\pm$ 0.05 & 1.48 $\pm$ 0.07 & 2.049 $\pm$ 0.026 & 2.10 $\pm$ 0.12\\
$\eta$ & 0.20 $\pm$ 0.04 & 0.61 $\pm$ 0.07 & 1.049 $\pm$ 0.080 & ---\\
$\eta'(958)$ & 0.03 $\pm$ 0.01 & 0.26 $\pm$ 0.10 & 0.152 $\pm$ 0.020 & ---\\
\hline
\multicolumn{5}{l}{\textbf{Scalar mesons:}}\\
$f_0(980)$ & 0.024 $\pm$ 0.006 & 0.05 $\pm$ 0.02 & 0.146 $\pm$ 0.012 & ---\\
$a_0^{\pm}$ & --- & --- & 0.27 $\pm$ 0.11 & ---\\
\hline
\multicolumn{5}{l}{\textbf{Vector mesons:}}\\
$\rho(770)^0$ & 0.35 $\pm$ 0.04 & 0.81 $\pm$ 0.08 & 1.231 $\pm$ 0.098 & ---\\
$\rho(770)^{\pm}$ & --- & --- & 2.049 $\pm$ 0.026 & ---\\
$\omega(782)$ & 0.30 $\pm$ 0.08 & --- & 1.016 $\pm$ 0.065 & ---\\
$K^{*}(892)^{\pm}$ & 0.27 $\pm$ 0.03 & 0.64 $\pm$ 0.05 & 0.715 $\pm$ 0.059 & ---\\
$K^{*}(892)^{0}$ & 0.29 $\pm$ 0.03 & 0.56 $\pm$ 0.06 & 0.738 $\pm$ 0.024 & ---\\
$\phi(1020)$ & 0.044 $\pm$ 0.003 & 0.085 $\pm$ 0.011 & 0.0963 $\pm$ 0.0032 & ---\\
\hline
\multicolumn{5}{l}{\textbf{Pseudovector mesons:}}\\
$f_1(1285)$ & --- & --- & 0.165 $\pm$ 0.051 & ---\\
$f_1(1420)$ & --- & --- & 0.056 $\pm$ 0.012 & ---\\
\hline
\multicolumn{5}{l}{\textbf{Tensor mesons:}}\\
$f_2(1270)$ & 0.09 $\pm$ 0.02 & 0.14 $\pm$ 0.04 & 0.166 $\pm$ 0.020 & ---\\
$f'_2(1525)$ & --- & --- & 0.012 $\pm$ 0.006 & ---\\
$K^{*}_{2}(1430)^{\pm}$ & --- & 0.09 $\pm$ 0.03 & --- & ---\\
$K^{*}_{2}(1430)^{0}$ & --- & 0.12 $\pm$ 0.06 & 0.084 $\pm$ 0.022 & ---\\
\hline
\multicolumn{5}{l}{\textbf{Baryons:}}\\
$p$ & 0.253 $\pm$ 0.016 & 0.640 $\pm$ 0.050 & 1.050 $\pm$ 0.032 & 1.41 $\pm$ 0.18\\
$\Lambda$ & 0.080 $\pm$ 0.007 & 0.205 $\pm$ 0.010 & 0.3915 $\pm$ 0.0065 & 0.39 $\pm$ 0.03\\
$\Sigma^0$ & 0.023 $\pm$ 0.008 & --- & 0.076 $\pm$ 0.011 & ---\\
$\Sigma^-$ & --- & --- & 0.081 $\pm$ 0.010 & ---\\
$\Sigma^+$ & --- & --- & 0.107 $\pm$ 0.011 & ---\\
$\Sigma^{\pm}$ & --- & --- & 0.174 $\pm$ 0.009 & ---\\
$\Xi^-$ & 0.0059 $\pm$ 0.0007 & 0.0176 $\pm$ 0.0027 & 0.0258 $\pm$ 0.001 & ---\\
$\Delta(1232)^{++}$ & 0.040 $\pm$ 0.010 & --- & 0.085 $\pm$ 0.014 & ---\\
$\Sigma(1385)^-$ & 0.006 $\pm$ 0.002 & 0.017 $\pm$ 0.004 & 0.0240 $\pm$ 0.0017 & ---\\
$\Sigma(1385)^+$ & 0.005 $\pm$ 0.001 & 0.017 $\pm$ 0.004 & 0.0239 $\pm$ 0.0015 & ---\\
$\Sigma(1385)^{\pm}$ & 0.0106 $\pm$ 0.0020 & 0.033 $\pm$ 0.008 & 0.0462 $\pm$ 0.0028 & ---\\
$\Xi(1530)^0$ & 0.0015 $\pm$ 0.0006 & --- & 0.0055 $\pm$ 0.0005 & ---\\
$\Omega^-$ & 0.0007 $\pm$ 0.0004 & 0.014 $\pm$ 0.007 & 0.0016 $\pm$ 0.0003 & ---\\
$\Lambda(1530)$ & 0.008 $\pm$ 0.002 & --- & 0.0222 $\pm$ 0.0027 & ---\\
\hline
\hline
\caption{Average hadron multiplicities per hadronic $e^{+}e^{-}$ annihilation event at $\sqrt{s} \approx$10, 29-35, 91, and 130-200 GeV. The rates given include decay products from resonances with $c\tau < 10$ cm, and include the corresponding anti-particle state \cite{Yao:2006px}.}
\end{longtable}
\end{minipage}\end{footnotesize}

\end{appendix}

\newpage

\markboth{List of Figures}{List of Figures}
\phantomsection
\addcontentsline{toc}{chapter}{List of Figures}
\listoffigures
\markboth{List of Tables}{List of Tables}
\phantomsection
\addcontentsline{toc}{chapter}{List of Tables}
\listoftables

\newpage

\addcontentsline{toc}{chapter}{References}
\renewcommand{\bibname}{References}

\newpage

\section*{Acknowledgments} 

I would like to thank everybody who supported me during the last year. Especially I would like to thank my advisor, Prof. Johanna Stachel, for giving me an  interesting thesis project, and for all the advices I got from her.\\ 
Furthermore I would like to thank Dr. Anton Andronic, Prof. Peter Braun-Munzinger and Prof. Krzysztof Redlich for the fantastic support during this work.\\
I have had many helpful discussions with Rachik Soualah, Kai Schweda, Fabian Rühle and Stefan Zimmer. I also acknowledge the support of Eric Mueller, Felix Groitl and Katharina Koch.\\ \\
Und vor allem möchte ich an dieser Stelle meinen Eltern danken. Wenn auch die Frage "und was wird man dann, wenn man Physik studiert hat" nie zu eurer Zufriedenheit beantworted wurde, danke dass ihr mir dieses Studium ermöglicht habt.

\clearpage\thispagestyle{empty}
\begin{center}\textbf{\large Erklärung}\end{center}

\noindent
Hiermit versichere ich, dass ich die vorliegende Arbeit selbstständig verfasst und keine anderen als die angegebenen Quellen und Hilfsmittel benutzt habe.

\vspace{4\baselineskip}

\noindent
Heidelberg, den \today \qquad 


\begin{thebibliography}{99}
\bibitem{Andronic:2008ev}
  A.~Andronic, F.~Beutler, P.~Braun-Munzinger, K.~Redlich and J.~Stachel,
  %``Thermal description of hadron production in e+e- collisions revisited,''
  Phys. Lett. B (2009), doi: 10.1016/j.physletb.2009.04.024
  arXiv:0804.4132 [hep-ph].
  %%CITATION = ARXIV:0804.4132;%%
%\cite{Andronic:2009sv}
\bibitem{Andronic:2009sv}
  A.~Andronic, F.~Beutler, P.~Braun-Munzinger, K.~Redlich and J.~Stachel,
  %``Statistical hadronization of heavy flavor quarks in elementary collisions:
  %successes and failures,''
  arXiv:0904.1368 [hep-ph].
  %%CITATION = ARXIV:0904.1368;%%
%\cite{Redlich:2009xx}
\bibitem{Redlich:2009xx}
  K.~Redlich, A.~Andronic, F.~Beutler, P.~Braun-Munzinger and J.~Stachel,
  %``Canonical Statistical Model and hadron production in $e^+e^-$
  %annihilations,''
  arXiv:0903.1610 [hep-ph].
  %%CITATION = ARXIV:0903.1610;%%
\bibitem{Beutler}
  F.~Beutler,
  Bachelor thesis (not puplished yet)
\bibitem{Yao:2006px}
  W.~M.~Yao {\it et al.}  [Particle Data Group],
  %``Review of particle physics,''
  J.\ Phys.\ G {\bf 33} (2006) 1.
  %%CITATION = JPHGB,G33,1;%%
\bibitem{Becattini:1995if}
  F.~Becattini,
  %``A Thermodynamical Approach To Hadron Production In E+ E- Collisions,''
  Z.\ Phys.\  C {\bf 69} (1996) 485.
  %%CITATION = ZEPYA,C69,485;%%
\bibitem{Cleymans:1990ia}
  J.~Cleymans, K.~Redlich and E.~Suhonen,
  %``Strangeness production in ion collisions using exact strangeness
  %conservation,''
  %%CITATION = UCT-TP-132-90;%%
\bibitem{Cleymans:1991yu}
  J.~Cleymans, E.~Suhonen and G.~M.~Weber,
  %``Exact baryon and strangeness conservation in hadronic gas models,''
  Z.\ Phys.\  C {\bf 53} (1992) 485.
  %%CITATION = ZEPYA,C53,485;%%
\bibitem{Cleymans:1997ib}
  J.~Cleymans, M.~Marais and E.~Suhonen,
  %``Exact baryon, strangeness, and charge conservation in hadronic gas
  %models,''
  Phys.\ Rev.\  C {\bf 56} (1997) 2747
  [arXiv:nucl-th/9705014].
  %%CITATION = PHRVA,C56,2747;%%
\bibitem{Hagedorn:1984uy}
  R.~Hagedorn and K.~Redlich,
  %``Statistical Thermodynamics In Relativistic Particle And Ion Physics:
  %Canonical Or Grand Canonical?,''
  Z.\ Phys.\  C {\bf 27} (1985) 541.
  %%CITATION = ZEPYA,C27,541;%%
\bibitem{numerical:recipes}
  William~H.~Press, Saul~A.~Teukolsky, William~T.~Vetterling, Brian~P.~Flannery,
  "Numerical Recipes in C++. The Art of Scientific Computing,"
\bibitem{Rafelski:2002}
  Jean~Letessier, Johann~Rafelski,
  "Hadrons and Quark Gluon Plasma"
\bibitem{Kaczmarek:2005zn}
  O.~Kaczmarek and F.~Zantow,
  %``The screening length in hot QCD,''
  PoS {\bf LAT2005} (2006) 177
  [arXiv:hep-lat/0510093].
  %%CITATION = POSCI,LAT2005,177;%%
\bibitem{Keranen:2001pr}
  A.~Keranen and F.~Becattini,
  %``Chemical factors in canonical statistical models for relativistic heavy
  %ion collisions,''
  Phys.\ Rev.\  C {\bf 65} (2002) 044901
  [Erratum-ibid.\  C {\bf 68} (2003) 059901]
  [arXiv:nucl-th/0112021].
  %%CITATION = PHRVA,C65,044901;%%
\bibitem{Collins:1974ky}
  J.~C.~Collins and M.~J.~Perry,
  %``Superdense Matter: Neutrons Or Asymptotically Free Quarks?,''
  Phys.\ Rev.\ Lett.\  {\bf 34} (1975) 1353.
  %%CITATION = PRLTA,34,1353;%%
\bibitem{Satz:1985vb}
  H.~Satz,
  %``The Transition From Hadron Matter To Quark - Gluon Plasma,''
  Ann.\ Rev.\ Nucl.\ Part.\ Sci.\  {\bf 35} (1985) 245.
  %%CITATION = ARNUA,35,245;%%
\bibitem{Politzer:1973fx}
  H.~D.~Politzer,
  %``RELIABLE PERTURBATIVE RESULTS FOR STRONG INTERACTIONS?,''
  Phys.\ Rev.\ Lett.\  {\bf 30} (1973) 1346.
  %%CITATION = PRLTA,30,1346;%%
\bibitem{Gross:1973id}
  D.~J.~Gross and F.~Wilczek,
  %``ULTRAVIOLET BEHAVIOR OF NON-ABELIAN GAUGE THEORIES,''
  Phys.\ Rev.\ Lett.\  {\bf 30} (1973) 1343.
  %%CITATION = PRLTA,30,1343;%%
\bibitem{Karsch:2003jg}
  F.~Karsch and E.~Laermann,
  %``Thermodynamics and in-medium hadron properties from lattice QCD,''
  arXiv:hep-lat/0305025.
  %%CITATION = HEP-LAT/0305025;%%
\bibitem{Laermann:2003cv}
  E.~Laermann and O.~Philipsen,
  %``Status of lattice QCD at finite temperature,''
  Ann.\ Rev.\ Nucl.\ Part.\ Sci.\  {\bf 53} (2003) 163
  [arXiv:hep-ph/0303042].
  %%CITATION = ARNUA,53,163;%%
\bibitem{Karsch:2001cy}
  F.~Karsch,
  %``Lattice QCD at high temperature and density,''
  Lect.\ Notes Phys.\  {\bf 583} (2002) 209
  [arXiv:hep-lat/0106019].
  %%CITATION = LNPHA,583,209;%%
\bibitem{Satz:2002ku}
  H.~Satz,
  %``Limits of confinement: The first 15 years of ultra-relativistic heavy ion
  %studies,''
  Nucl.\ Phys.\  A {\bf 715} (2003) 3
  [arXiv:hep-ph/0209181].
  %%CITATION = NUPHA,A715,3;%%
\bibitem{BraunMunzinger:2003zd}
  P.~Braun-Munzinger, K.~Redlich and J.~Stachel,
  %``Particle production in heavy ion collisions,''
  arXiv:nucl-th/0304013.
  %%CITATION = NUCL-TH/0304013;%%
\bibitem{Stock:1999hm}
  R.~Stock,
  %``The parton to hadron phase transition observed in Pb + Pb collisions at
  %158-GeV per nucleon,''
  Phys.\ Lett.\  B {\bf 456} (1999) 277
  [arXiv:hep-ph/9905247].
  %%CITATION = PHLTA,B456,277;%%
\bibitem{Abreu:1994rg}
  P.~Abreu {\it et al.}  [DELPHI Collaboration],
  %``Production characteristics of K0 and light meson resonances in hadronic
  %decays of the Z0,''
  Z.\ Phys.\  C {\bf 65} (1995) 587.
  %%CITATION = ZEPYA,C65,587;%%
\bibitem{Alexander:1996qj}
  G.~Alexander {\it et al.}  [OPAL Collaboration],
  %``Strange baryon production in hadronic Z0 decays,''
  Z.\ Phys.\  C {\bf 73} (1997) 569.
  %%CITATION = ZEPYA,C73,569;%%
\bibitem{Alexander:1996qi}
  G.~Alexander {\it et al.}  [OPAL Collaboration],
  %``Sigma+, Sigma0 and Sigma- hyperon production in hadronic Z0 decays,''
  Z.\ Phys.\  C {\bf 73} (1997) 587.
  %%CITATION = ZEPYA,C73,587;%%
\bibitem{Akers:1994ez}
  R.~Akers {\it et al.}  [OPAL Collaboration],
  %``Measurement of the production rates of charged hadrons in e+ e-
  %annihilation at the Z0,''
  Z.\ Phys.\  C {\bf 63} (1994) 181.
  %%CITATION = ZEPYA,C63,181;%%
\bibitem{Braunschweig:1988hv}
  W.~Braunschweig {\it et al.}  [TASSO Collaboration],
  %``Pion, Kaon And Proton Cross-Sections In E+ E- Annihilation At 34-Gev And
  %44-Gev Center-Of-Mass Energy,''
  Z.\ Phys.\  C {\bf 42} (1989) 189.
  %%CITATION = ZEPYA,C42,189;%%
\bibitem{Althoff:1984iz}
  M.~Althoff {\it et al.}  [TASSO Collaboration],
  %``A Detailed Study Of Strange Particle Production In E+ E- Annihilation At
  %High-Energy,''
  Z.\ Phys.\  C {\bf 27} (1985) 27.
  %%CITATION = ZEPYA,C27,27;%%
\bibitem{Braunschweig:1988wh}
  W.~Braunschweig {\it et al.}  [TASSO Collaboration],
  %``Strange Baryon Production In E+ E- Annihilation,''
  Z.\ Phys.\  C {\bf 45} (1989) 209.
  %%CITATION = ZEPYA,C45,209;%%
\bibitem{Yndurain}
  F.~J.~Yndurain, "Quantum chromodynamics: an introduction to Theory of Quarks and Gluons", Springer, New York, 1983.
\bibitem{Altarelli:1989hv}
  G.~Altarelli, R.~Kleiss and C.~Verzegnassi,
  %``Z PHYSICS AT LEP-1. PROCEEDINGS, WORKSHOP, GENEVA, SWITZERLAND, SEPTEMBER
  %4-5, 1989. VOL. 1: STANDARD PHYSICS,''
%\href{http://www.slac.stanford.edu/spires/find/hep/www?irn=2184923}{SPIRES entry}
{\it  Geneva, Switzerland: CERN (1989) 453 p. CERN Geneva - CERN 89-08 (89,rec.Dec.) 453 p}
\bibitem{Chekanov:1994fn}
  S.~V.~Chekanov,
  %``Intermittency in cluster models: Correlation and fluctuation approaches,''
  Acta Phys.\ Polon.\  B {\bf 25} (1994) 1583.
  %%CITATION = APPOA,B25,1583;%%
\bibitem{Knowles:1995kj}
  I.~G.~Knowles {\it et al.},
  %``QCD Event Generators,''
  arXiv:hep-ph/9601212.
  %%CITATION = HEP-PH/9601212;%%
\bibitem{Sjostrand:1993yb}
  T.~Sjostrand,
  %``High-energy physics event generation with PYTHIA 5.7 and JETSET 7.4,''
  Comput.\ Phys.\ Commun.\  {\bf 82} (1994) 74.
  %%CITATION = CPHCB,82,74;%%
\bibitem{James:1980yn}
  F.~James,
  %``Monte Carlo Theory And Practice,''
  Rept.\ Prog.\ Phys.\  {\bf 43} (1980) 1145.
  %%CITATION = RPPHA,43,1145;%%
\bibitem{Andersson:1983ia}
  B.~Andersson, G.~Gustafson, G.~Ingelman and T.~Sjostrand,
  %``Parton Fragmentation And String Dynamics,''
  Phys.\ Rept.\  {\bf 97} (1983) 31.
  %%CITATION = PRPLC,97,31;%%
\bibitem{Bethke:2004bp}
  S.~Bethke,
  %``QCD studies at LEP,''
  Phys.\ Rept.\  {\bf 403-404} (2004) 203
  [arXiv:hep-ex/0406058].
  %%CITATION = PRPLC,403-404,203;%%
\bibitem{Knowles:1997dk}
  I.~G.~Knowles and G.~D.~Lafferty,
  %``Hadronization in Z0 decay,''
  J.\ Phys.\ G {\bf 23} (1997) 731
  [arXiv:hep-ph/9705217].
  %%CITATION = JPHGB,G23,731;%%
\bibitem{Webber:1983if}
  B.~R.~Webber,
  %``A QCD Model For Jet Fragmentation Including Soft Gluon Interference,''
  Nucl.\ Phys.\  B {\bf 238} (1984) 492.
  %%CITATION = NUPHA,B238,492;%%
\bibitem{Montvay:1979py}
  I.~Montvay,
  %``Where Are The Jets?,''
  Phys.\ Lett.\  B {\bf 84} (1979) 331.
  %%CITATION = PHLTA,B84,331;%%
\bibitem{Andersson:1978vj}
  B.~Andersson, G.~Gustafson and C.~Peterson,
  %``A Semiclassical Model For Quark Jet Fragmentation,''
  Z.\ Phys.\  C {\bf 1} (1979) 105.
  %%CITATION = ZEPYA,C1,105;%%
\bibitem{Andersson:1984af}
  B.~Andersson, G.~Gustafson and T.~Sjostrand,
  %``Baryon Production In Jet Fragmentation And Upsilon Decay,''
  Phys.\ Scripta {\bf 32} (1985) 574.
  %%CITATION = PHSTB,32,574;%%
\bibitem{Andersson:1981ce}
  B.~Andersson, G.~Gustafson and T.~Sjostrand,
  %``A Model For Baryon Production In Quark And Gluon Jets,''
  Nucl.\ Phys.\  B {\bf 197} (1982) 45.
  %%CITATION = NUPHA,B197,45;%%
\bibitem{Becattini:2004td}
  F.~Becattini,
  %``What is the meaning of the statistical hadronization model?,''
  J.\ Phys.\ Conf.\ Ser.\  {\bf 5} (2005) 175
  [arXiv:hep-ph/0410403].
  %%CITATION = 00462,5,175;%%
\bibitem{Han:1965pf}
  M.~Y.~Han and Y.~Nambu,
  %``Three-triplet model with double SU(3) symmetry,''
  Phys.\ Rev.\  {\bf 139} (1965) B1006.
  %%CITATION = PHRVA,139,B1006;%%
\bibitem{Greenberg:1964pe}
  O.~W.~Greenberg,
  %``Spin And Unitary Spin Independence In A Paraquark Model Of Baryons And
  %Mesons,''
  Phys.\ Rev.\ Lett.\  {\bf 13} (1964) 598.
  %%CITATION = PRLTA,13,598;%%
\bibitem{Fritzsch:1973pi}
  H.~Fritzsch, M.~Gell-Mann and H.~Leutwyler,
  %``Advantages Of The Color Octet Gluon Picture,''
  Phys.\ Lett.\  B {\bf 47} (1973) 365.
  %%CITATION = PHLTA,B47,365;%%
\bibitem{Aitchison:1989}
  Aitchison and Hey, Gauge Theories in Particle Physics. Adam Hilger, 2nd edition (1989)
\bibitem{Halzen:1984}
  Halzen and Martin, Quarks and Leptons. Wiley (1984)
\bibitem{Renton:1990}
  P.~Renton, Electroweak Interactions. Cambridge University Press (1990)
\bibitem{Sjostrand:2006za}
  T.~Sjostrand, S.~Mrenna and P.~Skands,
  %``PYTHIA 6.4 physics and manual,''
  JHEP {\bf 0605} (2006) 026
  [arXiv:hep-ph/0603175].
  %%CITATION = JHEPA,0605,026;%%
\bibitem{Sjostrand:2007gs}
  T.~Sjostrand, S.~Mrenna and P.~Skands,
  %``A Brief Introduction to PYTHIA 8.1,''
  Comput.\ Phys.\ Commun.\  {\bf 178} (2008) 852
  [arXiv:0710.3820 [hep-ph]].
  %%CITATION = CPHCB,178,852;%%
\bibitem{Corcella:2000bw}
  G.~Corcella {\it et al.},
  %``HERWIG 6: An event generator for hadron emission reactions with
  %interfering gluons (including supersymmetric processes),''
  JHEP {\bf 0101} (2001) 010
  [arXiv:hep-ph/0011363].
  %%CITATION = JHEPA,0101,010;%%
\bibitem{Green}
  M.~G.~Green,S.~L.~Lloyed,P.~N.~Ratoff and D.~R.~Ward,
  "Electron-Positron Physics at the Z"
\bibitem{Higgs:1964ia}
  P.~W.~Higgs,
  %``Broken symmetries, massless particles and gauge fields,''
  Phys.\ Lett.\  {\bf 12} (1964) 132.
  %%CITATION = PHLTA,12,132;%%
\bibitem{Quigg:1999xg}
  C.~Quigg,
  %``Electroweak symmetry breaking and the Higgs sector,''
  Acta Phys.\ Polon.\  B {\bf 30} (1999) 2145
  [arXiv:hep-ph/9905369].
  %%CITATION = APPOA,B30,2145;%%
\bibitem{Redlich:1981xs}
  K.~Redlich and L.~Turko,
  %``Phase Transition In Hadronic Matter With Internal Symmetry,''
  %%CITATION = CERN-TH-3053;%%
\bibitem{Becattini:1996gy}
  F.~Becattini,
  %``Universality of thermal hadron production in p p, p anti-p and e+ e-
  %collisions,''
  arXiv:hep-ph/9701275.
  %%CITATION = HEP-PH/9701275;%%
\bibitem{Turko:1981nr}
  L.~Turko,
  %``Quantum Gases With Internal Symmetry,''
  Phys.\ Lett.\  B {\bf 104} (1981) 153.
  %%CITATION = PHLTA,B104,153;%%
\bibitem{Redlich:1979bf}
  K.~Redlich and L.~Turko,
  %``Phase Transitions In The Statistical Bootstrap Model With An Internal
  %Symmetry,''
  Z.\ Phys.\  C {\bf 5} (1980) 201.
  %%CITATION = ZEPYA,C5,201;%%
\bibitem{Hagedorn:1970gh}
  R.~Hagedorn,
  %``Remarks on the thermodynamical model of strong interactions,''
  Nucl.\ Phys.\  B {\bf 24} (1970) 93.
  %%CITATION = NUPHA,B24,93;%%
\bibitem{Shuryak:1973pv}
  E.~V.~Shuryak,
  %``The multibody production at mediate energy,''
  Phys.\ Lett.\  B {\bf 42} (1972) 357.
  %%CITATION = PHLTA,B42,357;%%
\bibitem{Rafelski:1980gk}
  J.~Rafelski and M.~Danos,
  %``The Importance Of The Reaction Volume In Hadronic Collisions,''
  Phys.\ Lett.\  B {\bf 97} (1980) 279.
  %%CITATION = PHLTA,B97,279;%%
\bibitem{Turko:2000if}
  L.~Turko and J.~Rafelski,
  %``Dynamics of multiparticle systems with non-Abelian symmetry,''
  Eur.\ Phys.\ J.\  C {\bf 18} (2001) 587
  [arXiv:hep-th/0003079].
  %%CITATION = EPHJA,C18,587;%%
\bibitem{Redlich:2003dw}
  K.~Redlich, F.~Karsch and A.~Tounsi,
  %``Group projection method in statistical systems,''
  arXiv:hep-ph/0302245.
  %%CITATION = HEP-PH/0302245;%%
\bibitem{Goldstone:1962es}
  J.~Goldstone, A.~Salam and S.~Weinberg,
  %``Broken Symmetries,''
  Phys.\ Rev.\  {\bf 127} (1962) 965.
  %%CITATION = PHRVA,127,965;%%
\bibitem{Glashow:1961tr}
  S.~L.~Glashow,
  %``Partial Symmetries Of Weak Interactions,''
  Nucl.\ Phys.\  {\bf 22} (1961) 579.
  %%CITATION = NUPHA,22,579;%%
\bibitem{Fliessbach}
  T.~Fliessbach,
  "Lehrbuch zur Theoretischen Physik IV: Statistische Physik"
\bibitem{Andronic:2005yp}
  A.~Andronic, P.~Braun-Munzinger and J.~Stachel,
  %``Hadron production in central nucleus nucleus collisions at chemical
  %freeze-out,''
  Nucl.\ Phys.\  A {\bf 772}, 167 (2006)
  [arXiv:nucl-th/0511071].
  %%CITATION = NUPHA,A772,167;%%
\bibitem{Becattini:2003wp}
  F.~Becattini, M.~Gazdzicki, A.~Keranen, J.~Manninen and R.~Stock,
  %``Study of chemical equilibrium in nucleus nucleus collisions at AGS and  SPS
  %energies,''
  Phys.\ Rev.\  C {\bf 69} (2004) 024905
  [arXiv:hep-ph/0310049].
  %%CITATION = PHRVA,C69,024905;%%
\bibitem{Becattini:2001fg}
  F.~Becattini and G.~Passaleva,
  %``Statistical hadronization model and transverse momentum spectra of  hadrons
  %in high energy collisions,''
  Eur.\ Phys.\ J.\  C {\bf 23} (2002) 551
  [arXiv:hep-ph/0110312].
  %%CITATION = EPHJA,C23,551;%%
\bibitem{Casher:1978wy}
  A.~Casher, H.~Neuberger and S.~Nussinov,
  %``Chromoelectric Flux Tube Model Of Particle Production,''
  Phys.\ Rev.\  D {\bf 20} (1979) 179.
  %%CITATION = PHRVA,D20,179;%%
\bibitem{Cleymans:1990mn}
  J.~Cleymans, K.~Redlich and E.~Suhonen,
  %``Canonical description of strangeness conservation and particle
  %production,''
  Z.\ Phys.\  C {\bf 51} (1991) 137.
  %%CITATION = ZEPYA,C51,137;%%
\bibitem{Broniowski:2000bj}
  W.~Broniowski and W.~Florkowski,
  %``Different Hagedorn temperatures for mesons and baryons from  experimental
  %mass spectra, compound hadrons, and combinatorial  saturation,''
  Phys.\ Lett.\  B {\bf 490} (2000) 223
  [arXiv:hep-ph/0004104].
  %%CITATION = PHLTA,B490,223;%%
\bibitem{Broniowski:2000hd}
  W.~Broniowski,
  %``Distinct Hagedorn temperatures from particle spectra: A higher one for
  %mesons, a lower one for baryons,''
  arXiv:hep-ph/0008112.
  %%CITATION = HEP-PH/0008112;%%
\bibitem{Broniowski:2004yh}
  W.~Broniowski, W.~Florkowski and L.~Y.~Glozman,
  %``Update of the Hagedorn mass spectrum,''
  Phys.\ Rev.\  D {\bf 70} (2004) 117503
  [arXiv:hep-ph/0407290].
  %%CITATION = PHRVA,D70,117503;%%
\bibitem{Peterson:1982ak}
  C.~Peterson, D.~Schlatter, I.~Schmitt and P.~M.~Zerwas,
  %``Scaling Violations In Inclusive E+ E- Annihilation Spectra,''
  Phys.\ Rev.\  D {\bf 27} (1983) 105.
  %%CITATION = PHRVA,D27,105;%%
\bibitem{Becattini:2008tx}
  F.~Becattini, P.~Castorina, J.~Manninen and H.~Satz,
  %``The Thermal Production of Strange and Non-Strange Hadrons in e+e-
  %Collisions,''
  arXiv:0805.0964 [hep-ph].
  %%CITATION = ARXIV:0805.0964;%%
\bibitem{agssps} P. Braun-Munzinger, J.  Stachel, J.P. Wessels and
  N. Xu, Phys. Lett. B {\bf 344} (1995) 43 [arXiv:nucl-th/9410026] and 
{\bf 365} (1996) 1 [arXiv:nucl-th/9508020].
\bibitem{satz} J. Cleymans, D. Elliott, H. Satz, and R.L. Thews, Z. Phys. C
{\bf 74} (1997) 319 [arXiv:nucl-th/9603004].
\bibitem{heppe} P. Braun-Munzinger, I. Heppe and J. Stachel, Phys. Lett. B
{\bf 465} (1999) 15 [arXiv:nucl-th/9903010].
\bibitem{cley} J. Cleymans and K. Redlich, Phys. Rev. C {\bf 60} (1999) 054908
[arXiv:nucl-th/9903063].
\bibitem{beca1} F. Becattini, J. Cleymans, A. Keranen, E. Suhonen, and
K. Redlich, Phys. Rev. C {\bf 64} (2001) 024901 [arXiv:hep-ph/0002267].
\bibitem{rhic} P. Braun-Munzinger, D. Magestro, K. Redlich, and
  J. Stachel, Phys. Lett. B {\bf 518} (2001) 41 [arXiv:hep-ph/0105229].
\bibitem{nu} N. Xu and M. Kaneta, Nucl. Phys. A {\bf 698} (2002) 306c.
\bibitem{beca2} F. Becattini, J. Phys. G {\bf 28} (2002) 1553.
\bibitem{rapp} R. Rapp and E. Shuryak, Phys. Rev. Lett. {\bf 86} (2001) 2980
[arXiv:hep-ph/0008326].
\bibitem{becgaz} F. Becattini, M. Gazdzicki, J. Manninen, Phys. Rev. C {\bf 73}
(2006) 044905 [arXiv:hep-ph/0511092].
\bibitem{Becattini:1997uf}
  F.~Becattini,
  %``Thermal hadron production in high energy collisions,''
  J.\ Phys.\ G {\bf 23} (1997) 1933
  [arXiv:hep-ph/9708248].
  %%CITATION = JPHGB,G23,1933;%%
\bibitem{heinz} U. Heinz, Nucl. Phys. A {\bf 685} (2001) 414 
[arXiv:hep-ph/0009170].
\bibitem{castorina} P. Castorina, D. Kharzeev, H. Satz, 
Eur. Phys. J. C {\bf 52} (2007) 187 [arXiv:0704.1426].
\bibitem{wetterich} P. Braun-Munzinger, J. Stachel, C. Wetterich,
  Phys. Lett. B {\bf 596} (2004) 61 [arXiv:nucl-th/0311005].
\bibitem{cle91} %canonical
J. Cleymans, K. Redlich, E. Suhonen, Z. Phys. C {\bf 51} (1991) 137.
\bibitem{pbm4} P. Braun-Munzinger, J. Cleymans, H. Oeschler, K. Redlich,  
Nucl. Phys. A {\bf 697} (2002) 902 [arXiv:hep-ph/0106066].
\bibitem{:2005ema}
  [ALEPH Collaboration and DELPHI Collaboration and L3 Collaboration and OPAL Collaboration],
  %``Precision electroweak measurements on the Z resonance,''
  Phys.\ Rept.\  {\bf 427} (2006) 257
  [arXiv:hep-ex/0509008].
  %%CITATION = PRPLC,427,257;%%
\bibitem{Wheaton:2004qb}
  S.~Wheaton and J.~Cleymans,
  %``THERMUS: A thermal model package for ROOT,''
  arXiv:hep-ph/0407174; J. Phys. G {\bf 31} (2005) S1069.
  %%CITATION = HEP-PH/0407174;%%
\bibitem{minuit} MINUIT, Function Minimization and Error Analysis, 
CERN Program Library Long Writeup D506, http://wwwasdoc.web.cern.ch/wwwasdoc/minuit/
\bibitem{phobos} B.B. Back et al., PHOBOS coll., arXiv:nucl-ex/0301017.
\bibitem{cley2} J. Cleymans, M. Stankiewicz, P. Steinberg, S. Wheaton, 
arXiv:nucl-th/0506027.
\bibitem{basile} M. Basile et al., Phys. Lett. B {\bf 92} (1980) 367,
  Phys. Lett. B {\bf 95} (1980) 311. 
\bibitem{minuit2} Fred James, Matthias Winkler, Minuit User's Guide, CERN, Geneva

\end{thebibliography}
\end{document}